\documentclass[a4paper, BCOR=9mm, DIV=14, 
11pt, headinclude
]{scrbook}

\usepackage[utf8]{inputenc}
\usepackage[headsepline, 
automark]{scrpage2}

\usepackage{layout}
\usepackage{amsmath}
\usepackage{amssymb}
\usepackage{amsfonts}
\usepackage{bbold}
\usepackage{multirow}
\usepackage[smalltableaux, centertableaux]{ytableau}
\usepackage{cite}
\usepackage{caption}
\usepackage{tabu}

\usepackage[
linkcolor=blue,
citecolor=blue,
urlcolor=blue,
]{hyperref}

\setkomafont{disposition}{\normalcolor\bfseries}

\ohead{\pagemark}
\ihead{\headmark}
\ofoot[]{}
\setheadsepline{0.6pt} 
\allowdisplaybreaks[2]

\newlength{\oldparskip}
\def\a{{\mathfrak{a}}}

\def\g{{\mathfrak{g}}}
\def\f{{\mathfrak{f}}}
\def\k{{\mathfrak{k}}}
\def\h{{\mathfrak{h}}}
\def\u{{\mathfrak{u}}}
\def\x{{\mathfrak{x}}}

\def\U{{\mathrm{U}}}
\def\su{{\mathfrak{su}}}
\def\SU{{\mathrm{SU}}}
\def\so{{\mathfrak{so}}}
\def\SO{{\mathrm{SO}}}

\def\SL{{\mathrm{SL}}}
\def\usp{{\mathfrak{usp}}}
\def\USp{{\mathrm{USp}}}
\def\OSp{{\mathrm{OSp}}}

\def\E{{\mathrm{E}}}

\def\AdS{\textrm{AdS}}

\newcommand{\cO}{\mathcal{O}}
\newcommand{\cT}{\mathcal{T}}

\newcommand{\cP}{\mathcal{P}}
\newcommand{\cC}{\mathcal{C}}
\newcommand{\cD}{\mathcal{D}}
\newcommand{\cL}{\mathcal{L}}
\newcommand{\cS}{\mathcal{S}}
\newcommand{\cK}{\mathcal{K}}
\newcommand{\cM}{\mathcal{M}}
\newcommand{\cN}{\mathcal{N}}

\newcommand{\cH}{\mathcal{H}}

\newcommand{\cF}{\mathcal{F}}

\newcommand{\cR}{\mathcal{R}}
\newcommand{\cV}{\mathcal{V}}
\newcommand{\cQ}{\mathcal{Q}}
\newcommand{\cX}{\mathcal{X}}

\newcommand{\dd}{\mathrm{d}}
\newcommand{\D}{\mathrm{D}}

\newcommand{\cc}{\mathrm{C}}

\newcommand{\id}{{\mathbb1}}
\newcommand{\bbN}{\mathbb{N}}

\newcommand{\al}[1]{{\alpha_{#1}}}
\newcommand{\hal}[1]{{\hat\alpha_{#1}}}
\newcommand{\tal}[1]{{\tilde\alpha_{#1}}}
\newcommand{\be}[1]{{\beta_{#1}}}
\newcommand{\hbe}[1]{{\hat\beta_{#1}}}
\newcommand{\tbe}[1]{{\tilde\beta_{#1}}}
\newcommand{\ga}[1]{{\gamma_{#1}}}
\newcommand{\hga}[1]{{\hat\gamma_{#1}}}
\newcommand{\tga}[1]{{\tilde\gamma_{#1}}}

\newcommand{\hde}[1]{{\hat\delta_{#1}}}
\newcommand{\tde}[1]{{\tilde\delta_{#1}}}

\DeclareMathOperator{\tr}{tr}

\begin{document}

\frontmatter

\begin{titlepage}
\KOMAoptions{twoside = false, BCOR = 0mm}
\thispagestyle{empty}
\begin{center}

\vskip 1cm

{\Large \bfseries Maximally Supersymmetric AdS Solutions in Gauged Supergravity \footnote{This article contains the Ph.D.~thesis of the author, submitted at the University of Hamburg in August 2017.
The full text of the thesis is also available at \url{http://ediss.sub.uni-hamburg.de/volltexte/2017/8801/}.}}
\vskip 1.2cm

{\bfseries  Severin L\"ust \footnote{From November 1, 2017: Centre de Physique Th\'eorique, \'Ecole Polytechnique, CNRS,
91128 Palaiseau, France ({\ttfamily severin.luest@polytechnique.edu}).}}

\vskip 0.8cm

{\em Fachbereich Physik der Universit\"at Hamburg, \\ Luruper Chaussee 149, 22761 Hamburg, Germany}
\vskip 0.3cm

{\em Zentrum f\"ur Mathematische Physik,
Universit\"at Hamburg,\\
Bundesstrasse 55, D-20146 Hamburg, Germany}
\vskip 0.3cm

\vskip 0.3cm

{\ttfamily severin.luest@desy.de}

\end{center}

\vskip 1cm

\begin{center} {\bfseries ABSTRACT } \end{center}

In this thesis we study maximally supersymmetric solutions of gauged supergravity theories, with special focus on anti-de Sitter solutions.
The latter are relevant in the context of the AdS/CFT correspondence.

In the first part we classify all maximally supersymmetric solutions of all gauged or deformed supergravity theories in \(D \geq 3\) space-time dimensions.
Without background fluxes the space-time background has to be either flat or AdS$_D$.
Solutions with non-trivial fluxes are only possible for a small class of theories and we derive a simple criterion for their existence.
These solutions coincide with those of the corresponding ungauged theories, therefore the known list of maximally supersymmetric solutions is exhaustive.

In the second part we exclusively study maximally supersymmetric AdS$_D$ solutions of gauged supergravities in dimensions \(D \geq 4\).
We show that such solutions can only exist if the gauge group after spontaneous symmetry breaking is of the form \(H_R \times H_\mathrm{mat}\).
This resembles the structure of the global symmetry groups of the holographically dual SCFTs, where \(H_R\) is interpreted as the R-symmetry.
Moreover, we discuss possible supersymmetry preserving continuous deformations.
The moduli spaces spanned by these deformations correspond to the conformal manifolds of the dual SCFTs.
Under the assumption that the scalar manifold of the supergravity is a symmetric homogeneous space we derive general conditions on the moduli.
In particular, we show that they have to be singlets with respect to \(H_R\).
Using these results we determine the AdS solutions of all gauged supergravities with more than 16 real supercharges.
We show that almost all of them do not have supersymmetry preserving deformations.
Only the AdS solutions of maximal supergravity in five dimensions have a non-trivial moduli space given by \(\SU(1,1)/\U(1)\).
Furthermore, we determine the AdS solutions of four-dimensional \(\cN = 3\) supergravities and seven-dimensional half-maximal supergravities and show that they do not admit supersymmetric moduli as well.
We confirm the holographic interpretation of the latter result and show that the existence of supersymmetric marginal deformations of six-dimensional $(1,0)$ SCFTs is forbidden by the representation theory of the underlying superconformal algebra.

\noindent

\vfill



%
%
%
%
%
%
%
%
%
%
\end{titlepage}

\thispagestyle{empty}

\thispagestyle{empty}
\clearpage
\thispagestyle{empty}

\tableofcontents
\thispagestyle{empty}
\addtocontents{toc}{\protect\thispagestyle{empty}}

\mainmatter\setcounter{page}{1}

\chapter{Introduction}\label{chap:introduction}


String theory is one of the most promising candidates for a sensible theory of quantum gravity as well as for a unified theory.
However, so far it has not been possible to find any experimental evidence for its realization in nature.
Nonetheless, besides its possible relevance for particle physics, string theory provides a lot of conceptual and fundamental insight into various branches and problems of theoretical physics.

A concept that turns out to be ubiquitous in string theory are dualities,
which are -- roughly speaking -- equivalences between theories that seem to be different.
Dualities are often very powerful tools as they allow for the analysis of a problem from different points of view and therefore yield additional insight into the involved theories.
One of the most interesting dualities arising from string theory is the AdS/CFT correspondence \cite{Maldacena:1997re, Witten:1998qj, Gubser:1998bc}.
This correspondence is a holographic duality, which means that it relates a theory containing gravity to a non-gravitational theory in one dimension less.
Holography is assumed to be one of the hallmarks of quantum gravity.%
\footnote{For a review of the holographic principle see e.g. \cite{Bousso:2002ju}.}

Modern theoretical physics is guided by the study of symmetries.
The symmetry which features most prominently in the AdS/CFT correspondence is the
conformal symmetry.
A conformal field theory (CFT) is characterized by its invariance with respect to all space-time transformations which locally preserve angles.
These conformal transformations are mathematically described by the conformal group \(\SO(d,2)\), where \(d\) denotes the number of space-time dimensions. 
An important consequence is that a CFT is scale independent, 
this means that 
it shows the same behavior at all length and energy scales.
Independent of their role in the AdS/CFT correspondence conformal field theories are a very interesting subject to study on their own as they often allow for exact analytic computations 
which are usually not possible for less symmetric (but more realistic) theories. 
Therefore, they can serve as
toy models for the conceptual understanding of quantum field theories.

Another theory which shares the same symmetries with a CFT is a gravitational theory on a  \((d+1)\)-dimensional anti-de Sitter (AdS) background space-time.
Anti-de Sitter space is a Lorentzian manifold with constant negative curvature.
Here the conformal group \(\SO(d,2)\) is realized as the isometry group of the AdS space-time.
The AdS/CFT correspondence conjectures that this agreement of symmetries is not only accidental but that there is indeed a duality between a suitable gravitational theory on an AdS background and a CFT.
The dual CFT is conjectured to live on the 
boundary of the AdS space on which the \(\SO(d,2)\) isometries act as conformal transformations.
This is the reason why the AdS/CFT correspondence is called a holographic duality.
All physics in the \((d+1)\)-dimensional volume (or bulk) of the AdS space-time is encoded in its \(d\)-dimensional boundary.

As a holographic duality the AdS/CFT correspondence in its most rigorous form is expected to be valid only in the regime of quantum gravity.
Therefore, we want to embed it into string theory,
which is also the framework where it has been discovered and explained first.
However, most consistent formulations of string theory require the introduction of another symmetry, namely supersymmetry, which relates particles of different spin. 
Moreover, 
string theory requires the existence of ten (or eleven) space-time dimensions.
It is therefore necessary to compactify some of these additional dimensions on a suitable internal space which we call \(Y\).
This means that we consider background space-times \(M_{10/11}\) of the form\footnote{In fact the product ansatz \eqref{eq:stringbackground} is not the most general ansatz preserving the \(SO(d,2)\) isometry.
It can be replaced by a warped product, i.e.~by a fibration of \(AdS_{(d+1)}\) over \(Y\).}
\begin{equation}\label{eq:stringbackground}
M_{10/11} = AdS_{(d+1)} \times Y_{(9/10-d)} \,,
\end{equation}
where \(Y\) is a compact manifold which is chosen in such a way that at least some of the supersymmetries of the ten (or eleven) dimensional string theory are preserved.
This is not possible for every \(d\) and highly restricts the geometry of \(Y\).
This setup is conjectured to be dual to a supersymmetric conformal field theory (SCFT) in $d$ dimensions.
Again the symmetry group of this SCFT should correspond to the isometry group of \(M_{10/11}\) which is given by \(\SO(d,2) \times H_R\) where \(H_R\) is the isometry group of \(Y\).
\(H_R\) takes the role of the R-symmetry of the SCFT.

It is now possible to discuss various limits of this duality.
The point of view we want to take here is to replace string theory by its low energy limit. 
This approximation is valid as long as the length of a string is small compared to the typical length scale  of the AdS background (often referred to as the AdS radius \(L\)), i.e.~for weakly curved backgrounds.
In this limit string theory effectively behaves as a classical theory of gravity which, if it is combined with supersymmetry, is called supergravity.
Moreover, the background \eqref{eq:stringbackground} is
a solution of the classical equations of motions of this theory.
Of course the replacement of string theory with classical supergravity has to be accompanied with an appropriate limit on the field theory side of the duality,
enforcing the dual SCFT to be strongly coupled and the rank \(N\) of its gauge group to be large.
Therefore, the supergravity limit can be used as a powerful tool to probe regions in the parameter space of an SCFT which are inaccessible by standard perturbation theory.

Given a theory that is invariant under a certain symmetry -- in our case supersymmetry and conformal invariance -- it is often an important question to ask whether and  how it is possible to modify or deform this theory without destroying its symmetry.
Especially, we expect two theories which are related by some duality to share the same set of deformations.
The objective of this thesis is to study symmetry preserving deformations of superconformal field theories
and of their dual AdS solutions
in the context of the AdS/CFT correspondence. 

One possible way of deforming a conformal field theory is to add some local operators \(\cO_i(x)\) to its Lagrangian \(\cL\), i.e.
\begin{equation}\label{eq:CFTdeformation}
\cL \rightarrow \cL + \lambda^i \cO_i \,,
\end{equation}
where \(\lambda^i\) are coupling constants which parametrize the deformation.%
\footnote{Note that a well-defined CFT does not necessarily require the existence of a Lagrangian description.
However, even for non-Lagrangian theories it is still possible to give a sensible meaning to the deformation \eqref{eq:CFTdeformation} by means of conformal perturbation theory, which is expected to be valid within a small but finite range of the coupling constants \(\lambda^i\) (see e.g. \cite{Kol:2002zt} for a short review).}
An important characteristic of an operator \(\cO\) is its conformal or scaling dimension \(\Delta_\cO\) which describes how the operator scales under space-time dilatations \(x^\mu \rightarrow \alpha x^\mu\).
This readily implies that the coupling constants scale as \(\lambda^i \rightarrow \alpha^{(d-\Delta_{\cO_i})} \lambda^i\), where \(d\) is the dimension of space-time.
Hence, the value of the scaling dimension in relation to the space-time dimension \(d\) determines the qualitative behavior of \(\lambda^i\) under the renormalization group (RG) flow and one commonly distinguishes between the following three cases:
\begin{enumerate}
\item[a)] $\Delta_{\cO_i} > d$: \(\lambda^i\) decreases when flowing to smaller energies and the deformation \eqref{eq:CFTdeformation} becomes less relevant in the IR: \(\cO_i\) is called an \emph{irrelevant} deformation.
\item[b)] $\Delta_{\cO_i} < d$: \(\lambda^i\) increases along an RG flow towards the IR and \eqref{eq:CFTdeformation} drives the theory away from a UV fixed point: \(\cO_i\) is called a \emph{relevant} deformation.
\item[c)] $\Delta_{\cO_i} = d$: \(\lambda^i\) is scale invariant, at least at leading order in perturbation theory. \(\cO_i\) is called a \emph{marginal} deformation.
\end{enumerate}
Clearly, a deformation that preserves conformal invariance must necessarily be marginal, since introducing a scale dependent parameter into the theory breaks its invariance under dilatations and therefore conformal invariance.
This is however not enough as the dimension of an operator might become renormalized at higher orders in \(\lambda^i\).
Therefore, we furthermore divide the marginal deformations into marginally irrelevant, marginally relevant and exactly marginal deformations according to the value of their renormalized (or anomalous) dimension.
It is precisely the \emph{exactly marginal} deformations we are interested in. 
Their conformal dimension is preserved under renormalization so adding them to the Lagrangian does not break conformal invariance.
We might say that a marginal deformation satisfies \(\Delta_{\cO_i} = d\) precisely at \(\lambda^i = 0\), while an exactly marginal deformation is marginal also at finite values of \(\lambda^i\).

This means that whenever it is possible to deform a CFT by exactly marginal deformations it is not anymore an isolated point in the space of theories but belongs to a continuous family of theories parametrized by the couplings \(\lambda^i\).
This phenomenon is captured by the notion of the \emph{conformal manifold} \(\cC\) which is defined to be the space spanned by the exactly marginal couplings \(\lambda^i\), i.e.
\begin{equation}
\cC = \bigl\{\lambda^i \, | \, \text{\(\cO_i\) is exactly marginal} \bigr\} \,.
\end{equation}
Moreover, the characteristic form of two point functions in CFTs induces a natural metric on \(\cC\) which gives it the structure of a Riemannian manifold \cite{Zamolodchikov:1986gt}.
It is called the Zamolodchikov metric and is given by
\begin{equation}
g_{ij}(\lambda) = |x|^{2 d} \bigl<\cO_i(x)\cO_j(0)\bigr>_{\lambda} \,.
\end{equation}
Conformal field theories are often understood or constructed as fixed points of an RG flow.
In this context the conformal manifold corresponds to a continuous connected family of fixed points, i.e.~a fixed line, surface or higher-dimensional equivalent. 

If the theory under consideration is not only conformal but also supersymmetric (i.e.~a SCFT) we are furthermore interested in those deformations \eqref{eq:CFTdeformation} which do not only preserve conformal invariance but also do not break supersymmetry.
This 
means that we impose the additional constraint
\begin{equation}
\bigl[Q, \cO_i\bigr] = \partial_\mu (\dots) \,,
\end{equation}
where \(Q\) denotes the supercharges of the SCFT and \(\partial_\mu (\dots)\) the total derivative of a well-defined operator (which might be vanishing).
These deformations are called \emph{supersymmetric (exactly) marginal}.
In an SCFT 
the existence of such deformations is often heavily restricted;
especially for theories with a large number of supersymmetries it is not uncommon that there are no supersymmetric marginal deformations at all.

Let us now turn to the supergravity side of the AdS/CFT correspondence.
Here, an exactly marginal deformation corresponds to a continuous parameter of the classical solution \eqref{eq:stringbackground}, such that a variation of this parameter does not change the \(AdS_{(d+1)}\)-part of the solution and therefore keeps its \(\SO(d,2)\) isometry intact.
Of course other possible fields and especially the metric of the internal space \(Y\) might depend non-trivially on the deformation parameter.
Such a deformation parameter is called a \emph{modulus}.
If we are additionally interested in supersymmetric marginal deformations, we must further restrict to only those deformations which do not break the supersymmetry of the solution.
This in turn implies that the corresponding parameters do not alter the isometries of \(Y\) either.
The space spanned by all (supersymmetric) deformation parameters, called the \emph{moduli space} \(\cM_{AdS}\), is the dual object to the conformal manifold \(\cC\).
It can be understood as a continuous family of supergravity solutions, all featuring an AdS\(_{(d+1)}\) factor and preserving the same amount of supersymmetry. 

However, due to the non-linear nature of the involved equations the direct analysis of (exactly marginal) deformations of supergravity solutions can be a rather involved task.
One possible approach to this problem is to expand the solution perturbatively in the deformation parameters and to solve the resulting equations order by order.
However, at higher orders this also becomes increasingly difficult and therefore this approach often does not allow for an exact treatment (see e.g.~\cite{Aharony:2002hx}).
Another possibility is to neglect most of the difficult \(Y\)-dependence of the solution and to work exclusively in the framework of \((d+1)\)-dimensional supergravity.
This is the approach we want to follow here.

As explained above, the holographically dual supergravity background is a classical solution of a ten or eleven dimensional supergravity of the form \eqref{eq:stringbackground}.
In the spirit of Kaluza-Klein theory such a background can be equivalently described in terms of a lower-dimensional theory in \((d+1)\) dimensions.
This is achieved by expanding all higher-dimensional fields in terms of eigenmodes of the appropriate differential operators on the internal space \(Y\).
As a result the lower-dimensional theory will include an infinite tower of massive fields.
However, after a suitable truncation of this spectrum it is possible to keep only a finite subset of the massive modes and the resulting theory can be described in terms of a gauged \((d+1)\)-dimensional supergravity theory.
Its gauge group is contained in the isometry group \(H_R\) of the internal space \(Y\).
It is now a relevant question which properties of the original higher-dimensional solution are preserved in its lower-dimensional description.
Of course we are especially interested in the moduli space of the \((d+1)\)-dimensional AdS background and its relation to the conformal manifold of the dual SCFT.

The topic of this thesis is to study general gauged supergravity theories and their supersymmetric AdS solutions as a subject on their own without any reference to a possible higher dimensional origin.
One should keep in mind that it is a priori not clear to which extend this approach can yield a sensible dual supergravity description of a SCFT and its marginal deformations.
Going from the full ten or eleven-dimensional solution to a lower-dimensional description in terms of a gauged supergravity requires the truncation of infinitely many modes and therefore always comes with a loss of information.
In the worst case some of the marginal deformations could correspond to truncated modes and might therefore not be visible in the gauged supergravity.
On the other hand this approach could also be considered to be more general.
It is not known if every gauged supergravity possesses a higher-dimensional origin, hence also some of their AdS solutions might not be directly related to a solution of the form \eqref{eq:stringbackground} and could therefore belong to a more general class of solutions.

We want to elaborate a bit more on the dual description of the conformal manifold on the gravity side of the AdS/CFT correspondence.
For this purpose we recall the field-operator map \cite{Witten:1998qj}.
It assigns to each scalar field of mass \(m\) living on an AdS\(_{(d+1)}\) background a scalar operator \(\cO\) of the boundary CFT.
The dimension \(\Delta_\cO\) of \(\cO\) is related to the mass of the scalar field via
\begin{equation}
\Delta_{\cO} = \frac{d}{2} + \sqrt{\frac{d^2}{4} + m^2 L^2} \,,
\end{equation}   
where \(L\) denotes the AdS radius, i.e.~the characteristic length scale of the background.
Therefore, we see that a scalar operator of dimension \(\Delta_\cO = d\), i.e.~a marginal deformation, corresponds to a massless scalar field \(\phi\).
The asymptotic value of \(\phi\) near the AdS boundary and consequently its background or vacuum expectation value determines the value of the corresponding coupling constant \(\lambda\) in \eqref{eq:CFTdeformation}. 
When does such a massless scalar field correspond not only to a marginal but also to an exactly marginal deformation?
According to our considerations above, this is precisely the case whenever a change of the background value of \(\phi\) does not destroy the conformal invariance of the boundary theory and therefore leaves the \(\SO(d,2)\) isometry of the AdS background intact.
This means that the scalar potential 
must not depend on \(\phi\).
In analogy to our previous discussion we could say that a massless scalar field corresponds to an exactly marginal deformation precisely if it remains massless under a change of its background value.

In an AdS solution the background value of the scalar fields must be independent of the space-time coordinates and hence must be a local minimum of the potential.
Consequently, if the potential in the neighbourhood of its minimum is independent of some scalar fields,
these fields parametrize a continuous family of minima and can be regarded as continuous deformation parameters of the solution.
Therefore, the moduli space \(\cM_{AdS}\) is nothing but a continuous family of minima of the scalar potential.


Clearly, it is not guaranteed that an arbitrary minimum of the potential of a supergravity theory corresponds to a solution which preserves some or all of 
the supersymmetries. 
Generically, this is not the case.
The same holds true for an arbitrary scalar deformation.
Therefore, if we are interested in supersymmetric AdS solutions and their moduli spaces we have to impose additional conditions, namely the vanishing of the supersymmetry variations of all fermionic fields.
Under an infinitesimal supersymmetry transformation, described by an infinitesimal parameter \(\epsilon\), the fermions 
vary schematically as
\begin{equation}\label{eq:schematicvariations}
\delta \psi_\mu = \nabla\!_\mu \epsilon + A_0(\phi) \gamma_\mu \epsilon \,,\qquad \delta \chi = A_1(\phi) \epsilon \,,
\end{equation}
where \(\psi_\mu\) denotes the gravitini and \(\chi\) all other fermionic fields.
Moreover, the objects \(A_0(\phi)\) and \(A_1(\phi)\), called shift matrices, generically depend on all the scalar fields present in the theory.
For \eqref{eq:schematicvariations} to vanish the spinorial parameter \(\epsilon\) has to be a Killing spinor, which means that it satisfies the Killing spinor equation \(\nabla_\mu \epsilon = \alpha \gamma_\mu \epsilon\), where \(\alpha\) is a constant proportional to the scalar curvature of the background space-time.
Therefore, the background values \(\phi_0\) of the scalar fields in a maximally supersymmetric AdS solution (i.e.~a solution which does not break any of the supersymmetries) must satisfy
\begin{equation}\label{eq:susyconditions}
A_0 (\phi_0) \sim L^{-1} \,,\qquad A_1(\phi_0) = 0 \,.
\end{equation}
Conveniently, these conditions already guarantee that \(\phi_0\) is a minimum of the potential.
Analogously, a supersymmetric deformation of the AdS solution corresponds to a continuous family of solutions of \eqref{eq:susyconditions}

With increasing numbers of supersymmetries the conditions \eqref{eq:susyconditions} become more and more restrictive, and -- similarly as for SCFTs -- often completely forbid the existence of supersymmetric moduli. 
There are two main cases we want to focus on in this thesis.
Firstly, for supergravities with more than eight (real) supersymmetries the scalar fields are always coordinates on a symmetric homogeneous space of the form \(G/H\) where \(G\) and \(H\) are Lie groups.
This allows us to analyze \eqref{eq:susyconditions} by purely group theoretical methods.
Secondly, if there are even more than sixteen supersymmetries, the field content of the theory is completely fixed and there is no freedom left in the choice of \(G\) and \(H\), which further simplifies the analysis considerably.

Let us now give a short outline and summary of this thesis.
After a review of some generic features of (gauged) supergravities 
we begin with a general discussion of maximally supersymmetric backgrounds for all gauged und ungauged supergravity theories in \(D \geq 3\) space-time dimensions.
This analysis is guided by the analysis of Killing spinor equations.
They have to admit an independent solution for each preserved supercharge which for the case of unbroken supersymmetry considerably constrains the allowed space-time backgrounds.
This allows us to give a complete classification of all maximally supersymmetric backgrounds.
We distinguish between the following two cases.

If the metric and the scalars are the only fields with a non-trivial background value, i.e.~without background fluxes, the Killing spinor equations take a very simple form such that they can be integrated directly.
The allowed space-time backgrounds are locally maximally symmetric and thus are locally isometric either to flat Minkowski space-time \(M_D\) or to anti-de Sitter space AdS\(_D\).
The former case also includes toroidal compactifications of the form \(M_d \times T^{(D-d)}\) and is possible in ungauged as well as gauged supergravity.
AdS\(_D\) solutions, however, require a non-trivial potential which gives rise to a negative cosmological constant.
Therefore, they can only exist in gauged or otherwise deformed supergravity.

To obtain more complicated solutions one has to allow for non-vanishing background fluxes in the gravitational multiplet.
As fluxes we understand non-trivial background values of the field strengths of $p$-form gauge potentials.
However, in case there are any spin-1/2 fermions present in the gravitational multiplet, these fluxes generically break supersymmetry (at least partially).
This breaking of supersymmetry can only be avoided by gauge potentials with (anti-)self-dual field strengths since they drop out of the supersymmetry variations of chiral spin-1/2 fermions.
We conclude that maximally supersymmetric solutions with fluxes are only possible for a small set of theories where there are either no spin-1/2 fermions 
in the gravitational multiplet or where the theory is chiral and allows for (anti-)self-dual fluxes.
Moreover, we argue that for all theories which satisfy this criterion the solutions of a gauged theory correspond to the solutions of the corresponding ungauged theory.
These have all been determined and classified \cite{Tod:1983pm, Gauntlett:2002nw, FigueroaO'Farrill:2002ft, Gutowski:2003rg, Chamseddine:2003yy} and are either of the Freund-Rubin form \(AdS_d \times S^{(D-d)}\) \cite{Freund:1980xh} or H\textit{pp}-wave solutions \cite{KowalskiGlikman:1984wv,KowalskiGlikman:1985im}, up to local isometry.
The only exception occurs in five space-time dimensions where more exotic solutions are possible \cite{Gauntlett:2002nw}.

In the next step we turn to our main subject of interest and focus specifically on maximally supersymmetric AdS\(_D\) solutions in gauged supergravity theories in dimensions \(D \geq 4\).
In our previous analysis we found general conditions on the fermionic shift matrices for the existence of such a background.
These conditions impose very generic constraints on the admissible gauge groups \(G^g\).
The most characteristic feature of the gauge group is that it always contains a reductive subgroup \(H^g_R\) which is solely generated by the vector fields in the gravitational multiplet, i.e.~the graviphotons.
\(H^g_R\) is uniquely determined to be the maximal subgroup of the R-symmetry group \(H_R\) such that it can be gauged by the graviphotons and such that the gravitino mass matrix is invariant with respect to \(H^g_R\).
Furthermore, in the vacuum the gauge group \(G^g\) must be spontaneously broken to a reductive subgroup \(H^g = H^g_R \times H^g_\mathrm{mat}\).
The second factor \(H^g_\mathrm{mat}\) is unconstrained by the conditions on the shift matrices but can only be gauged by vector multiplets.
Under the AdS/CFT correspondence this vacuum gauge group is interpreted as the global symmetry group of the dual SCFT, whereas \(H^g_R\) corresponds to the SCFT's R-symmetry group and \(H^g_\mathrm{mat}\) to an additional flavor symmetry.

We are eventually interested in the supersymmetric deformations of AdS\(_D\) solutions, i.e.~their moduli spaces.
A necessary condition for a scalar field to be a supersymmetric modulus is that the first order variations of the fermionic shift matrices with respect to this scalar field vanish.
This implies that the scalar field is massless.
However, consistency requires that there is one massless scalar field per spontaneously broken gauge group generator.
We show from the generic structure of gauged supergravities that there is indeed one such massless field for each non-compact generator of the gauge group \(G^g\).
These fields are Goldstone bosons and constitute the additional degrees of freedom of those gauge fields which obtain a mass during spontaneous symmetry breaking.
They can therefore not be counted as candidates for (supersymmetric) moduli.
To make our analysis more concrete we turn to the special case where the scalar field space of the supergravity theory under consideration is a symmetric homogeneous space \(\cM = G/H\) for some Lie groups \(G\) and \(H\).
Here we find that every modulus must necessarily be a singlet with respect to the previously introduced \(H^g_R\).
We finally restrict to theories with more than 16 real supercharges.
In this case the only allowed supermultiplet is the gravitational multiplet, which simplifies the analysis considerably.
Here, the moduli space must be a homogeneous space as well and we can give a general recipe for its determination.

Using these results we discuss all maximally supersymmetric AdS$_D$ solutions in gauged supergravities with more than 16 real supercharges in space-time dimensions \(D = 4,5\text{ and }7\).
Due to the absence of vector multiplets only compact gauge groups are allowed.
Therefore the entire gauge group must be given by \(H^g_R\) and is uniquely determined.
The individual results precisely agree with the R-symmetry groups of the respective dual SCFTs.
Almost all of these solutions do not admit supersymmetric moduli,
which follows from the absence of scalar fields that transform as singlets under \(H^g_R\).
The only exception occurs for maximal gauged supergravity in five dimensions.
Here the moduli space is given by \(\SU(1,1)/\U(1)\) and has a well-known holographic interpretation as the complex gauge coupling of four-dimensional \(\cN = 4\) super Yang-Mills theory.

We also study the AdS solutions of gauged half-maximal supergravities in seven dimensions 
and of gauged \(\cN =3\) supergravities in four dimensions
as examples for theories with 16 or less real supercharges.
Due to the possible existence of vector multiplets their analysis is slightly more complicated.
As most of the cases with more than 16 real supercharges also these solutions do not admit for supersymmetric moduli.


We finally draw our attention to the field theory side of the AdS/CFT correspondence and study \(\cN=(1,0)\) superconformal field theories in six dimensions,
which are holographically dual to the discussed supersymmetric AdS\(_7\) backgrounds.
From the representation theory of the underlying superconformal algebra \(\mathfrak{osp}(6,2|2)\) we show that all supersymmetric marginal deformations are forbidden by unitarity bounds.
Consequently, no conformal manifold exists, which is in perfect agreement with our results on the moduli space of supersymmetric AdS\(_7\) solutions in gauged supergravity.

This thesis is organized as follows:
In chapter~\ref{chap:supergravity} we summarize some general facts about gauged supergravity theories in arbitrary dimensions.
Here our objective is to develop a unifying notation. 
In chapter~\ref{chap:classification} we give a complete classification of all maximally supersymmetric solutions of all gauged and ungauged supergravity theories.
We find that AdS solutions are ubiquitous.
In chapter~\ref{chap:ads} we specifically discuss AdS\(_D\) solutions in gauged \(D\)-dimensional supergravity.
Firstly, we develop some general properties of their gauge groups and moduli spaces.
Secondly, we give a general recipe for the computation of AdS moduli spaces for theories with more than 16 real supercharges.
In chapter~\ref{chap:adsmoduli} we use the previously developed algorithm to compute the moduli spaces of all theories with more than 16 real supercharges.
We find that most of these solutions do not admit any moduli either.
Moreover, we also discuss the AdS solutions of half-maximal gauged supergravity in seven dimensions 
and of \(\cN = 3\) gauged supergravity in four dimensions.
We find that they do not allow for moduli as well.
In chapter~\ref{chap:SCFT} we show that for \(\cN = (1,0)\) superconformal field theories in six dimensions the existence of supersymmetric marginal deformations is forbidden by superconformal representation theory.
In chapter~\ref{chap:conclusions} we conclude.

In appendix~\ref{app:conventions} we outline the notations and conventions used throughout this thesis.
In appendix~\ref{app:susy} we collect the general form of the supergravity Lagrangian and the supersymmetry transformation laws of the involved fields. Furthermore, we compute explicit expressions for the Killing vectors and their moment maps in terms of the fermionic shift matrices.
In appendix~\ref{app:integrability} and appendix~\ref{app:ads} we present some technical proofs and computations needed in chapter~\ref{chap:classification} and chapter~\ref{chap:ads}, respectively.
In appendix~\ref{app:SCFT} we review the six-dimensional $(1,0)$ superconformal algebra and discuss the restrictions on Lorentz invariant descendant operators.

The results presented in chapter~\ref{chap:classification} and chapter~\ref{chap:SCFT} have been previously published in \cite{Louis:2015mka, Louis:2016tnz}.
The main analysis of maximally supersymmetric AdS$_D$ solutions in chapters~\ref{chap:ads} and~\ref{chap:adsmoduli} has recently appeared in
\cite{Lust:2017aqj}.



\chapter{Basic Notions of Supergravity}\label{chap:supergravity}
In this chapter we discuss some basic concepts and properties of (gauged) supergravity theories.
We try to be as generic as possible and do not focus on a specific space-time dimension or number of supercharges.
The main purpose of this chapter is to set the stage for the analyses in the subsequent chapters and to introduce a unifying notation which allows us to discuss all cases more or less simultaneously, avoiding a cumbersome case-by-case analysis.%
\footnote{For a review of gauged supergravities see e.g.~\cite{Weidner:2006rp, Samtleben:2008pe, Trigiante:2016mnt}.
For a more detailed discussion of the geometrical structures underlying supergravities see e.g.~\cite{Cecotti:2015wqa}.}

\section{The ungauged theory}
A supergravity theory in \(D\) space-time dimensions always contains a gravitational multiplet.
The generic field content of this multiplet includes the metric \(g_{MN}\)  (\(M,N = 0, \dots, D-1\)), \(\cN\) gravitini \(\psi^i_M\) (\(i = 1, \dots, \cN\)), a set of \((p-1)\)-form fields or gauge potentials \(A^{(p-1)}\), a set of spin-\(\frac12\) fermions \(\chi^{\hat a}\) as well as a set of scalar fields \(\phi\).
Note that not all of these component fields
necessarily have to be
part of a given gravitational multiplet but we 
gave the most general situation.
Moreover, the theory might be coupled to additional multiplets, for example vector, tensor or matter multiplets.
If they are present, these multiplets always contain some spin-$\frac12$ fermions which we collectively call \(\chi^{\tilde a}\).
On the bosonic side they can have additional \((p-1)\)-form fields \(A^{(p-1)}\) among their components, as well as scalar fields which we universally call $\phi$.

We denote all form-fields from the gravitational multiplet as well as those from the other multiplets collectively by \(A^{I_p}\), where the index \(I_p\) labels all fields of the same rank $(p-1)$.
The reason for this is that there often exist duality transformations which mix fields from different multiplets and make it therefore impossible to distinguish from which multiplet a certain bosonic field originates.
Moreover, we need to introduce the corresponding field strengths \(F^{I_p}\) which are differential forms of rank \(p\).
In some situations it will prove convenient to consider also the scalar fields \(\phi\) as $0$-form fields, so we often denote them by \(A^{I_1}\), and their field strengths by \(F^{I_1}\).

We collectively denote all spin-$\frac12$ fermions as \(\chi^a\), but we often want to distinguish the fermions which are part of the gravity multiplet from all the other fermions by calling the former \(\chi^{\hat a}\) and the latter \(\chi^{\tilde a}\).
This is possible because there is no symmetry or duality relating fermions from different types of multiplets.
The fermions \(\psi^i_M\) and \(\chi^a\) can always be arranged in representations of a group \(H\),
\begin{equation}\label{eq:H}
H = H_R \times H_\mathrm{mat} \,,
\end{equation}
where \(H_R\) is the R-symmetry group,
i.e.~the automorphism group of the supersymmetry algebra,
 and \(H_\mathrm{mat}\) is a compact group which -- loosely speaking -- rotates multiplets of the same kind into each other.
Notice that all fields from the gravitational multiplet (i.e.~the gravitini \(\psi^i_M\) and the \(\chi^{\hat a}\)) are necessarily inert under \(H_\mathrm{mat}\) transformations, they can only transform non-trivially under \(H_R\).

Using these ingredients the general bosonic Langrangian takes a relatively simple form and reads
\begin{equation}\label{eq:bosonicaction}
e^{-1} \cL_B = -\frac{R}{2} 
- \frac{1}{2} \sum_{p \geq 1} M^{(p)}_{I_{p} J_{p}}\!\left(\phi\right)\, F^{I_p} \wedge \ast F^{J_p} + e^{-1} \cL_\mathrm{top} \,.
\end{equation}
The last part \(\cL_\mathrm{top}\) does not depend on the space-time metric and is therefore topological, a common example for such a term is a Chern-Simons term.
It is not necessarily part of every supergravity theory.
The matrices \(M^{(p)}_{I_pJ_p}(\phi)\) depend generically on all scalar fields and have to be symmetric and positive definite.
Therefore, they can be diagonalized by introducing vielbeins \(\cV^{\alpha_p}_{I_p}\), i.e.
\begin{equation}\label{eq:kinmatrix}
M^{(p)}_{I_pJ_p} = \delta_{\alpha_p\beta_p} \cV^{\alpha_p}_{I_p}  \cV^{\beta_p}_{J_p}  \,.
\end{equation}
Of course the vielbeins \(\cV^{\alpha_p}_{I_p}\) are scalar dependent as well.
We can use them to convert between the indices \(I_p\) and \(\alpha_p\).
It is convenient to introduce the abbreviations
\begin{equation}
F^{\alpha_p} = F^{I_{p}} \cV^{\alpha_p}_{I_p} \,.
\end{equation}
The benefit of working in this frame is that it allows us to couple the bosonic fields to the fermions, which is crucial for supergravity.
In fact the \(F^{\alpha_p}\) now transform under the same group \(H\) as the fermions but possibly in different representations.
Moreover, the invariance of the theory with respect to such \(H\)-transformations requires that \(\delta_{\alpha_p\beta_p}\) is \(H\)-invariant.
This means that if we denote an element of the Lie algebra \(\h\) of \(H\) in the respective matrix representation by \({J_\al{p}}^\be{p}\), it needs to satisfy
\begin{equation}\label{eq:Hdelta}
{J_{(\al{p}}}^\ga{p} \delta_{\be{p})\ga{p}} = 0 .
\end{equation}
Later on it will be important to distinguish which of the form fields enter the supersymmetry variations of the gravitini.
For this purpose we go one step further and also split the indices \(\alpha_p\) according to
\begin{equation}\label{indexsplit}
\alpha_p = \left(\hat \alpha_p, \tilde \alpha_p\right) \,,
\end{equation}
in the same way as we split the index \(a = \left(\hat a, \tilde a\right)\) labelling the spin-$\frac12$ fermions. 
We then denote by \(F^{\hat\alpha_p}\) 
the field strengths in the gravitational multiplet 
(e.g.\ the graviphotons for \(p = 2\)) and by  \(F^{\tilde\alpha_p}\)
the field strengths which arise in all other multiplets that might be present.
Also \(F^{\hat\alpha_p}\) do not transform under \(H_\mathrm{mat}\) but only non-trivially under the R-symmetry \(H_R\). 
Note that this split depends on the scalar fields via the vielbeins $\cV$
and thus is background dependent.

In the general bosonic Lagrangian \eqref{eq:bosonicaction} we have written the kinetic term of the scalar fields on equal footing with all other 
form fields.
However, the scalar field sector is of particular relevance for the construction of supergravities, it is therefore appropriate to introduce a separate notation for its description.
Therefore, we often denote the scalar fields by \(\phi^r\) instead of \(A^{I_1}\) and their kinetic matrix by \(g_{rs}(\phi)\) instead of \(\cM_{I_1 J_1}(\phi)\).
Moreover, their field strengths \(F^{I_1}\) are given by the derivatives \(\dd \phi^r\), so their kinetic term can be expressed as
\begin{equation}\label{eq:sigmamodel}
\cL_\mathrm{kin,scal} = -\frac{e}{2} g_{rs}(\phi) \dd\phi^r \wedge \ast \dd\phi^s \,.
\end{equation}
This is the Lagrangian of a (non-linear) sigma model.
We interpret the scalar fields as maps from the space-time manifold \(\Sigma\) into some target-space manifold \(\cM\) with Riemannian metric \(g\), i.e.
\begin{equation}
\phi \colon \Sigma \rightarrow \cM \,.
\end{equation}
From the discussion above it follows that the other fields (besides being space-time differential forms) must be sections of some vector bundles over \(\cM\) with bundle metrics \(M^{(p)}_{I_pJ_p}\) and structure group \(H\).
Using this language the \(\cV^{\alpha_p}_{I_p}\) are nothing but local orthonormal frames on these bundles.
Sometimes we also want to introduce a local frame \(e^{\alpha_1}\) on \(\cM\), i.e.~\(
g_{rs} = \delta_{\alpha_1\beta_1} e^{\alpha_1}_r e^{\beta_1}_s 
\),
such that \eqref{eq:sigmamodel} reads
\begin{equation}\label{eq:scalarvielbeins}
\cL_\mathrm{kin,scal} = \frac{e}{2} \delta_{\alpha_1\beta_1} \cP^{\alpha_1} \wedge \ast \cP^{\beta_1} \,, \qquad \text{with} \qquad \cP^{\alpha_1} = \phi^\ast e^{\alpha_1} = e^{\alpha_1}_r \dd \phi^r \,,
\end{equation}
where \(\phi^\ast\) denotes the pullback with respect to \(\phi\).

In a supersymmetric theory bosonic and fermionic fields are mapped into each other via supersymmetry transformations, so also the fermions should be sections of some vector bundles over \(\cM\).
In many cases these bundles correspond to the tangent bundle \(T\cM\) or are subbundles of \(T\cM\).

Let us make this more specific for the example of the gravitini, which are the fermions that are present in every supergravity theory.
They are sections of a vector bundle 
\begin{equation}
\cR \rightarrow \cM \,,
\end{equation}
with structure group \(H_R\).
On this bundle (or better on the associated principal bundle) there exists a local connection form \(\theta\),
i.e.~a \(\h_R\)-valued 1-form on \(\cM\), where \(\h_R\) denotes the Lie-algebra of \(H_R\).
The corresponding curvature 2-form \(\Omega\)
is given by
\begin{equation}\label{eq:Rcurvature}
\Omega = \dd \theta + \theta \wedge \theta \,.
\end{equation}
This induces a covariant derivative \(\cD_M \psi^i_n\) which transforms covariantly under scalar-depedent \(H_R\)-transformations,
\begin{equation} \label{eq:covariantderiv}
\cD_M \psi^i_N = \nabla\!_M \psi^i_N - \left(\cQ^R_M\right)^i_j \psi^j_N \,,
\end{equation}
where \(\nabla_M\) is the space-time Levi-Cevita connection and \(\left(\cQ_M\right)^i_j\) is the pullback of the connection form \(\theta\), expressed in the appropriate \(\h_R\)-representation, i.e.
\begin{equation}\label{eq:Rconnection}
\cQ^R = \phi^\ast \theta \,.
\end{equation}
The corresponding curvature or field strength is obtained from the commutator of two covariant derivatives.
Explicitly, we have
\begin{equation}\label{eq:Dcomm}
\left[\cD_M, \cD_N \right] \epsilon^i = \tfrac{1}{4} R_{MNPQ}\Gamma^{PQ}\, \epsilon^i - \left(\cH^R_{MN}\right)^i_j \epsilon^j \,,
\end{equation}
where \(R_{MNPQ}\) is the space-time Riemann curvature tensor and \(\cH^R\) is the pullback of the curvature form \(\Omega\), i.e.~\(\cH^R = \phi^\ast \Omega\).

In a similar way we can introduce covariant derivatives for the other fermionic fields.
They transform in general not only under \(H_R\) but also under \(H_\mathrm{mat}\), or in other words they are sections of a vector bundle \(\cX \rightarrow \cM\) with structure group \(H\).
Analogous to our previous construction, we define
\begin{equation}\label{eq:chicovderiv}
\cD_M \chi^a = \nabla\!_M \chi^a - (\cQ_M)^a_b \chi^b= \nabla\!_M \chi^a - (\cQ^R_M)^a_b \chi^b - (\cQ^\mathrm{mat}_M)^a_b \chi^b \,,
\end{equation}
where \((\cQ_M)^a_b\) is the pull-back of the connection form on \(\cX\), expressed in the appropriate \(H\)-representation.
Since \(H\) is the product of \(H_R\) and \(H_\mathrm{mat}\) it splits into \(\cQ_M^R\) and \(\cQ_M^\mathrm{mat}\), where the former agrees with \eqref{eq:Rconnection}.
This indicates that in general \(\cR\) is a subbundle of \(\cX\). 
We finally want to note that according to the split \(a = (\hat a, \tilde a)\) we have \((\cQ_M)^{\hat a}_{\tilde a} = (\cQ_M)^{\tilde a}_{\hat a} = 0\) and \((\cQ_M^\mathrm{mat})^{\hat a}_{\hat b} = 0\).
The last identity is due to the fact that the components of the gravity multiplet do not transform with respect to \(H_\mathrm{mat}\).

We are now in the position to give the supersymmetry variations of the fermions.%
\footnote{The supersymmetry variations of the bosons (as well as of the fermions) are summarized in appendix~\ref{app:susyvariations}.}
They are of special importance in the following chapter, where we study maximally supersymmetric solutions.
In general they also contain terms of higher order in the fermionic fields.
However, we omit these terms as they vanish identically in the purely bosonic solutions we are interested in.
Under an infinitesimal supersymmetry transformation described by the spinorial parameter \(\epsilon^i = \epsilon^i(x^M)\), the gravitini transform as
\begin{equation}\label{eq:ungaugedgravitinovariation}
\delta \psi^i_M = \cD_M  \epsilon^i  + \left(\cF_M\right)^i_j \epsilon^j \ ,
\end{equation} 
where \(\cD_M\) is the covariant derivative introduced in \eqref{eq:covariantderiv}.
The second term in \eqref{eq:ungaugedgravitinovariation} contains the various field strengths and is given by
\begin{equation}\label{eq:cFM}
\big(\cF_M\big)^i_j  =  \tfrac{1}{2D-4}\sum_{p \geq 2} \big(B^{(p)}_{ \hat\alpha_p}\big)^i_j\,
F^{\hat\alpha_p}_{{N_1}\dots {N_p}} {T_{(p)}^{{N_1}\dots {N_p}}}{}_M \,,
\end{equation}
where the \(B^{(p)}\) are constant matrices correlating the different \(H_R\)-representations.
(See Appendix~\ref{app:susyvariations} for a more detailed discussion of their properties.)  
The matrices \({T_{(p)}^{{N_1}\dots {N_p}}}{}_M\) are a certain combination of \(\Gamma\)-matrices and are defined in \eqref{eq:T}.

The supersymmetry variations of the spin-$\frac12$ fermions are even simpler and take the generic form
\begin{equation}
\delta \chi^a = \cF^a_i \epsilon^i \,,
\end{equation}
where \(\cF^a_i\) contains the various field strengths.
The crucial observation is that the variations of the fermions \(\chi^{\hat a}\) which are part of the gravity multiplet can contain only the field strengths \(F^{\hat \alpha_p}\), while the variations of the \(\chi^{\tilde a}\) depend only on \(F^{\tilde \alpha_p}\).
Explicitly \(\cF^a_i\) is given by
\begin{equation}\label{eq:cFhat}
\cF^{\hat a}_i = \sum_{p \geq 1} \sum_{\hat\alpha_p}\big(C^{(p)}_{ \hat\alpha_p}\big)^{\hat a}_i \, F^{\hat\alpha_p}_{N_1\dots N_p} \Gamma^{N_1\dots N_p} \epsilon^i \,,
\end{equation}
and
\begin{equation}\label{eq:cFtilde}
\cF^{\tilde a}_i = \sum_{p \geq 1} \sum_{\tilde\alpha_p}\big(C^{(p)}_{ \tilde\alpha_p}\big)^{\tilde a}_i \, F^{\tilde\alpha_p}_{N_1\dots N_p} \Gamma^{N_1\dots N_p} \epsilon^i \,.
\end{equation}
As in the gravitino variations \(C^{(p)}\) are constant matrices. 
Contrary to \eqref{eq:cFM}, the sums in \eqref{eq:cFhat} and
\eqref{eq:cFtilde} start already at \(p = 1\)
and thus include  the fields strengths  of the scalar fields
\(F^{\alpha_1}_M = \cP^{\alpha_1}_M\) which
do not enter the gravitino variations \eqref{eq:ungaugedgravitinovariation}.
Notice that the matrices \(C^{(1)}\) constitute an isomorphism between \(\cR \otimes \cX\) and the tangent bundle \(T\cM\), which highly constrains the geometry of \(\cM\) \cite{Cecotti:2015wqa}.

We finally want to mention that supersymmetry imposes a non-trivial condition on the curvature of the R-symmetry bundle \(\cR\).
For global supersymmetry this condition reads \(\cH^R = 0\) and requires therefore that \(\cR\) is flat.
However, for supergravity \(\cR\) must have a non-trivial curvature.
Indeed, we compute in appendix~\ref{app:susycalculations} that \(\cH^R\) has to satisfy
\begin{equation}\label{eq:hscalar}
\cH^R = - \tfrac14 C^\dagger_{\alpha_1} C_{\beta_1} P^{\alpha_1} \wedge P^{\beta_1} \,,
\end{equation}
or equivalently that \(\Omega = - \tfrac14 C^\dagger_\al{1} C_\be{1} e^\al{1} \wedge e^\be{1}\), so \(\Omega\) must be non-vanishing at every point of \(\cM\).
Moreover, it follows from \eqref{eq:hscalar} that
\begin{equation}\begin{aligned}\label{eq:Omegaparallel}
\cD_r \Omega_{st} &\equiv \nabla\!_r \Omega_{st} + \bigl[\theta_r, \Omega_{st} \bigr] \\ 
&= - \tfrac12 C^\dagger_{[\alpha_1} C_{\beta_1]} \Bigl[\bigl(\nabla\!_r e^\al{1}_s\bigr) e^\be{1}_t + e^\al{1}_s  \bigl(\nabla\!_r e^\be{1}_t\bigr)\Bigr] \\
&\qquad - \tfrac12 \Bigr[ {\left(\theta_r\right)_\al{1}}^\ga{1} C^\dagger_{\gamma_1} C_{\beta_1}  + C^\dagger_{\alpha_1} C_{\gamma_1}  {\left(\theta_r\right)_\be{1}}^\ga{1} \Bigl] e^\al{1}_{[s} e^\al{1}_{t]} \\
&= - \tfrac12 C^\dagger_{[\alpha_1} C_{\beta_1]} \Bigl[\bigl(\cD_r e^\al{1}_s\bigr) e^\be{1}_t + e^\al{1}_s  \bigl(\cD_r e^\be{1}_t\bigr)\Bigr] = 0 \,,
\end{aligned}\end{equation}
where we first used the general property \eqref{eq:appBCproperty} of the matrices \(C_\al{1}\).
In the second step we used that
\({\left(\theta_r\right)_\be{1}}^\al{1}\) corresponds to the spin-connection on \(T \cM\)
and therefore \(\cD_r e^\al{1}_s = \nabla\!_r e^\al{1}_s + {\left(\theta_r\right)_\be{1}}^\al{1} e^\be{1}_s = 0\), where \(\nabla\) denotes the Levi-Civita connection on \(T\cM\).%
\footnote{See also \eqref{eq:spinconnection} and the discussion there.}
Equation \eqref{eq:Omegaparallel} does not only imply the Bianchi identy%
\footnote{For a \(\h^g_R\)-valued \(p\)-form \(\alpha\) and \(q\)-form \(\beta\) one defines
\(
\bigl[\alpha\wedge\beta\bigr] \equiv \alpha \wedge \beta - (-1)^{pq} \beta \wedge \alpha 
\).}
\begin{equation}
\cD \Omega = \dd \Omega + \bigl[\theta\wedge\Omega\bigr]= 0 \,,
\end{equation}
but also that \(\Omega\) is parallel with respect to \(\cD\).
A set of parallel tensors on \(\cM\) 
in turn restricts the holonomy of \(\cM\).

\section{Gauging}\label{sec:gauging}

A generic supergravity theory is often invariant under a global symmetry group \(G\).
Let us denote the generators of \(G\) by \(t_\rho\), with \(\rho = 1, \dots , \dim(G)\).
They satisfy 
\begin{equation}\label{eq:Gstrconst}
\bigl[t_\rho, t_\sigma] = {f_{\rho\sigma}}^{\tau} t_\tau \,,
\end{equation}
where \({f_{\rho\sigma}}^{\tau}\) are the structure constants of the Lie algebra \(\g\) of \(G\).

We now want to convert a subset of these symmetries, corresponding to a subgroup \(G^g \subseteq G\), from global to local symmetries. This procedure is called \emph{gauging}.
Making a symmetry local is only possible if there exist appropriately transforming gauge fields, i.e.~1-form or vector fields \(A^I\),%
\footnote{For the sake of simplicity, from now on we often write for gauge fields \(A^I\) instead of \(A^{I_2}\).}
 such that we can replace ordinary derivatives \(\partial_\mu\) by covariant derivatives \(D_\mu\),
\begin{equation}\label{eq:gencovderiv}
D_\mu = \partial_\mu - A^I_\mu X_I \,,
\end{equation}
whereas the \(X_I\) generate the respective subalgebra \(\g^g \subseteq \g\).
However, in supergravity the presence of gauge fields as well as their transformation behavior with respect to the global symmetry group \(G\) cannot be chosen freely but is usually restricted by supersymmetry.
This obstruction makes the gauging procedure more subtle.
To be more specific, let us denote the \(\g\)-representation of the gauge fields corresponding to the index \(I\) by \(\mathbf{v}\).
Clearly, the gauging can only be successful if the adjoint representation of \(\g^g\) can be found in the decomposition of \(\mathbf{v}\) into \(\g^g\)-representations.
 
The problem of finding a gaugable subgroup \(G^g\) of \(G\) can be tackled systematically by means of the \emph{embedding tensor formalism} \cite{Nicolai:2000sc,Nicolai:2001sv,deWit:2002vt} (see e.g.~\cite{Samtleben:2008pe} for a review).
Here one describes the embedding of \(\g^g\) into \(\g\) in terms of a constant map \(\Theta \colon \mathbf{v} \rightarrow \g\).
Explicitly, this embedding reads
\begin{equation}\label{eq:gaugegenerators}
X_I = {\Theta_I}^\rho t_\rho \,,
\end{equation}
where \({\Theta_I}^\rho\) is called the \emph{embedding tensor}.
If we denote the generators of \(\g\) in the gauge field representation \(\mathbf{v}\) by \({(t_\rho)_I}^J\) and accordingly introduce \({X_{IJ}}^K = {(X_I){}_J}^K = {\Theta_I}^\rho {(t_\rho)_J}^K\), the condition that the \(X_I\) span a closed subalgebra of \(\g\) reads
\begin{equation}\label{eq:quadconstr}
\bigl[X_I, X_J] = -{X_{IJ}}^K X_K \,.
\end{equation}
Note that \({X_{IJ}}^K\) can only be regarded as the structure constants of \(\g^g\) under the above contraction with \(X_K\), on its own they do not even have to be antisymmetric in their lower indices.
This is the case because the \(X_I\) are not necessarily all linearly independent since the rank of \(\g^g\) might be smaller than the dimension of \(\mathbf{v}\).
The condition \eqref{eq:quadconstr} is equivalent to the \(\g^g\)-invariance of \(\Theta\), or explicitly
\({\Theta_I}^\rho \bigl( {(t_\rho)_J}^K {\Theta_K}^\sigma + {f_{\rho\tau}}^\sigma {\Theta_J}^\tau\bigr) = 0\).
Hence, it is called the \emph{quadratic constraint}.

However, not every embedding which is actually compatible with the quadratic constraint can be realized in a given supergravity.
Supersymmetry imposes a second condition on the embedding tensor, called the \emph{linear constraint}.
By construction \(\Theta\) transforms under \(\g\) in the product representation \(\mathbf{\overline{v}} \otimes \g\), which can be decomposed into a direct sum of irreducible \(\g\)-representations.
Not all of these irreducible representations describe a gauging which can be consistently realized in a supergravity theory.
Some of the irreducible representations in \(\mathbf{\overline{v}} \otimes \g\) are therefore not allowed and have to be set equal to zero.
Schematically, the linear constraint reads
\begin{equation}
\mathbb{P}\!_{lc} \, \Theta = 0 \,,
\end{equation}
where \(\mathbb{P}\!_{lc}\) is an operator that projects onto the forbidden \(\g\)-representations.
In a similar fashion we could also write the quadratic constraint as
\begin{equation}
\mathbb{P}\!_{qc} \, \Theta \otimes \Theta = 0 \,,
\end{equation}
with some appropriate projection operator \(\mathbb{P}\!_{qc}\).

A generic object \(\cO\) transforms under a local and infinitesimal gauge transformation para\-metrized by \(\lambda^I(x)\) according to
\begin{equation}\label{eq:gengaugetransf}
\delta \cO = \lambda^I X_I \cO = \lambda^I {\Theta_I}^\rho t_\rho \cO\,,
\end{equation}
where \(t_\rho\) are here the generators of \(G\) in the respective representation of \(\cO\).
In order for the covarariant derivative \(D_\mu \cO\) \eqref{eq:gencovderiv} to transform in the same way (i.e.~covariantly) the gauge fields \(A^I\) need to transform according to
\begin{equation}\label{eq:Atransf}
\delta A^I = D \lambda^I = \dd \lambda^I + {X_{JK}}^I A^J \lambda^K \,.
\end{equation} 
This transformation behavior requires an appropriate modification of the corresponding field strength 2-forms \(F^I\) such that they transform covariantly as well, i.e.
\begin{equation}\label{eq:Ftransf}
\delta F^I = - \lambda^J {X_{JK}}^I F^K \,.
\end{equation}
Note that this is precisely the same as \eqref{eq:gengaugetransf} for an object transforming in the gauge field representation \(\mathbf{v}\).
Due to the fact that the \({X_{IJ}}^K\) are not in one-to-one correspondence with the structure constants of \(\g^g\), finding covariantly transforming field strengths \(F^I\) is more subtle than in standard Yang-Mills theory.
The precise form of \(F^I\), however, is not important for the following discussion, so we do not need to comment further on this point.
Analogously, of course also the field strengths \(F^{I_p}\) of the other higher-rank form fields (if present) need to be modified appropriately.

Let us now turn to a discussion of the scalar field sector.
The sigma model Lagrangian \eqref{eq:sigmamodel} is invariant under all transformations of the scalar fields which leave the metric \(g_{rs}\) invariant.
In other words the global symmetry group \(G\) must be contained in the isometry group \(\mathrm{Iso}(\cM)\) of \(\cM\).
To be more specific, an infinitesimal transformation \(\phi^r \rightarrow \phi^r + \lambda^\rho k^r_\rho\) leaves \eqref{eq:sigmamodel} invariant if the \(k^r_\rho\) are Killing vectors of \(g_{rs}\), i.e. \(\nabla_{(r} k_{s)\rho} = 0\), and if the \(k_\rho\) generate a subgroup \(G\) of \(\mathrm{Iso}(\cM)\), i.e. \(\left[k_\rho, k_\sigma\right] = - {f_{\rho\sigma}}^\tau k_\tau\), where \({f_{\rho\sigma}}^\tau\) are the structure constants of the Lie algebra \(\g\) of \(G\), cf. \eqref{eq:Gstrconst}.
We now want to gauge some of these symmetries, so according to our above considerations we select a subgroup \(G^g \subset G\) via
\begin{equation}
k_I = {\Theta_I}^\rho k_\rho \,,
\end{equation}
such that
\begin{equation}\label{eq:killingcommutator}
\bigl[k_I, k_J\bigr] = {X_{IJ}}^K k_K \,,
\end{equation}
where \({X_{IJ}}^K\) is defined in the same way as in \eqref{eq:quadconstr}.
In the end we want to construct a Lagrangian which is invariant under local \(G^g\) transformations
\begin{equation}\label{eq:scalargaugetransf}
\phi^r(x) \rightarrow \phi^r(x) + \lambda^I(x) k^r_I(\phi) \,,
\end{equation}
where the infinitesimal parameters \(\lambda^I(x)\) are allowed to depend on the space-time coordinates explicitly.
Such transformations induce additional terms in the derivative \(\dd\phi^r\) which have to be compensated by the introduction of covariant derivatives \(D\phi^r\),
\begin{equation}\label{eq:scalarcovderiv}
D\phi^r = \dd \phi^r - A^I k^r_I \,,
\end{equation}
where the \(A^I\) transform according to \eqref{eq:Atransf}.
The form of \eqref{eq:scalargaugetransf} and \eqref{eq:scalarcovderiv} indicates again that the Killing vectors take the role of the general gauge group generators \(X_I\) on the scalar field sector.
Analogously the vielbeins \(\cP^\al{1}\) get replaced by
\begin{equation}\label{eq:gaugedP}
\hat \cP^\al{1} = \cP^\al{1} + A^I \cP^\al{1}_I \,,\qquad \cP^\al{1}_I  = k^r_I e^\al{1}_r \,.
\end{equation}
It is often beneficial to use \(\cP^\al{1}_I\), which are the Killing vectors expressed in the local frame \(e^\al{1}\)\eqref{eq:sigmamodel}, instead of working directly with \(k^r_I\).

The complete supersymmetric Lagrangian consists not only of the sigma model part \eqref{eq:sigmamodel}, but also features all the other fields living in vector bundles over \(\cM\).
Therefore, a symmetry of the complete theory must be more than just an isometry of the scalar manifold \(\cM\).
We furthermore demand that Killing vectors are compatible with the various bundle structures.
For the R-symmetry bundle \(\cR\) these conditions read%
\footnote{Our discussion follows \cite{DAuria:1990qxt, Andrianopoli:1996cm}.}
\begin{equation}\label{eq:W}
\cL_I \Omega = \bigl[\Omega, W_I\bigr] \,, \qquad \cL_I \theta = \cD W_I \equiv \dd W_I + \bigl[\theta, W_I\bigr] \,.
\end{equation}
Here \(\cL_I\) denotes the Lie derivative in the direction of \(k_I\), i.e. \(\cL_I = \cL_{k_I}\),\footnote{
The Lie derivative describes how a scalar field dependent object varies under a variation of the scalar fields.
For example, under an infinitesimal gauge transformation \eqref{eq:scalargaugetransf} parametrized by \(\epsilon^I(x)\), a geometrical object \(T\) defined on \(\cM\) transforms according to \(\delta_\epsilon T = \epsilon^I \cL_I T\).}
 and \(\theta\) and \(\Omega\) are the connection and curvature form on \(\cR\), see \eqref{eq:Rcurvature}.
The \(W_I\) are (local) \(\h_R\)-valued functions on \(\cM\) which are required to satisfy the condition
\begin{equation}\label{eq:Wcocycle}
\cL_I W_J - \cL_J W_I + \bigl[W_I,W_J\bigr] = {X_{IJ}}^K W_K \,.
\end{equation}
To find the correct modification of the covariant derivative \eqref{eq:covariantderiv} of the gravitini and supersymmetry parameters we need to introduce the generalized moment maps \(\cQ^R_I\), which are locally defined by
\begin{equation}\label{eq:genmomentmap}
\cQ^R_I = \iota_I \theta - W_I
\end{equation}
It follows directly from the definition of the curvature form \(\Omega\) \eqref{eq:Rcurvature} and from \eqref{eq:Wcocycle} that
\begin{equation}\label{eq:Qderiv}
\cD \cQ^R_I = - \iota_I \Omega \,,
\end{equation}
which is often taken as the definition of \(\cQ^R_I\).
Moreover, it follows from \eqref{eq:Wcocycle} that the Lie derivative of the moment maps with respect to the Killing directions is given by
\begin{equation}\label{eq:momentmapliederiv}
\cL_I \cQ^R_J = - \bigl[W_I, \cQ^R_J \bigr] + {X_{IJ}}^K \cQ^R_K \,,
\end{equation}
which implies that they satisfy the equivariance condition
\begin{equation}\label{eq:equivariance}
\bigl[\cQ^R_I, \cQ^R_J\bigr] = - {X_{IJ}}^K \cQ^R_K + \Omega\!\left(k_I, k_J\right) \,.
\end{equation}
The transformation property \eqref{eq:momentmapliederiv} shows that \(\cQ^R_I\) is the correct object to build a gauged version \(\hat\cD\) of the covariant derivative \(\cD\) introduced in \eqref{eq:covariantderiv}.
Explicitly, we define
\begin{equation}\label{eq:gaugedQ}
\hat\cD_M \epsilon^i = \nabla_M \epsilon^i - (\hat\cQ^R_M)^i_j \epsilon^j \,,\qquad\text{with}\qquad \hat\cQ^R = \cQ^R + A^I \cQ^R_I \,.
\end{equation}
This covariant derivative transforms properly if \(\epsilon^i\) transform under a gauge transformation as
\begin{equation}
\delta \epsilon^i = -\lambda^I (W_I)^i_j \epsilon^j \,,
\end{equation}
where \((W_I)^i_j\) is the \(\h_R\)-compensator \eqref{eq:W} expressed in the appropriate representation of \(H_R\). 
Analogously to \eqref{eq:Dcomm}, the commutator of two gauged covariant derivatives \(\hat\cD\) is given by
\begin{equation}\label{eq:Dcommgauged}
\bigl[\hat\cD_M, \hat\cD_N \bigr] \epsilon^i = \tfrac{1}{4} R_{MNPQ}\Gamma^{PQ}\, \epsilon^i - \bigl(\hat\cH^R_{MN}\bigr)^i_j \epsilon^j \,,
\end{equation}
where the curvature or field strength \(\hat\cH^R\) now also contains a term that depends on the field strengths \(F^I\) of the gauge fields \(A^I\),
\begin{equation}\label{eq:hrgenerators}
\hat\cH^R = \cH^R + F^{I} \cQ^R_{I} \,.
\end{equation}
Let us finally mention that even in the absence of scalar fields, i.e.~if \(\cM\) is degenerated to a point, it is still often consistent to assign a non-trivial (constant) value to \(\cQ^R_I\), known as a Fayet-Iliopoulos term \cite{Fayet:1974jb, Fayet:1975yi}.

In a similar fashion to the construction above we need to modify the \(H\)-covariant derivative \eqref{eq:chicovderiv} of the other fields and introduce
\begin{equation}\label{eq:chigaugedcovderiv}
\hat\cD_M \chi^a = \nabla\!_M \chi^a - (\hat\cQ_M)^a_b \chi^b = \nabla\!_M \chi^a - (\cQ_M)^a_b \chi^b - A^I (\cQ_I)^a_b \chi^b \,.
\end{equation}
Notice that the moment maps \(\cQ_I\) split in general according to
\begin{equation}
\cQ_I = \cQ^\cR_I + \cQ_I^\mathrm{mat} \,,
\end{equation}
where \(\cQ^R\) is the R-symmetry moment map which we have constructed above.

We want to illustrate these concepts for the tangent bundle \(T\cM\) which is by construction an \(H\)-bundle as well.
Here the connection form \(\theta\) is given by the Levi-Civita connection.
With respect to the local frame \(e^\al{1}\) it is defined as the solution of
\begin{equation}\label{eq:spinconnection}
\dd e^\al{1} + {\theta_\be{1}}^\al{1} \wedge e^\be{1} = 0 \,.
\end{equation}
Accordingly the covariant derivative of the Killing vectors \(\cP^\al{1}_I\) reads
\begin{equation}
\cD \cP^\al{1}_I = \dd \cP^\al{1}_I - {\theta_\be{1}}^\al{1} \cP^\be{1}_I \,,
\end{equation}
and the moment maps \(\cQ_I\) in the respective \(H\)-representation are given by \cite{Bandos:2016smv}
\begin{equation}\label{eq:TMQ}
{\bigl(\cQ_I\bigr)_\al{1}}^\be{1} = - \cD_\al{1} \cP^\be{1}_I \,.
\end{equation}
A moment map introduced in this way indeed satisfies the defining property \eqref{eq:Qderiv}.
This follows from the general fact that the second covariant derivative of a Killing vector is given by a contraction of the same Killing vector with the Riemann tensor (see e.g. \cite{Weinberg:1972kfs}).
Moreover, \eqref{eq:TMQ} implies that 
\begin{equation}\label{eq:Pcovderiv}
\cD_I \cP^\al{1}_J = {X_{IJ}}^K \cP^\al{1}_K - {\bigl(\cQ_I\bigr)_\be{1}}^\al{1} \cP^\be{1}_J \,,
\end{equation}
which in turn shows in combination with \eqref{eq:genmomentmap} that \(\cP^\al{1}_I\) transforms under a gauge transformation in the appropriate way, i.e.
\begin{equation}
\cL_I \cP^\al{1}_J = {X_{IJ}}^K \cP^\al{1}_K + {\bigl(W_I\bigr)_\be{1}}^\al{1} \cP^\be{1}_J \,.
\end{equation}

Let us again come back to the gauge field sector.
As we have seen above the field strengths \(F^I\) are not inert under gauge transformations but transform according to \eqref{eq:Ftransf}.
Therefore the gauge invariance of the kinetic term in \eqref{eq:bosonicaction} demands an analogous transformation law for the matrix \(M_{IJ}(\phi)\), i.e.
\begin{equation}
\cL_I M_{JK} = 2 {X_{I (J}}^L M_{K) L} \,,
\end{equation}
consistent with \(M_{IJ}\) transforming in the \(\left(\mathbf{\overline{v}} \otimes \mathbf{\overline{v}}\right)_\mathrm{sym}\) representation.
Correspondingly, the vielbeins \(\cV^\al{2}_I\) transform according to
\begin{equation}\label{eq:generalvielbeinvariation}
\cL_I \cV^\al{2}_J = {X_{IJ}}^K \cV^\al{2}_K + {(W_I)_\be{2}}^\al{2} \cV^\be{2}_J \,. 
\end{equation}
The additional term with the \(H\)-compensator \(W_I\) is due to the fact that \(\cV^\al{2}_I\) lives in an \(H\)-bundle over \(\cM\).
Similar considerations hold for the other \(p\)-form fields.

In addition to the replacement of \(\cD\) with \(\hat\cD\) the gauging of the theory requires the modification of the fermionic supersymmetry variations by shift matrices \(A^i_{0\,j}\) and \(A^a_{1\,i}\).
These matrices in general depend on the scalar fields and the specific form of the gauging.
We will derive some explicit relations between \(A_0\) and \(A_1\) and the Killing vectors and moment maps in appendix~\ref{app:susycalculations}.
Altogether, the supersymmetry variations of the fermions read
\begin{subequations}\label{eq:generalfermionicvariations}
\begin{equation}\label{eq:gravitinovariation}
\delta \psi^i_M = \hat\cD_M  \epsilon^i  + \left(\cF_M\right)^i_j \epsilon^j + A^i_{0\,j} \epsilon^j\ ,
\end{equation} 
\begin{equation}\label{eq:spin12variation}
\delta \chi^a = \cF^a_i \epsilon^i + A^a_{1\,i} \epsilon^i \,,
\end{equation}
\end{subequations}
where \(\cF_M\) and \(\cF\) are the same objects as defined in \eqref{eq:cFM}, \eqref{eq:cFhat} and \eqref{eq:cFtilde}, depending on the gauge covariant field strengths.
In addition, the shift matrices also act as fermionic mass-matrices, we give their explicit form in \eqref{eq:appfermionmass}.
Moreover, supersymmetry requires also the existence of a non-trivial scalar potential which can be expressed in terms of \(A_0\) and \(A_1\).
It is given by
\begin{equation}\label{eq:generalpotential}
\delta^i_j V = - 2(D-1)(D-2) \bigl(A^\dagger_0\bigr)^i_k A^k_{0\,j} + 2 \bigl(A_1^\dagger\bigr)^i_a A^a_{1\,j} \,,
\end{equation}
and \(V\) can be obtained by taking the trace on both sides.
Of course, for the gauging procedure to be consistent the potential must be invariant with respect to local \(G^g\) transformations, i.e.~\(\cL_I V = 0\).

We finally want to mention that in some cases there exist deformations 
which can not be expressed as the gauging of a global symmetry.
These deformation can give rise to fermion shift matrices and to a scalar potential as well.
Prominent examples are the superpotential of four-dimensional \(\cN = 1\) supergravity or massive type IIA supergravity in ten dimensions.

\section{Coset geometry}\label{sec:coset}
In this section we discuss the application of the previously introduced concepts to theories where the target space \(\cM\) is a symmetric space.\footnote{We follow the discussion of \cite{Bandos:2016smv, Trigiante:2016mnt}.}
This is necessarily the case for all theories with more than 8 real supercharges.
For these theories we can write \(\cM\) as a coset
\begin{equation}
\cM = \frac{G}{H} \,,
\end{equation}
where \(G\) is a non-compact Lie group and \(H\) its maximally compact subgroup.
\(H\) coincides with the group introduced in \eqref{eq:H}.
The points of \(\cM\) are the equivalence classes in \(G\) with respect to the right multiplication of \(H\), i.e. \(g \sim g h\) for some \(h \in H\), and thus the left-cosets \(gH\) with \(g \in G\).
Note that the map \(g \mapsto gH\) induces on \(G\) a natural structure as an \(H\)-principal bundle over \(G/H\), which is precisely the kind of structure we need for supergravity.

The Lie algebra \(\g\) of \(G\) can be decomposed as
\begin{equation}\label{eq:gdecomp}
\g = \h \oplus \k \,,
\end{equation}
where the direct sum is to be understood only as a direct sum of vector spaces.
Here \(\h\) denotes the Lie algebra of \(H\) and \(\k\) spans the remaining directions of \(\g\).
Since \(\h\) is a subalgebra of \(\g\) it is by definition closed with respect to the Lie-bracket, i.e.~\(\left[\h,\h\right] \subseteq \h\).
If \(\g\) is a reductive Lie algebra (this means it is the direct sum of only simple or abelian Lie-algebras) we can always find a decomposition of \(\g\) such that
\begin{equation}\label{eq:reductive}
\left[\h,\k\right] \subseteq \k \,.
\end{equation}
In this case also the coset space \(G/H\) is called \emph{reductive}.
In particular, this means that \(\k\) transforms in an \(\h\)-representation with respect to the adjoint action.
Moreover, we call \(G/H\) \emph{symmetric} if it is reductive and
\begin{equation}\label{eq:symmetric}
\left[\k,\k\right] \subseteq \h \,.
\end{equation}
All coset spaces that we encounter will be symmetric.
It is sometimes convenient to give an explicit basis for \(\h\) and \(\k\).
In this case we denote the generators of \(\h\) by \(J^A\) and the generators of \(\k\) by \(K^\alpha\).
In this basis the conditions \eqref{eq:reductive} and \eqref{eq:symmetric} in terms of the structure constants read
\begin{equation}
{f_{\alpha A}}^B = {f_{\alpha\beta}}^\gamma = 0 \,.
\end{equation}

Let \(\phi: \Sigma \rightarrow \cM \) be the scalar fields describing a sigma model on \(\cM\), and let \(\phi^r\) be the scalar fields in local coordinates.
Each value of \(\phi\) corresponds to a coset and can be therefore described by a coset representative \(L(\phi) \in G\).
Acting on \(L(\phi)\) from the left with some element \(g \in G\) yields another element in \(G\) that generically lies in a different coset, represented by \(L(\phi')\).
As \(g L(\phi)\) and \(L(\phi')\) are in the same \(H\)-coset, they must only differ by the right action of some \(h(\phi,g) \in H\) and therefore
\begin{equation}\label{eq:gcosettransf}
g L(\phi) = L(\phi') h(\phi, g) \,.
\end{equation}
To formulate the sigma model action we introduce the Maurer-Cartan form
\begin{equation}\label{eq:MCform}
\omega = L^{-1} \dd L \,,
\end{equation}
which takes values in \(\g\) and satisfies the Maurer-Cartan equation \(\dd \omega + \omega \wedge \omega = 0\).
We split \(\omega\) according to the decomposition \eqref{eq:gdecomp} of \(\g\),
\begin{equation}\label{eq:maurercartan}
\omega = \cP + \cQ \,,\qquad\text{such that}\qquad \cP \in \k\,, \cQ \in \h \,,
\end{equation}
or explicitly \(\cP = \cP^\alpha K_\alpha\) and \(\cQ = \cQ^A J_A\).
We use \(\cP\) to formulate the kinetic term of a sigma model on \(\cM\).
Its Lagrangian reads
\begin{equation}\label{eq:cosetlagrangian}
\cL_\mathrm{kin,scal} = -\frac{e}{2} \tr\left(\cP\wedge \ast \cP\right) = -\frac{e}{2} g_{\alpha \beta} \cP^\alpha \wedge \ast \cP^\beta \,,
\end{equation}
where \(g_{\alpha\beta} = \tr\left(K_\alpha K_\beta\right)\) is the restriction of the Killing form of \(\g\) on \(\k\).
Notice that it is always possible to find a basis of generators \(K_\alpha\) such that \(g_{\alpha\beta} = \delta_{\alpha\beta}\).
In this frame the \(\cP^\alpha\) directly correspond to the vielbeins introduced in \eqref{eq:scalarvielbeins}.
This Lagrangian is invariant under a global \(G\)-transformation \eqref{eq:gcosettransf}.
Indeed, \(\cP\) and \(\cQ\) transform as
\begin{equation}\begin{aligned}
\cP(\phi') &= h \cP(\phi) h^{-1} \,, \\
\cQ(\phi') &= h \cQ(\phi) h^{-1} + h \dd h^{-1} \,,
\end{aligned} \end{equation}
which shows the invariance of \eqref{eq:cosetlagrangian}.
Moreover, \(\cQ\) has the transformation behavior of a \(H\)-connection, it is the local connection form of the principal \(H\)-bundle over \(\cM\) and can be used to define an \(H\)-covariant derivative.
The action of this covariant derivative on the coset representative \(L\) is given by
\begin{equation}\label{eq:cosetreprderiv}
\cD L = d L - L \cQ = L \cP \,,
\end{equation}
where the second equality follows from the definition of \(\cP\) and \(\cQ\), see \eqref{eq:MCform} and \eqref{eq:maurercartan}.
The Maurer-Cartan equation expressed in terms of \(\cP\) and \(\cQ\) reads
\begin{equation}\begin{aligned} 
\cD \cP &= \dd \cP + \cQ \wedge \cP + \cP \wedge \cQ = 0 \,, \\
\cH &= \dd \cQ + \cQ \wedge \cQ = - \cP \wedge \cP \,.
\end{aligned}\end{equation}
The first equation can be rewritten as \(\dd \cP^\alpha + {f_{A\beta}}^\alpha \cQ^A \wedge \cP^\beta = 0\). 
This is Cartan's structure equation for the vielbein \(\cP^\alpha\) and shows that \(\cQ^\alpha_\beta = \cQ^A {f_{A\beta}}^\alpha\) is a connection on the tangent bundle \(T\cM\) compatible with the metric \(g_{\alpha\beta}\).
The second equation is nothing but \eqref{eq:hscalar}, as can be seen by expressing it as \(\cH = - {f_{\alpha\beta}}^A J_A \cP^\alpha \wedge \cP^\beta\).
Moreover, it shows that the holonomy group of \(\cM\) is given by \(\mathrm{Hol}(\cM) = H\).

Let us finally discuss the isometries of \(\cM\) and the gauged version of the above construction.
The metric \(g_{\alpha\beta}\) is invariant under the left action of \(G\), therefore every element of \(G\) (acting on \(\cM\) from the left) corresponds to an isometry of \(\cM\) and therefore its isometry group must be (at least contained in) \(G\).
We start with a discussion of the action of an infinitesimal isometry on the coset representative \(L\), described by the left action of 
\begin{equation}
g = 1 + \epsilon^\rho t_\rho \,,
\end{equation}
where \(t_\rho\in \g\),
This induces a transformation of the scalars \(\phi\) along the corresponding Killing vector \(k_\rho\),
\begin{equation}
\phi' = \phi + \epsilon^\rho k_\rho \,.
\end{equation}
According to \eqref{eq:gcosettransf} we need a compensating \(H\)-transformation
\begin{equation}
h(\phi, g) = 1 - \epsilon^\rho W_\rho \,,
\end{equation}
where \(W_\rho \in \h\).
Inserting this into \eqref{eq:gcosettransf} and collecting all terms at linear order in the parameter \(\epsilon^\rho\) yields
\begin{equation}
L^{-1} t_\rho L = \iota_\rho \cP + \cQ_\rho \,,
\end{equation}
where the moment map \(\cQ_\rho\) is given by
\begin{equation}
\cQ_\rho = \iota_\rho \cQ - W_\rho \,.
\end{equation}
Notice that this agrees precisely with the general form of the moment map as defined in \eqref{eq:genmomentmap}.
To describe a gauged sigma model on \(\cM\) 
we proceed along the lines of the general discussion and select a subalgebra \(\g^g\) of \(\g\) using the embedding tensor formalism.
The generators \(X_I\) of \(\g^g\) are given in terms of \(t_\rho\) by \eqref{eq:gaugegenerators}.
We then introduce the gauged version of the Maurer-Cartan form \eqref{eq:maurercartan}
\begin{equation}
\hat\omega = L^{-1} \left(d + A^I X_I\right) L \,.
\end{equation}
It is by construction invariant under a local transformation of the form \(\delta L = \epsilon^I(x) X_I L\) if we demand \(A^I\) to transform according to \eqref{eq:Atransf}.
We learn from our previous considerations that for the gauged versions of the vielbein \(\cP\) and the connection \(Q\) this yields
\begin{equation}\begin{aligned}
\hat\cP &= \cP + A^I \cP_I \,, \\
\hat\cQ &= \cQ + A^I \cQ_I \,,
\end{aligned}\end{equation}
which is exactly the same as \eqref{eq:gaugedP} and \eqref{eq:gaugedQ}, so \(\hat\cP\) and \(\hat\cQ\) indeed are the correct quantities to describe the gauged sigma model on \(\cM = G/H\).

Instead of working with the generators \(X_I\) themselves, it is often more convenient to work with their contracted or dressed version
\begin{equation}\label{eq:dressedgenerator}
\cT_I = L^{-1} X_I L = \cP_I + \cQ_I \,,
\end{equation}
and \(\cP_I\) and \(\cQ_I\) are the \(\k\)-part and \(\h\)-part of \(\cT_I\).
Since the coset representative \(L\) is invertible, \(\cT_I\) carries the same amount of information as \(X_I\) and clearly satisfies the same commutator algebra.
One can go one step further and also dress the remaining index \(I\) with the vielbein \(\cV^I_\al{2}\) to obtain
\begin{equation}\label{eq:Ttensor}
\cT_\al{2} = \cV^I_\al{2} \cT_I \,.
\end{equation}
This object is often called the T-tensor \cite{deWit:1981sst, deWit:1982bul}.
In the same way as the embedding tensor \(\Theta\) decomposes into irreducible representation of \(\g\), the T-tensor can be decomposed into irreducible representations of \(\h\).
Again, the linear constraint restricts which representations can appear in a consistently gauged supergravity.
The allowed representations for \(\cT_\al{2}\) can be obtained by branching the allowed \(\g\)-representation of \(\Theta\) into \(\h\)-representations.

The T-tensor -- or equivalently its components \(\cQ_\al{2}\) and \(\cP_\al{2}\) -- features in the construction of the fermionic shift matrices \(A_0\) and \(A_1\).
Denoting the \(\h\)-representations of the gravitini and the spin-1/2 fermions by \(\mathbf{s}\) and \(\mathbf{x}\), respectively, \(A_0\) and \(A_1\) a priori transform in the tensor product representations \(\mathbf{s} \otimes \mathbf{\overline{s}}\) and \(\mathbf{x} \otimes \mathbf{\overline{s}}\).%
\footnote{The \(\h\) representation of \(A_0\) is furthermore often restricted since the gravitino mass term \(A^i_{0\,j} \bar\psi_{M i} \Gamma^{MN} \psi^j_N\) can impose an (anti-)symmetry property on \(A_0\).}
The components of \(\cT_\al{2}\) that transform in a representation which appears in these tensor products agree with the respective components of \(A_0\) and \(A_1\).
In appendix~\ref{app:susycalculations} we further elaborate on the relation between the T-tensor and the fermionic shift matrices and give explicit expressions for \(\cQ_\al{2}\) and \(\cP_\al{2}\) in terms of \(A_0\) and \(A_1\).

We finally want to point out that also the vielbeins \(\cV^\al{p}_{I_p}(\phi)\) of the kinetic matrices \(\cM_{I_p J_p}\) \eqref{eq:kinmatrix} are nothing but the coset representative \(L(\phi)\) taken in the respective representations of \(G\) and \(H\).
In this sense we can express any scalar field dependence solely in terms of the coset representative (and its derivatives). 




\chapter{Classification of Maximally Supersymmetric Solutions}\label{chap:classification}

In this chapter we discuss classical supersymmetric solutions of supergravity and give a complete classification of all maximally supersymmetric solutions in gauged and ungauged supergravities.
This chapter is based on \cite{Louis:2016tnz}.

A solution of a (super)gravitational theory is a classical field configuration solving the theories equations of motion.
The equations of motion are a set of second order differential equations and include the Einstein field equations.
Thus, the space-time metric and the corresponding space-time manifold on which the metric lives are an essential part of every supergravity solution.
Moreover, such solutions can often be used as a background or vacuum configuration for an (effective) quantum field theory and fluctuations around the solutions can be interpreted as quantum fields.
Therefore, in this context the terms \emph{solution}, \emph{background} and \emph{vacuum} are often used interchangeably. 
Analogously, the values of the various fields in a given solution are sometimes called \emph{background} or \emph{vacuum expectation values}.

As postulated in the previous chapter, the Langrangian of a supergravity theory is invariant under local supersymmetry transformations which are parametrized by one or multiple space-time dependent, spinorial parameters \(\epsilon(x)\).
We often denote the total number of independent real components of \(\epsilon(x)\) by \(q\).

Given a classical solution it is consequently possible to ask if some (and how many) of these supersymmetries are preserved as symmetries of the solution, i.e.~as symmetries of the associated background field configuration.
This requires the existence of a spinor field \(\left<\epsilon(x)\right>\) on the previously mentioned background space-time manifold such that the solution does not vary under a supersymmetry transformation with respect to \(\left<\epsilon(x)\right>\).
In general, \(\left<\epsilon(x)\right>\) will not be uniquely determined but can dependent on some number \(q_0\) of independent real constant parameters.
In this case we say that the solutions preserves \(q_0\) (of maximally \(q\)) supersymmetries.
Note that even though \(\left<\epsilon(x)\right>\) might have a non-trivial profile over space-time, the preserved supersymmetries have to be understood as global symmetries as they only dependent on constant parameters.

Instead of starting with a solution of the equations of motion and checking if it preserves some supersymmetry, it is in practice usually more convenient to use the supersymmetry variations in the first place for finding supersymmetric solutions.
The supersymmetry variations -- often also called Killing spinor equations -- are first order differential equations for the fields as well as for \(\epsilon(x)\) and usually already imply the equations of motion. 
Analogously their solutions \(\left<\epsilon(x)\right>\) are called Killing spinors.
Their existence heavily restricts the admissible space-time geometries,
for example their bilinears of the form \(\left<\bar\epsilon \Gamma\!_M \epsilon\right>\) can give rise to Killing vectors and therefore to isometries of the background space-time.
In the past the analysis of Killing spinors has been used very successfully in many cases to classify supersymmetric solutions \cite{Tod:1983pm, Behrndt:1997ny, Gauntlett:2002nw, FigueroaO'Farrill:2002ft, Gauntlett:2002fz, Gutowski:2003rg, Bellorin:2005zc, Cacciatori:2007vn, Bellorin:2007yp, Gran:2008vx}.

To be more specific, let us denote all bosonic fields collectively by \(B(x)\) and the fermionic fields by \(F(x)\),
such that supersymmetry transformations take the schematic form%
\footnote{We have collected the general form of all supersymmetry transformations in appendix~\ref{app:susyvariations}, neglecting all terms of higher order in the fermions.} \cite{Freedman:2012zz}
\begin{equation}
\delta_\epsilon B(x) = \bar F(x) f_1(B(x)) \epsilon(x) + \cO(F^3) \,,\qquad 
\delta_\epsilon F(x) = f_2(B(x)) \epsilon(x) + \cO(F^2) \,.
\end{equation}
Here we only want to discuss purely bosonic solutions, this means solutions for which all fermionic fields are set identically equal to zero, i.e.\footnote{See \cite{Houston:2016nbk, Houston:2017qlr} for some recent work on fermionic supersymmetric solutions.}
\begin{equation}
\left<F(x)\right> = 0 \,.
\end{equation}
Under this assumption the supersymmetry variations \(\left<\delta_\epsilon B(x)\right>\) of the bosonic fields vanish trivially, we are therefore only left with finding solutions \(\left<B(x)\right>\) and \(\left<\epsilon(x)\right>\) of
\begin{equation}\label{eq:deltaF}
\delta_\epsilon F(x) = B(x) \epsilon(x) = 0 \,.
\end{equation}
Adopting the notation of the previous chapter, the fermionic fields present in a supergravity theory are the gravitini \(\psi^i_M\) as well as the spin-1/2 fermions \(\chi^a\)
and the condition \eqref{eq:deltaF} becomes
\begin{equation}\label{eq:backgroundvariations}
\left<\delta_\epsilon \psi^i_M\right> = \left<\delta_\epsilon \chi^a\right> = 0 \,,
\end{equation}
with the supersymmetry variations given in \eqref{eq:generalfermionicvariations}.

Our goal is to determine all maximally supersymmetric solutions, i.e.~all bosonic field configurations for which \eqref{eq:backgroundvariations} is satisfied for the maximal number of \(q\) independent Killing spinor fields \(\epsilon(x)\).
Our strategy relies on the following simple case-by-case analysis.
In the first step we exclude all background fluxes, this means the only fields for which we allow non-trivial values are the space-time metric as well as possibly the scalar fields.
This simplifies the variations \eqref{eq:backgroundvariations} considerably such that the resulting Killing spinor equations can be integrated directly.
In this case the only allowed background space-times are flat \(D\)-dimensional Minkowski space \(M_D\) or \(D\)-dimensional anti-de Sitter space \(AdS_D\).

To obtain more complicated solutions one has to allow for non-trivial fluxes, i.e non-zero values of the $p$-form field strengths of some of the other bosonic fields that might be present in the gravity multiplet.
However, generically such fluxes break supersymmetry, at least partially.
Only for a small set of theories (which are listed in table~\ref{tab:pformfluxes}) it is actually possible to switch on fluxes without breaking supersymmetry.
Moreover, we argue that for all these theories the maximally supersymmetric solutions are in one-to-one agreement with the solutions of the corresponding ungauged theories.
These solutions have already been determined and classified in the literature,
which eventually allows us to give a complete list of all maximally supersymmetric solutions (cf.~table~\ref{tab:adsbackgrounds}).

According to the above outline this chapter is organized as follows.
In the first part~\ref{sec:susybackgrounds} we only study solutions without fluxes, whereas in the second part~\ref{sec:susyflux} we discuss solutions with non-trivial background fluxes.
Moreover, we relegate a technical computation 
to appendix~\ref{app:integrability}.

\section{Supersymmetric solutions without fluxes}\label{sec:susybackgrounds}

Let us first analyze the situation where all background fluxes vanish and hence
eqs.~\eqref{eq:gravitinovariation} and \eqref{eq:spin12variation}
simplify. 
If all supercharges are preserved, 
\(\delta \chi^a = 0\) implies
via \eqref{eq:spin12variation}
that\footnote{The equation only has to hold in the background,
i.e.\ the condition reads $\langle {A_1}\rangle=0$.
However, in order to keep the notation manageable 
we generically omit the brackets henceforth.}
\begin{equation}
A_1 = 0\ .
\end{equation}
On the other hand, the vanishing of the gravitino variation \eqref{eq:gravitinovariation}
\begin{equation}\label{eq:susyvariation}
\delta \psi^i_M = \hat\cD_M \epsilon^i + A_{0\,j}^i \Gamma_M \epsilon^j = 0 \,,
\end{equation}
says that $\epsilon^i$ has to be a Killing spinor.
Its existence implies a strong constraint on the space-time manifold
which can be derived by 
acting with another covariant derivative, anti-symmetrizing
and using \eqref{eq:Dcommgauged}. 
This  implies
\begin{equation}\label{eq:riemann}
\biggl[\left(\tfrac{1}{4} {R_{MN}}^{PQ} \delta^i_k+ 2 A^i_{0\,j} A^j_{0\,k} \delta^P_M \delta^Q_N \right) \Gamma_{PQ} + 2\left(\hat\cD_{[M} A_0 \right)^i_k\Gamma_{N]} \biggr]\epsilon^k = 0 \ ,
\end{equation} 
where we also used that $\hat\cH^R_{MN}$ vanishes in backgrounds without any fluxes
and where 
the covariant derivative of \(A_0\) is defined as
  \(\hat\cD_M A_0 = \partial_{M} A_0 - \bigl[\hat\cQ_{M}, A_0\bigr]\).
In a background which preserves all supercharges the expression in the
bracket in \eqref{eq:riemann} has to vanish at each order in the \(\Gamma\)-matrices independently.
From the term linear in \(\Gamma\) we learn that \(A_0\) is covariantly
constant.
The part quadratic in \(\Gamma\) then says that \(A^2_0\) needs to be proportional to the identity matrix and must be constant since
\begin{equation}
\partial_M A_0^2 = \hat\cD_M A_0^2 = 0 \,.
\end{equation}
Moreover, it implies that in a given supergravity the maximally supersymmetric backgrounds
have to be maximally symmetric space-times with
a Riemann tensor given by
\begin{equation}
R_{MNPQ} = - \frac{4}{\cN}\ \mathrm{tr}\!\left(A_0^2\right) \left(g_{MP} g_{NQ} - g_{MQ}g_{NP}\right) \,.
\end{equation}
From the canonical Einstein equations one readily infers that in such backgrounds the cosmological constant \(\Lambda\) is given by
\begin{equation}\label{LambdaA}
\Lambda = - \frac{2}{\cN}\, (D-1)(D-2)\, \mathrm{tr}\!\left(A^2_0\right) \,,
\end{equation}
and the background value of the 
scalar potential is given by \(\left<V\right> = \Lambda\).
Note that this is consistent with the expression \eqref{eq:generalpotential} for $V$.
For $A_0\neq0$ we thus have an  AdS background $\Sigma_D=\AdS_D$ while for 
$A_0=0$ the background is flat. 
So altogether fully supersymmetric backgrounds without background fluxes have to be one of the following cases
\begin{equation}\label{eq:firstresult}
 \Sigma_D=\AdS_D   \qquad \textrm{or} \qquad  
 \Sigma_D = \mathrm{Mink}_D
\, ,
\end{equation} 
up to local isometries.
We see in particular that without fluxes supersymmetric backgrounds with an $\AdS_{d}$ factor cannot exist for $d<D$.

Before we proceed let us note that in a given $D$-dimensional 
gauged supergravity
the existence of the $\Sigma_D=\AdS_D$ background requires the existence 
of a solution with
\begin{equation}\label{AdSsolution}
A_0^2= -\tfrac{\Lambda}{2(D-1)(D-2)}\, \mathbb{1}\ ,\qquad A_1 = 0\ .
\end{equation}
Explicit solutions have indeed been constructed in a variety of supergravities
(see, for example, \cite{Hristov:2009uj,deAlwis:2013jaa,Louis:2014gxa,Louis:2015mka,Louis:2015dca,Louis:2016qca} and references therein). 
However, from \cite{Nahm:1977tg} it is known that AdS superalgebras only exist for 
$D<8$ and in $D=6$ only for the non-chiral $\cN = (1,1)$ supergravity. 
In the other cases no solution of \eqref{AdSsolution} can exist.
The analysis of the conditions \eqref{AdSsolution} is the main objective of chapters~\ref{chap:ads} and~\ref{chap:adsmoduli}.

\section{Supersymmetric solutions with fluxes}\label{sec:susyflux}

In this section we extend our previous analysis in that we
consider backgrounds with non-trivial
fluxes and reanalyze the implications for the possible
space-time manifolds.
In this case  the vanishing supersymmetry variations of the spin-1/2
fermions  given in
\eqref{eq:spin12variation} 
immediately impose additional constraints.
As we will see, they are particularly strong for the fermions 
$\chi^{\hat a}$ in the gravitational multiplet.
Since the $\Gamma$-matrices and their antisymmetric products are
linearly independent, \(\delta \chi^a=0\) enforces
\begin{equation}\label{eq:spin12conditions}
A_1 = 0 \qquad \textrm{and}\qquad F^{(p)} = 0\ , 
\end{equation}
for all possible values of \(p\).\footnote{In even dimensions \(D\)
  all antisymmetric products of gamma matrices are linearly
  independent while in odd dimensions only those up to rank
  \((D-1)/2\) are linearly independent as can bee seen from
  \eqref{eq:gammahodgeodd}. This however is strong enough to enforce \eqref{eq:spin12conditions}.}
This seems to imply that no background fluxes can be turned on.
However, this conclusion  can be evaded either if  there simply are
no  spin-1/2 fermions in the gravity multiplet or 
if there is an (anti-)self-dual field strength in a chiral theory. 

In the first case there is no condition on the fluxes $F^{\hat\alpha_p}$
which appear in the gravitino variation \eqref{eq:gravitinovariation} and
\eqref{eq:cFM} but only on the fluxes $F^{\tilde\alpha_p}$ which feature
in \eqref{eq:cFtilde}.
The second exception follows from the definition of the chirality operator  
\(\Gamma_\ast\)
(given in \eqref{eq:gamma5}) which implies that 
in even dimensions \(D\) the Hodge-dual of a \(p\)-form \(F^{(p)}\) satisfies 
\begin{equation}\label{eq:Fdual}
\ast F^{(p)} \cdot \Gamma = -(-1)^{p(p-1)/2} i^{D/2 + 1} \left(F^{(p)} \cdot \Gamma\right) \Gamma_\ast \ ,
\end{equation}
where we abbreviated
\(F^{(p)} \cdot \Gamma=F_{N_1\dots N_p}^{(p)} \Gamma^{N_1\dots N_p}\)
(and used \eqref{eq:gammahodgeeven}).
Note that the prefactor is real in dimensions \(D = 2 \mod 4\), which are precisely those dimensions in which chiral theories can exist.
In these dimensions one finds for an (anti-)self-dual \(D/2\)-form \(F_{\pm} = \pm \ast F_{\pm}\) that
\begin{equation}\label{eq:FGselfdual}
F_{\pm} \cdot \Gamma = \left(F_{\pm} \cdot \Gamma\right) P_{\pm} \,,
\end{equation}
where \(P_\pm = \frac{1}{2}\left(\id \pm \Gamma_\ast\right)\).
In the chiral supergravities in $D=6, 10$ \cite{Schwarz:1983wa,Schwarz:1983qr,
Nishino:1984gk,Awada:1985er}  the supergravity multiplet contains 
two or four-form fields, respectively, with self-dual field strengths \(F_+^{\hat\alpha_{D/2}}\).
In these theories the gravitini and consequently also the supersymmetry parameters \(\epsilon^i\)
are left-handed. Therefore, a term of the form \((F^{\hat\alpha_{D/2}}_+\cdot \Gamma)\, \epsilon^{i-}\) cannot appear in \eqref{eq:spin12variation} which indeed shows that a non-vanishing background value for a self-dual field strength does not break supersymmetry in these theories.
Nevertheless, \(F_+^{\hat\alpha_{D/2}}\) still enters the variation of the gravitini since a different contraction with \(\Gamma\)-matrices appears in \eqref{eq:cFM}.
Hence, maximally supersymmetric solutions with non-trivial background flux are possible.
 
The previous considerations in this section enable us to conclude
that solutions which preserve all supercharges of a given supergravity
and which are 
different from the ones
described 
in the previous section 
can only exist if at least one of the following two conditions hold:

\emph{Either the gravity multiplet contains $p$-form gauge fields but 
no spin-$\frac12$ fermions~$\chi^{\hat a}$ 
or the theory is chiral and (some of) 
the gauge potentials in the gravity multiplet 
satisfy an (anti-) self-duality condition such that they drop out of 
$\delta\chi^{\hat a}$.}

In table~\ref{tab:pformfluxes} we list all possible supergravities in dimensions \(D \geq 3\) which satisfy these conditions, together with the possible background fluxes.%
\footnote{It is in fact easy to see that such theories cannot exist in \(D=3\) dimensions.
Since three-dimensional gravity is non-dynamical, the graviton, and via supersymmetry also the gravitini, do not carry any on-shell degrees of freedom.
So whenever the gravity multiplet contains vector or scalar fields (which are dual in three dimensions) it must also contain spin-1/2 fields as supersymmetric partners.}
We now proceed by analyzing the supersymmetry variation of the gravitini \eqref{eq:gravitinovariation} for these theories 
in more detail.

\begin{table}[htb]
\centering
\begin{tabular}{|c|c|c|c|c|}
\hline
dimension & supersymmetry & q 
& possible flux & ref. \\
\hline
\(D = 11\) &  \(\cN = 1\) & 32 & \(F^{(4)}\) & \cite{FigueroaO'Farrill:2002ft} \\
\(D = 10\) & IIB & 32 & \(F_+^{(5)}\) & \cite{FigueroaO'Farrill:2002ft} \\
\(D = 6\) & \(\cN = (2,0)\) & 16 & \(5 \times F_+^{(3)}\) & \cite{Chamseddine:2003yy} \\
\(D = 6\) & \(\cN = (1,0)\) & 8 & \(F_+^{(3)}\) & \cite{Gutowski:2003rg} \\
\(D = 5\) & \(\cN = 2\) & 8 & \(F^{(2)}\) & \cite{Gauntlett:2002nw} \\
\(D = 4\) & \(\cN = 2\) & 8 & \(F^{(2)}\) &  \cite{Tod:1983pm} \\
\hline
\end{tabular}
\caption{Supergravity theories which allow for a background flux that does not break supersymmetry. $q$ denotes the number of real supercharges. In the last column we give the reference for the classification of maximally supersymmetric solutions.}
\label{tab:pformfluxes}
\end{table}

Taking a covariant derivative of \eqref{eq:gravitinovariation} and using \eqref{eq:Dcommgauged} we arrive at the integrability condition
\begin{equation}\begin{split}\label{eq:fluxgaugeintegrability}
\biggl(&\frac{1}{4}R_{MNPQ} \Gamma^{PQ} \delta^i_j - \left(\hat\cH_{MN}\right)^i_j + 2 \left(\hat\cD_{[M} \cF_{N]} + \hat\cD_{[M} A_0 \Gamma_{N]}\right)^i_j \\
&\qquad\qquad+ \left[\left(\cF_M + A_0 \Gamma_M\right)^i_k\left(\cF_N + A_0 \Gamma_N\right)^k_j - (M \leftrightarrow N)\right]\biggr) \epsilon^j = 0 \,.
\end{split}\end{equation}
In a maximally supersymmetric background this has to vanish at each order in the \(\Gamma\)-matrices independently.
As we show in appendix~\ref{app:integrability} for all the theories in table~\ref{tab:pformfluxes} the only term at zeroth order in $\Gamma$ is \(\hat\cH^R_{MN}\)
and thus we find that
\begin{equation}\label{eq:rfsvac}
\hat\cH^R_{MN} = 0 \,.
\end{equation}
Furthermore, due to \eqref{eq:spin12conditions} all scalar fields have vanishing field strengths,
\(F^{\hat\alpha_1} = F^{\tilde\alpha_1} = 0\),
and therefore, using  \eqref{eq:hscalar}, \(\cH^R_{MN}\) automatically vanishes. From \eqref{eq:hrgenerators} we then learn that 
\eqref{eq:rfsvac} implies
\begin{equation}\label{eq:hgaugecondition}
F^{I_2}_{MN} \cQ^R_{I_2} = 0 \,.
\end{equation}

In a next step we show that \eqref{eq:hgaugecondition} implies
that there can be either no background fluxes at all or that
alternatively both \(A_0\) and \(\cQ^R_{I_2}\) vanish in the
background.  
To see this 
we derive in appendix~\ref{app:susycalculations} that the supersymmetry 
conditions \(A_1 = 0\) of \eqref{eq:spin12conditions} enforce \(\cQ^R_\al{2}= \cV^{I_2}_\al{2} \cQ^R_{I_2}\) to be of the generic form
\begin{equation}\label{eq:hgauge}
\cQ^R_{\hat\alpha_2} = (D-3) 
 \bigl\{A_0, B_{\hat\alpha_2}\bigr\} \,,\qquad \cQ^R_\tal{2} = 0 \,.
\end{equation}
Let us study the implications of \eqref{eq:hgauge} for the supergravities 
of table~\ref{tab:pformfluxes}. We already showed that
the theories which are not in this list cannot have non-vanishing background fluxes so that \eqref{eq:hgaugecondition} is trivially satisfied and does not 
impose any conditions on \(A_0\). Similarly, for
the first three theories in the table~\ref{tab:pformfluxes} it is known that 
deformations by a non-vanishing \(A_0\) do not exist.
In addition no massless vector fields appear 
in the gravitational or in any other multiplet.
Hence 
\(\cQ^R_{I_2}\) does
not exist
and the theories are always ungauged, consistent with \eqref{eq:hgauge}.
On the other hand the possible background fluxes of higher rank field strengths
are not restricted. Similarly,
the six-dimensional \(\cN = (1,0)\) theories
cannot be deformed by \(A_0 \neq 0\) and do not feature any vector
fields in the gravity multiplet.
In principle it is possible to gauge these theories by
coupling them to vector multiplets. However, in the maximally
supersymmetric background this is forbidden due to \eqref{eq:hgauge}
and therefore also here \(\cQ^R_{I_2} = 0\) holds. 
This was explicitly shown in \cite{Akyol:2010iz}.

The analysis of the two remaining supergravities in the list, the four- and five-dimensional \(\cN = 2\) theories, is slightly more involved.
Both can be deformed by \(A_0 \neq 0\) and both have one single gauge field, 
the graviphoton \(A^{\hat\alpha_2}\), in the gravity multiplet.
Consequently there is also only one single matrix
\(B_{\hat\alpha_2}\).
As the graviphoton is an R-symmetry singlet, \(B_{\hat\alpha_2}\)
has to be proportional to the identity.
Therefore \eqref{eq:hgauge} gives
\begin{equation}
F^{I_2}_{MN} \cQ^R_{I_2} 
\sim  F_{MN} A_0 \,,
\end{equation}
where $F_{MN}$ is the field strength of the graviphoton.
As a consequence, \eqref{eq:hgaugecondition} implies that 
either \(F_{MN}\)  or \(A_0\) has to vanish in the background.
For $N=2$ theories in \(D=4\) this has been explicitly shown for
pure gauged supergravity in \cite{Caldarelli:2003pb} and for arbitrary
gauging in \cite{Hristov:2009uj}. For pure gauged supergravity in
\(D=5\) this has been obtained in \cite{Gauntlett:2003fk} and related
results for arbitrary gaugings in \cite{Bellorin:2008we}.
In contrast to their results our analysis here is completely general and does not rely on the concrete formulation of the gauged supergravities.
\pagebreak[1]

Let us summarize our results so far. There are two different branches of maximally supersymmetric solutions:
\begin{itemize}
\item[i)] \(A_0 \neq 0\).

In this case all background fluxes must necessarily vanish and the background space-time is \(AdS_D\) as described in section~\ref{sec:susybackgrounds}.

\item[ii)] \(A_0 = 0\).

In this case non-vanishing background fluxes are allowed but \(\hat\cQ^R_M\)
vanishes in the background. As a consequence the fermionic
supersymmetry transformation \eqref{eq:gravitinovariation}
take exactly the same form as for the ungauged theory and
hence the maximally supersymmetric solutions coincide
with the solutions of the ungauged theories. 

\end{itemize}

The solutions of the ungauged theories  have been classified for 
all supergravities listed in table~\ref{tab:pformfluxes}
and this classification  can thus  be used for  case ii).
These solutions can be found in the references given in table~\ref{tab:pformfluxes}.
%
Let us shortly review the main results.
For vanishing \(A_0\) and \(Q_M\) the integrability condition \eqref{eq:fluxgaugeintegrability} simplifies considerably and reads
\begin{equation}\label{eq:fluxintegrability}
\frac{1}{4}R_{MNPQ} \Gamma^{PQ} \delta^i_j + 2 \left(\nabla_{[M} \cF_{N]}\right)^i_j
+ 2 \left(\cF_{[M}\right)^i_k\left(\cF_{N]}\right)^k_j = 0 \ .
\end{equation}
Expanding in powers of the \(\Gamma\)-matrices and collecting all  terms quadratic in \(\Gamma\) we observe that 
the Riemann tensor of the space-time background is expressed solely in terms of the background flux \(F^{\hat\alpha_p}\) and its derivatives.
Furthermore, all supergravities listed in table~\ref{tab:pformfluxes}
have solutions  with the property
\begin{equation}\label{eq:nablaf}
\nabla F^{\hat\alpha_p} = 0 \,.
\end{equation}
Only in the five-dimensional \(\cN = 2\) supergravity one finds
solutions of \eqref{eq:fluxintegrability}
which do not satisfy \eqref{eq:nablaf} \cite{Gauntlett:2002nw}.
In all other cases  
there are no additional solutions or in other words 
\emph{all} solutions share the property
\eqref{eq:nablaf}. 
For these solutions
also the Riemann tensor is parallel, i.e.\ \(\nabla_M R_{NPQR} = 0\), which says that the space-time is locally symmetric.
The locally symmetric spaces with Lorentzian signature are classified 
\cite{Cahen1970, FigueroaO'Farrill:2002ft}.\footnote{They have to be 
locally isometric to 
a product of a Riemannian symmetric space times a Minkowskian, dS, 
AdS or H\textit{pp}-wave geometry.}
Furthermore, in \cite{Gutowski:2003rg,Chamseddine:2003yy,FigueroaO'Farrill:2002ft}
it was shown that $F^{\hat\alpha_p}$ can be written as
\begin{equation}\label{eq:decomp}
F^{\hat\alpha_p} = v^{\hat\alpha_p} F \qquad \text{or} \qquad F^{\hat\alpha_p} = v^{\hat\alpha_p}\left(F + \ast F\right) \,,
\end{equation}
where \(v^{\hat\alpha_p}\) is constant and \(F\) is decomposable, i.e.\
 it can always be expressed as the wedge-product of \(p\) one-forms.
The second decomposition holds for a self-dual \(F^{\hat\alpha_p}\).\footnote{Notice that in \(D=4\) dimensions \(F^{\hat\alpha_2}\) itself is not necessarily decomposable.
Instead we have to split it into a complex self-dual and anti-self-dual part and use the appropriate form of the second decomposition in \eqref{eq:decomp}.}
Excluding the trivial case where \(F = 0\) and where the background is flat, there are therefore only two cases to be distinguished:
\begin{enumerate}
\item \(F\) is \emph{not} a null form (i.e.\ $F^2\neq0$).

These are the well-known solutions of Freund-Rubin type \cite{Freund:1980xh} for which the space-time is the product of an AdS space and a sphere such that \(F\) is a top-form on one of the two factors, i.e.
\begin{equation}
\cM_D = \AdS_p \times S^{(D-p)} \qquad\text{or}\qquad \cM_D = \AdS_{(D-p)} \times S^p \,.
\end{equation}
We explicitly list all these solutions in table~\ref{tab:adsbackgrounds}. 
Notice that besides the pure \(\AdS_D\) solutions discussed in section~\ref{sec:susybackgrounds}
these are the only possible maximally supersymmetric solutions with an \(\AdS\)-factor. All other \(\AdS\) solutions in supergravity will necessarily break supersymmetry.

\item \(F\) is a null form (i.e.\ $F^2=0$).

These solutions are homogeneous \textit{pp}-waves (H\textit{pp}-waves)
first discovered by Kowalski-Glikman \cite{KowalskiGlikman:1984wv,
  KowalskiGlikman:1985im} and therefore often referred to as KG solutions.
They can be obtained from the respective \(\AdS \times S\) solutions by a Penrose limit \cite{Penrose1976, Gueven:2000ru, Blau:2002dy, Blau:2002mw}.
\end{enumerate}

As we have already mentioned above, this list of solutions is exhaustive if one excludes the five-dimensional \(\cN = 2\) supergravity.
In the latter theory there can be more exotic solutions with \(F\) not parallel or decomposable and consequently also the background space-time \(\cM_D\) not locally symmetric.
These exceptional solutions are classified in \cite{Gauntlett:2002nw}
and are a G\"odel-like universe and the near-horizon limit of the
rotating BMPV  black hole \cite{Breckenridge:1996is}.%
\footnote{In \cite{Gauntlett:2002nw} three additional solutions have been found but were left unidentified, it was shown in \cite{Fiol:2003yq} that they also belong to the family of near-horizon BMPV solutions. See also \cite{Chamseddine:2003yy}.} The latter family of solutions contains the \(AdS_2 \times S^3\) and \(AdS_3 \times S^2\) solutions as special cases.
Even though there are maximally supersymmetric solutions which are not locally symmetric, they all happen to be homogeneous space-times \cite{Cahen1970, AlonsoAlberca:2002wr, Chamseddine:2003yy}.
It is also interesting to note that the maximally supersymmetric solutions of the theories with 8 real supercharges in \(D=4,5,6\) dimensions are related via dimensional reduction or oxidation \cite{LozanoTellechea:2002pn, Chamseddine:2003yy}.


\begin{table}[htb]
\centering
\begin{tabular}{|c|c|c|cl|l|c|}
\hline
dim. & SUSY & q 
& \multicolumn{2}{|c|}{\(AdS \times S\)} &  H\textit{pp}-wave & others \\
\hline
\multirow{2}{*}{\(D = 11\)} & \multirow{2}{*}{\(\cN = 1\)} & \multirow{2}{*}{32} & \(\AdS_4 \times S^7\) &\multirow{2}{*}{\cite{Freund:1980xh}}& \multirow{2}{*}{\(\mathrm{KG}_{11}\) \cite{KowalskiGlikman:1984wv}} & \multirow{2}{*}{-}  \\
&&&\(\AdS_7 \times S^4\) &
&& \\ [0.9ex]
\(D = 10\) & IIB & 32 & \(\AdS_5 \times S^5\)& \cite{Schwarz:1983qr,Schwarz:1983wa} & \(\mathrm{KG}_{10}\) \cite{Blau:2001ne} & - \\[0.9ex]
\multirow{2}{*}{\(D=6\)} & \(\cN = (2,0)\) & 16 & \multirow{2}{*}{\(\AdS_3 \times S^3\)} &\multirow{2}{*}{\cite{Gibbons:1994vm}}& \multirow{2}{*}{\(\mathrm{KG}_{6}\)\cite{Meessen:2001vx} } & \multirow{2}{*}{-} \\
& \(\cN = (1,0)\) & 8 & && & \\[0.9ex]
\multirow{2}{*}{\(D = 5\)} & \multirow{2}{*}{\(\cN = 2\)} & \multirow{2}{*}{8} & \(\AdS_2 \times S^3\)&\multirow{2}{*}{\cite{Gibbons:1994vm,Chamseddine:1996pi}} & \multirow{2}{*}{\(\mathrm{KG}_{5}\) \cite{Meessen:2001vx}} & G\"odel-like \cite{Gauntlett:2002nw}, \\
&&& \(\AdS_3 \times S^2\)  &&& NH-BMPV \cite{Cvetic:1998xh,Gauntlett:1998fz} \\[0.9ex]
\(D = 4\) & \(\cN = 2\) & 8 & \(\AdS_2 \times S^2\) &\cite{Bertotti:1959pf,Robinson:1959ev}& \(\mathrm{KG}_{4}\) \cite{KowalskiGlikman:1985im} & - \\
\hline
\end{tabular}
\caption{All possible maximally supersymmetric solutions with
  non-trivial flux; $q$ denotes the number of real supercharges, cf.\
\cite{AlonsoAlberca:2002dw}.}
\label{tab:adsbackgrounds}
\end{table}

\chapter{AdS Solutions and their Moduli Spaces}\label{chap:ads}

In the previous chapter we classified all maximally supersymmetric supergravity solutions and found that there are two separate classes of AdS solutions.
Firstly, whenever a theory allows for non-supersymmetry breaking fluxes, it always has a solution of the form \(AdS \times S\) such that the flux is a top-form on either the AdS or the sphere factor.
However, these solutions appear only sporadically in a small class of theories.
On the other hand, gauged supergravities generically admit pure AdS solutions.
The existence of such solutions is restricted by conditions on the fermionic shift matrices \eqref{AdSsolution}.

In this chapter we focus exclusively on the latter class of AdS solutions and analyze the implications of the conditions \eqref{AdSsolution}.
For technical reasons this analysis is restricted to dimensions \(D \geq 4\).
As mentioned in chapter~\ref{chap:supergravity} and explicitly computed in appendix~\ref{app:susycalculations}, the shift matrices \(A_0\) and \(A_1\) depend on the moment maps \(\cQ^R_I\) and Killing vectors \(\cP_I\). 
Therefore, a constraint on \(A_0\) and \(A_1\) also restricts the possible gauge groups.
Consequently, a maximally supersymmetric AdS solution is not possible for arbitrary gaugings.
We argue that after spontaneous symmetry breaking the gauge group must alway be of the form
\begin{equation}\label{eq:AdSgaugegroup}
H^g = H^g_R \times H^g_\mathrm{mat} \,,
\end{equation}
where \(H^g_R\) and \(H^g_\mathrm{mat}\) are products of abelian and compact semi-simple Lie groups (both of them can be trivial).
The group \(H^g_R\) is uniquely determined by the conditions on the shift matrices and thus is completely fixed. (Of course \(H^g_R\) depends on the space-time dimension and the number of supersymmetries of the respective supergravity.)
On the other hand \(H^g_\mathrm{mat}\) is mostly unconstrained and only subject to general restrictions of supergravity gaugings (compare with the discussion in chapter~\ref{sec:gauging}).
However, \(H^g_\mathrm{mat}\) requires for the existence of vector multiplets and therefore can not exist for highly supersymmetric theories where the only allowed supermultiplet is the gravitational one.

This nicely resembles the structure of the holographically dual SCFTs.
A gauge symmetry of the AdS background translates via the AdS/CFT dictionary \cite{Gubser:1998bc, Witten:1998qj} to a global symmetry of the boundary CFT.\footnote{If we denote the conserved current of a global symmetry of the boundary CFT by \(J\) it couples via \(\int_{\partial AdS} A \wedge \ast J\) to the gauge field \(A\) of a local symmetry in the bulk.}
The first factor \(H^g_R\) in \eqref{eq:AdSgaugegroup} corresponds to the R-symmetry of the SCFT.
As a subgroup of the full superconformal group the R-symmetry must always be present and cannot be chosen freely.\footnote{Note, however, that there are SCFTs without an R-symmetry, as for example three-dimensional \(\cN=1\) theories. In this case also the gauge group factor \(H^g_R\) of the dual supergravity solution is trivial.}
Moreover, many SCFTs are allowed to posses additional global symmetries which commute with the R-symmetry.
They are called flavor symmetries and correspond to the second factor \(H^g_\mathrm{mat}\).

Note that in principle the conditions on the moment maps and Killing vectors which can be derived from \eqref{AdSsolution} are only necessary conditions for the existence of a maximally supersymmetric AdS solution.
Some supergravities allow for deformations that cannot be described as the gauging of a global symmetry and in some cases an AdS solution is only possible if these additional deformations are turned on.%
\footnote{A prime example are \(\cN = 1\) supergravities in four dimensions where a non-trivial superpotential is necessarily required for the existence of an supersymmetric AdS background.
Also the AdS solutions of half-maximal supergravity in seven dimensions discussed in chapter~\ref{sec:equal16} require additional massive deformations.}
On the other hand there are other theories that do not admit for AdS solutions at all, even though they can be gauged or otherwise deformed.
In particular, this is the case for all theories in dimensions \(D > 7\).

In the second part of this chapter we want to turn over to the moduli spaces of AdS solutions, this means we want to analyze if the conditions \eqref{AdSsolution} allow for continuous families of solutions. 
As explained in the introduction such moduli spaces correspond to the conformal manifolds (i.e.~the spaces spanned by exactly marginal deformations) of the dual SCFTs.
For this purpose we compute the variations of the shift matrices \(A_0\) and \(A_1\) (and hence of the AdS conditions \eqref{AdSsolution}) with respect to the scalar fields.
If these variations vanish to first order along some direction in the scalar manifold, the corresponding scalar field is massless and is therefore dual to a marginal deformation.
However, there is a slight complication as some of the massless scalar fields can arise as Goldstone bosons in connection with a spontaneous breaking of the gauge group.
These modes carry the additional degrees of freedoms of the now massive gauge fields and therefore do not count as physically independent fields.
The remaining directions in which the variations of \(A_0\) and \(A_1\) vanish are candidates for continuous deformation parameters, i.e.~moduli.
A true modulus is not only massless but also has no higher order contributions to the potential.
Analogously we demand its variation of the shift matrices to vanish at all orders.
This resembles the distinction between marginal and exactly marginal deformations on the dual SCFT side.

We do not attempt to discuss the moduli spaces of AdS solutions at the same level of generality as their gauge groups.
Instead, we only focus on a particular subset of theories where the scalar manifold is a symmetric homogeneous space of the form \(\cM = G/H\).
Moreover, we assume that there are no other deformations than gaugings.
This implies that the shift matrices can be entirely expressed in terms of the moment maps \(\cQ^R_I\) and the Killing vectors \(\cP_I\) and therefore varying the conditions on \(\cQ^R_I\) and \(\cP_I\) derived from \eqref{AdSsolution} has the same impact as varying \(A_0\) and \(A_1\) directly.
The variations of \(\cQ^R_I\) and \(\cP_I\) in turn can be expressed in a group theoretical language.
In particular, we find that every modulus transforms necessarily as a singlet with respect to the subgroup \(H^g_R\) of the total gauge group.
This is not only consistent with their interpretation as the supergravity dual of supersymmetric marginal deformations but also often constrains the existence of moduli considerably.
Nonetheless, a general analysis is still rather difficult.
Therefore, we outline the characteristic implications of our general conditions separately for theories with different numbers of supercharges.
The discussion simplifies the most for
 theories with more than 16 real supercharges.
Here the only allowed supermultiplet is the gravity multiplet.
The absence of other multiplets, in particular vector multiplets, makes the involved structures and equations considerably easier compared to theories with less supersymmetry.
One effect is that the complete gauge group is now only given by \(H^g_R\).
This makes it straightforward to show that the vanishing of a variation at first order implies that it vanishes also at all orders and that the moduli space (if existent) is a coset space as well.
We use these results in chapter~\ref{chap:adsmoduli} to determine the AdS moduli spaces for all such theories explicitly.

This chapter is organized as follows.
In section~\ref{sec:adsgaugings} we analyze the allowed gauge groups for maximally supersymmetric AdS solutions and show that the vacuum gauge group is always of the form \eqref{eq:AdSgaugegroup}.
In section~\ref{sec:adsmoduli} we discuss a version of the Higgs mechanism and compute conditions on the moduli spaces of AdS solutions under the assumption that the scalar manifold is a symmetric homogeneous space.
We discuss the characteristic implications of these conditions for theories with different amounts of supersymmetry.

\section{The gauge group of AdS solutions}\label{sec:adsgaugings}
As we found in the previous chapter a maximally supersymmetric AdS solution is only possible at points of the scalar manifold where the shift matrices \(A_0\) and \(A_1\) satisfy \eqref{AdSsolution}
\begin{equation}\label{eq:adsconditions}
(A_0)^2 = - \frac{\Lambda}{2(D-1)(D-2)} \id \,,\quad
A_1 = 0 \,,
\end{equation}
where \(\Lambda\) is the negative cosmological constant.
Note that it follows from the general form of the scalar potential V given in \eqref{eq:generalpotential} that \(A_1 = 0\) already implies \((A_0)^2 \sim \id\).
Therefore, demanding \(A_0 \neq 0\) and \(A_1 = 0\) is enough to guarantee that also the first equation in \eqref{eq:adsconditions} is solved for some value of \(\Lambda\).

These conditions in turn enforce constraints on the possible gauge groups of the theory.
Let us introduce the dressed moment maps \eqref{eq:genmomentmap} and Killing vectors \eqref{eq:gaugedP},
\begin{equation}\label{eq:dressedQP}
\cQ^R_\al{2} = \cV^I_\al{2} \cQ^R_I \,,\qquad \cP_\al{2} = \cV^I_\al{2} \cP_I \,,
\end{equation}
where \(\cV^I_\al{2}\) are the vielbeins of the vector field kinetic matrix \eqref{eq:kinmatrix}.\footnote{Note the similarity with the definition of the T-tensor in \eqref{eq:Ttensor}.}
In appendix~\ref{app:susycalculations} we derive how to express \(\cQ_\al{2}^R\) and \(\cP_\al{2}\) in terms of \(A_0\) and \(A_1\).
For vanishing \(A_1\) the resulting equations \eqref{eq:QRA0A1} and \eqref{eq:PA1} read
\begin{equation}
\cQ^R_\al{2} = (D-3) \bigl\{A_0, B_\al{2}\bigr\} \,,\qquad\text{and}\qquad \cP_\al{2} B_\be{2} \delta^{\al{2}\be{2}} = 0 \,,
\end{equation}
where \(B_\al{2}\) are the same matrices as appearing in the supersymmetry variations of the gravitini \eqref{eq:cFM}.
As in the previous chapter we want to employ the split of \(\al{2}\) into \(\hal{2}\) and \(\tal{2}\) \eqref{indexsplit},
where \(\hal{2}\) labels those fields strengths which enter the gravitini variations and \(\tal{2}\) their orthogonal complement.
Consequently, the \(B_\hal{2}\) are a set of linearly independent matrices, while on the other hand \(B_\tal{2} = 0\)
and we find the following general conditions for a maximally supersymmetric AdS solution in terms of the dressed moment maps and Killing vectors, 
\begin{equation}\begin{gathered}\label{eq:AdSconditionsQP}
\cQ^R_{\hal{2}} = (D-3) \bigl\{A_0, B_{\hal{2}}\bigr\} \,, \\
\cQ^R_{\tal{2}} = \cP_\hal{2} = 0 \,.
\end{gathered}\end{equation}
Of course these equations are only to be understood as restrictions on the background values of \(\cQ^R_\al{2}\) and \(\cP_\al{2}\), at an arbitrary point of the scalar manifold they do not need to be satisfied.

Let us analyze the implications of the equations \eqref{eq:AdSconditionsQP} on the gauge group \(G^g\). 
As discussed in chapter~\ref{sec:gauging} the generators of \(G^g\) are denoted by \(X_I\) \eqref{eq:gaugegenerators} and their action on the scalar manifold is described in terms of the Killing vectors \(\cP_I\) or equivalently by the dressed Killing vectors \(\cP_\al{2}\) \eqref{eq:dressedQP}.
Contrary to \(\cP_\tal{2}\) the background values of the Killing vectors \(\cP_\tal{2}\) are unrestricted by \eqref{eq:AdSconditionsQP}, none the less some (or all) might also be vanishing.
For this reason we again split the index \(\tal{2}\) into \(\tal{2}'\) and \(\tal{2}''\) such that the background values of \(\cP_{\tal{2}'}\) are all non-vanishing and linearly independent and such that in the background \(\cP_{\tal{2}''} = 0\).
Let us furthermore collectively denote all Killing vectors with vanishing background value by \(\cP_{\alpha_2^0} = (\cP_\hal{2}, \cP_{\tal{2}''})\).

The Killing vectors \(\cP_{\alpha_2^0}\) with vanishing background value (or equivalently the generators \(X_{\alpha_2^0}\)) generate a subgroup 
\begin{equation}
H^g \subseteq G^g
\end{equation} 
of the gauge group.
To see this we express the commutator \eqref{eq:killingcommutator} of Killing vectors \(\cP_{\alpha_2^0}\) according to our split of indices as 
\begin{equation}\label{eq:P0comm}
\bigl[\cP_{\alpha_2^0}, \cP_{\beta_2^0}\bigr] = {X_{\alpha_2^0 \beta_2^0}}^{\gamma_2^0} \cP_{\gamma_2^0} + {X_{\alpha_2^0 \beta_2^0}}^{\gamma_2'} \cP_{\gamma_2'} \,,
\end{equation}
where \({X_{\al{2}\be{2}}}^\ga{2} = \cV^I_\al{2} \cV^J_\be{2} \cV^\ga{2}_K {X_{IJ}}^K\).
In the background only \(\cP_{\gamma_2'}\) on the right hand side of \eqref{eq:P0comm} does not vanish, which enforces \({X_{\alpha_2^0 \beta_2^0}}^{\gamma_2'} = 0\).
Moreover, inserting \(\cP_{\alpha_2^0} = 0\) into \eqref{eq:genmomentmap} gives \(\cQ_{\alpha_2^0} = - W_{\alpha_2^0}\) and from \eqref{eq:generalvielbeinvariation} we find
\begin{equation}\label{eq:backgroundQ}
{X_{\alpha_2^0 \be{2}}}^\ga{2} = {\bigl(\cQ_{\alpha_2^0}\bigr)_\be{2}}^\ga{2} \,.
\end{equation}
However, since \(\cQ_{\alpha_2^0}\) is an element of \(\h\) it satisfies \({\bigl(\cQ_{\alpha_2^0}\bigr)_\tbe{2}}^\hga{2} = {\bigl(\cQ_{\alpha_2^0}\bigr)_\hbe{2}}^\tga{2} = 0\)
and therefore we find for the commutators \eqref{eq:quadconstr} of the corresponding gauge group generators
\begin{equation}\begin{aligned}\label{eq:Hcomm}
\bigl[X_\hal{2}, X_\hbe{2}\bigr] &= -{\bigl(X_\hal{2}\bigr)_\hbe{2}}^\ga{2} X_\ga{2} = 
- {\bigl(\cQ_\hal{2}\bigr)_\hbe{2}}^\hga{2} X_\hga{2}  \,, \\
\bigl[X_{\tal{2}''}, X_{\tbe{2}''}\bigr] &= - {\bigl(X_{\tal{2}''}\bigr)_{\tbe{2}''}}^\ga{2} X_\ga{2} = 
- {\bigl(\cQ_{\tal{2}''}\bigr)_{\tbe{2}''}}^\tga{2} X_\tga{2}  \,.
\end{aligned}\end{equation}
Moreover \eqref{eq:AdSconditionsQP} implies that \(\cQ_{\tal{2}''}\) cannot have any hatted indices and thus
\begin{equation}\label{eq:HRHmatcomm}
\bigl[X_{\tal{2}''}, X_\hal{2}\bigr] = - {\bigl(X_{\tal{2}''}\bigr)_\hal{2}}^\be{2} X_\be{2} = 
0 \,.
\end{equation}
(Note that equations \eqref{eq:backgroundQ} - \eqref{eq:HRHmatcomm} are understood to be evaluated in the background.)
Together \eqref{eq:Hcomm} and \eqref{eq:HRHmatcomm} show that \(H^g\) factorizes into two mutually commuting subgroups, i.e.
\begin{equation}
H^g = H^g_R \times H^g_\mathrm{mat} \,,
\end{equation}
 where \(H^g_R\) is generated by \(X_\hal{2}\) and \(H^g_\mathrm{mat} \subseteq H_\mathrm{mat}\) is generated by \(X_{\tal{2}''}\).
Note that even though the Killing vectors \(\cP_{\alpha_2^0}\) vanish in the background they can still generate a nontrivial group \(H^g\).
In particular, the equivariance condition \eqref{eq:equivariance} becomes
\begin{equation}
\bigl[\cQ_{\alpha_2^0}, \cQ_{\beta_2^0}\bigr] = {f_{\alpha_2^0 \beta_2^0}}^{\gamma_2^0} \cQ_{\gamma_2^0} \,,
\end{equation}
and therefore non-vanishing moment maps imply a non-trivial gauge group \(H^g \subseteq H\).
The fact that \(H^g\) is a subgroup of \(H\) and that it is generated by the moment maps \(\cQ_{\alpha_2^0}\) allows us to restrict \(H^g\) further.
The expression \eqref{eq:backgroundQ} for the generators \(X_{\alpha_{2}^0}\) of \(H^g\) in combination with the general property \eqref{eq:Hdelta} of every element of \(\h\) yields
\begin{equation}
{\bigl(X_{\alpha_2^0}\bigr)_{(\be{2}}}^{\delta_2} \delta_{\ga{2})\delta_2} = 0 \,.
\end{equation}
Therefore, an equivalent invariance property must hold true also for the structure constants \({f_{\alpha_2^0 \beta_2^0}}^{\gamma_2^0}\) of the Lie algebra  \(\h^g\) of \(H^g\), i.e.
\begin{equation}
{f_{\alpha_2^0 (\beta_2^0}}^{\delta_2^0} \delta_{\gamma_2^0)\delta_2^0} = 0 \,.
\end{equation}
The presence of the invariant symmetric positive-definite matrix \(\delta_{\al{2}\be{2}}\) implies that \(\h^g\) is reductive, i.e.~that it is the direct sum of an abelian Lie algebra and a semi-simple Lie algebra, and that the semi-simple factors in \(H^g\) are compact, see e.g.~\cite{Weinberg:1996kr} for a proof.

So far we have not included the first equation of \eqref{eq:AdSconditionsQP} into our analysis.
This condition completely determines the commutators \(\bigl[X_\hal{2}, X_\hbe{2}\bigr] = {X_{\hal{2}\hbe{2}}}^\hga{2} X_\hga{2}\) of the generators of \(H^g_R\) via
\begin{equation}\label{eq:HgRcomm}
{X_{\hal{2}\hbe{2}}}^\hga{2} = \bigl(\cQ_\hal{2}\bigr)_\hbe{2}^\hga{2} = \bigl(\cQ^R_\hal{2}\bigr)_\hbe{2}^\hga{2} \,.
\end{equation}
However, it still leaves some freedom for the embedding of \(H^g_R\) into \(H\) because it does not determine \({X_{\hal{2}\tbe{2}}}^\tga{2} = {\bigl(\cQ_\hal{2}\bigr)_\tbe{2}}^\tga{2}\).
Let us denote the subgroup of \(H_R\) which is generated by \(\cQ^R_\hal{2}\) by \(\hat H^g_R\).
It follows from the equivariance condition \eqref{eq:equivariance} that also \(\bigl[\cQ^R_\hal{2}, \cQ^R_\hbe{2}\bigr] = {X_{\hal{2}\hbe{2}}}^\hga{2} \cQ^R_\hga{2}\).
Therefore \(H^g_R\) and \(\hat H^g_R\) share the same commutator relations and are isomorphic (at least at the level of their Lie algebras).
Nonetheless, as subgroups of \(H\) they do not need to be identical since \(H^g_R\) is not necessarily a subgroup of just \(H_R\) but might be embedded diagonally into \(H = H_R \times H_\mathrm{mat}\).
This is the case if \({X_{\hal{2}\tbe{2}}}^\tga{2}\) is non-vanishing.

Given an explicit expression for the matrices \(B_\hal{2}\) we could now compute \(\cQ^R_\hal{2}\) from the prescription \eqref{eq:AdSconditionsQP} and thus determine \(\hat H^g_R\).
This calculation is demonstrated for a couple of examples in the next chapter.
However, without any reference to an explicit realization of \(B_\hal{2}\) we can already say a lot about \(H^g_R\) just from the general properties of \(B_\hal{2}\).
In appendix~\ref{app:representationtheory} we show that the \(\cQ^R_\hal{2}\) given by \eqref{eq:AdSconditionsQP} generate a subgroup \(\hat H^g_R \subseteq H_R\) under which \(A_0\) is invariant, i.e.~\(\bigl[\cQ^R_\al{2}, A_0\bigr] = 0\).
To be more specific, let us denote by \(\x\) the maximal subalgebra of \(\h_R\) such that \([\x, A_0] = 0\)
and let us decompose the representation \(\mathbf v\) of \(\h_R\) which  corresponds to the index \(\hal{2}\) into irreducible representations of \(\x\).
The Lie algebra \(\hat\h^g_R\) of \(\hat H^g_R\) must be a subalgebra of \(\x\) such that the adjoint representation of \(\hat\h^g_R\) appears in the decomposition of \(\mathbf v\) into representations of \(\x\).

Let us finally talk about the spontaneous breaking of the gauge group \(G^g\) in the AdS vacuum.
In the background the gauged vielbeins \eqref{eq:gaugedP} read \(\hat \cP = \cP + A^{\tal{2}'} \cP_{\tal{2}'}\).
Inserting this expression into the scalar kinetic term \eqref{eq:scalarvielbeins} produces the mass term
\begin{equation}\label{eq:gaugemass}
\cL_\mathrm{mass} = \tfrac12 \delta_{\al{1}\be{2}}\cP^\al{1}_{\tal{2}'}\cP^\be{1}_{\tbe{2}'} A^{\tal{2}'} \wedge \ast A^{\tbe{2}'} \,.
\end{equation}
Because the \(\cP_{\tal{2}'}\) are linearily independent this generates mass terms for all gauge fields \(A^{\tal{2}'}\), while all the other gauge fields \(A^\hal{2}\) and \(A^{\tal{2}''}\) remain massless.
In other words the mass term \eqref{eq:gaugemass} breaks \(G^g\) spontaneously to \(H^g\), i.e.
\begin{equation}
G^g \rightarrow H^g_R \times H^g_\mathrm{mat} \,.
\end{equation}
This result is physically satisfactory as it shows that the gauge group must be broken to a product of abelian and compact semi-simple subgroups.
Moreover, as discussed in the beginning of this chapter, we can interpret \(H^g_R\) as the R-symmetry group of the holographically dual SCFT and \(H_\mathrm{mat}\) as some additional flavor symmetry.


For theories where the scalar manifold is a symmetric space \(\cM = G/H\) the gauge group \(G^g\) must be a subgroup of \(G\).
The generators of \(G^g\) can be expressed in terms of the T-tensor \(\cT_\al{2}\) \eqref{eq:Ttensor}.
The AdS conditions \eqref{eq:AdSconditionsQP} dictate that they are of the general form
\begin{equation}\begin{aligned}
\cT_\hal{2} &= \cQ^R_\hal{2} + \cQ^\mathrm{mat}_\hal{2} \,, \\
\cT_{\tal{2}'} &=  \cP_{\tal{2}'} + \cQ^\mathrm{mat}_{\tal{2}'} \,, \\
\cT_{\tal{2}''} &=  \cQ^\mathrm{mat}_{\tal{2}''} \,,
\end{aligned}\end{equation}
where we employed our previous split of \(\tal{2}\) into \(\tal{2}'\) and \(\tal{2}''\).
The generators \(\cT_{\tal{2}'}\) can possibly lead to a non-compact or non-reductive gauge group \(G^g\), but according to our previous discussion they are spontaneously broken in the vacuum.

In the next section we will be especially interested in theories where the only multiplet is the gravitational multiplet.
For these theories there is no \(H_\mathrm{mat}\) and no gauge fields \(A^\tal{2}_M\).
Consequently the only generators of \(G^g\) are given by%
\footnote{Notice, that for the four-dimensional \(\cN = 6\) theory there could be in principle an additional generator \(\cT_0 = \cP_0\)
but we show in chapter~\ref{sec:greater16} that \(\cP_0 = 0\).}
\begin{equation}\label{eq:maximalgenerators}
\cT_\hal{2} = \cQ^R_\hal{2} + \cP_\hal{2} = \cQ^R_\hal{2} \,,
\end{equation}
and therefore
\begin{equation}\label{eq:maximalGg}
G^g = H^g = H^g_R \,,
\end{equation}
i.e.~the complete gauge group must be compact and is uniquely determined by the AdS conditions \eqref{eq:AdSconditionsQP}.

Let us finally mention that these results can be straightforwardly translated to maximally supersymmetric Minkowski solutions as well as to maximally supersymmetric solutions with non-trivial flux.
Both classes of solutions require not only \(A_1 = 0\) but also \(A_0 = 0\).
This in turn implies via \eqref{eq:AdSconditionsQP} that \(\cQ^R_\al{2} = 0\).
Hence here \(H^g_R\) must be trivial.

\section{The moduli space}\label{sec:adsmoduli}

We now turn to the moduli spaces of AdS solutions, i.e.~we want to discuss if there are any directions in the scalar field space which are undetermined by the conditions \eqref{eq:adsconditions}.
Let us denote a point in the scalar manifold at which \eqref{eq:adsconditions} is satisfied by \(\left<\phi\right>\) and vary it according to
\begin{equation}\label{eq:phivariation}
\phi = \left<\phi\right> + \delta\phi \,,
\end{equation}
where \(\delta\phi\) is an infinitesimal variation or in other words an infinitesimal tangent vector, i.e.~\(\delta\phi \in T_{\left<\phi\right>}\cM\).
Our goal is to determine if there are any variations \(\delta\phi\) 
under which the AdS conditions \eqref{eq:adsconditions} do not change, i.e.~we are looking for solutions of
\begin{equation}\label{eq:adsmoduli}
\bigl< \partial_{\delta\phi} A^2_0 \bigr> = \bigl< \partial_{\delta\phi} A_1 \bigr> = 0 \,.
\end{equation}
However, the vanishing of the first derivative with respect to \(\delta\phi\) is a priori only a necessary condition for \(\delta\phi\) to be a modulus.
For the existence of a true modulus, i.e.~a continuous deformation parameter of the AdS solution, \(A_0^2\) and \(A_1\) have to be invariant not only under an infinitesimal variation \eqref{eq:phivariation} but also under finite variations.
Equivalently, a modulus is characterized by the vanishing of not only the first derivative with respect to \(\delta\phi\) but also of all higher-order derivatives,
\begin{equation}\label{eq:allordervariation}
\bigl< \partial^n_{\delta\phi} A^2_0 \bigr> = \bigl< \partial^n_{\delta\phi} A_1 \bigr> = 0 \,, \qquad \forall n \geq 1 \,,
\end{equation}
assuming analyticity in \(\phi\).
This resembles the distinction between marginal and exactly marginal deformations of SCFTs.

As mentioned in the discussion below equation \eqref{eq:adsconditions}, the vanishing of \(A_1\) already implies \(A_0^2 \sim \id\).
Hence it is conceivable that also the vanishing of the variations of \(A_0^2\) is guaranteed by the vanishing of \(A_1\) and its variations.
Indeed, there is a relation of the form \(\cD A_0 \sim A_1\), called gradient flow equation \cite{DAuria:2001rlt}, between the (covariant) derivative of \(A_0\) and the value of \(A_1\).
We rederive the precise form of the gradient flow equation, adopted to our notation, 
in 
appendix~\ref{app:susycalculations}.
It reads
\begin{equation}\label{eq:gradientflow}
\cD_{\al{1}} A_0 = \tfrac{1}{2(D-2)} \bigl(A^\dagger_1 C_{\alpha_1} + C^\dagger_{\alpha_1} A_1 \bigr) \,,
\end{equation}
where \(C_\al{1}\) are the same matrices as in the supersymmetry variations \eqref{eq:cFhat} and \eqref{eq:cFtilde}.
At every point in the scalar manifold where \(A_1 = 0\) we therefore automatically have \(\cD_{\delta\phi} A_0 = 0\) for all variations \(\delta\phi \in T_{\left<\phi\right>} \cM\) and thus
\begin{equation}
\partial_{\delta\phi} A^2_0 = \cD_{\delta\phi} A^2_0 = (\cD_{\delta\phi} A_0) A_0 + A_0 (\cD_{\delta\phi} A_0) = 0 \,,
\end{equation}
where the replacement of the ordinary derivative of \(A_0^2\) with its covariant derivative is allowed due to \(A_0^2 \sim \id\).
Analogously, the vanishing of all higher-order variations of \(A_1\) implies the vanishing of all higher-order variations of \(A_0^2\), i.e.
\begin{equation}
\partial^n_{\delta\phi} A_1 = 0 \,,\quad \forall n \geq 0 \qquad \Rightarrow \qquad \partial^n_{\delta\phi} A_0^2 = 0 \,,\quad \forall n \geq 1 \,.
\end{equation}
It is therefore sufficient to study the variations of \(A_1\).

Note that the gradient flow equation \eqref{eq:gradientflow} together with \eqref{eq:generalpotential} also guarantees that every solution of \eqref{eq:adsconditions} is indeed a critical point of the potential \(V\), i.e.
\begin{equation}
\bigl<\partial_{\delta\phi} V \bigr> = 0 \,,\qquad \forall\,  \delta\phi \in T_{\left<\phi\right>}\cM \,,
\end{equation}
and therefore a solution of the equations of motion.


Let us temporarily neglect the problem of finding exact solutions \(\delta\phi\) of \eqref{eq:allordervariation} at all orders, but let us for the moment only focus on the leading order variation.
This means we are looking for solutions of
\begin{equation}\label{eq:A1moduli}
\bigl<\cD_{\delta\phi} A_1 \bigr> = \bigl<\partial_{\delta\phi} A_1 \bigr> = 0 \,,
\end{equation}
where \(\cD\) and \(\partial\) can be identified due to \(A_1 = 0\).
If \(\delta\phi\) solves \eqref{eq:A1moduli} it is straightforward to show that
\begin{equation}
\bigl< \partial^2_{\delta\phi} V \bigr> = 0 \,,
\end{equation}
and therefore \(\delta\phi\) corresponds to a massless excitation.
As discussed in the introduction a massless scalar fields gets mapped via the AdS/CFT correspondence to an operator of conformal dimension \(\Delta = d\) on the \(d\)-dimensional boundary SCFT.
This again illustrates that a solution of \eqref{eq:A1moduli} is dual to a supersymmetric marginal deformation.
On the other hand a solution of \eqref{eq:allordervariation} fulfills \(\bigl<\partial^n_{\delta\phi} V \bigr> = 0\) (\(\forall n \geq 1\)) and thus corresponds to an exactly marginal deformation.

From now on we assume that all derivatives are evaluated at \(\phi = \left<\phi\right>\) and stop indicating this explicitly to simplify the notation.

In the previous section we found that the general AdS conditions on \eqref{eq:adsconditions} constrain the background values of the dressed moment maps \(\cQ^R_\al{2}\) and Killing vectors \(\cP_\al{2}\) to be of the form \eqref{eq:AdSconditionsQP}.
Therefore, a solution of \eqref{eq:A1moduli} must necessarily satisfy
\begin{equation}\label{eq:QPmoduli}
\cD_{\delta\phi} \cQ^R_\al{2} = \cD_{\delta\phi} \cP_\hal{2} = 0 \,.
\end{equation}
In many cases gaugings are the only possible deformations of a supergravity and \(A_0\) and \(A_1\) can be expressed exclusively in terms of \(\cQ^R_\al{2}\) and \(\cP_\hal{2}\).
Under these circumstances \eqref{eq:QPmoduli} is also a sufficient condition for \eqref{eq:A1moduli}.
In the remainder of this chapter we want to assume that this is indeed the case.
However, if there are other contributions to the shift matrices, e.g.~by a non-trivial superpotential, \eqref{eq:A1moduli} and \eqref{eq:QPmoduli} are not equivalent.

In the previous section we have seen that  the gauge group \(G^g\) gets spontaneously broken if there are Killing vectors \(\cP_\al{2}\) with non-vanishing background values. 
According to Goldstones theorem we expect that for each broken generator there exists one massless scalar field, called a Goldstone boson.
Indeed, a gauged supergravity theory is constructed in such a way that its action and hence also the potential \(V\) are \(G^g\)-invariant.
The shift matrices \(A_0\) and \(A_1\), however, since they couple to the fermions, are only gauge invariant up to a compensating \(H\)-transformation, described by the \(H\)-compensator \(W_I\) \eqref{eq:W}.
This \(H\)-transformation drops out in the expression for \(V\) in terms of \(A_0\) and \(A_1\) \eqref{eq:generalpotential} due to the involved trace.
Consequently an inifinitesimal gauge transformation parametrized by \(\lambda^I\) which acts on the scalar fields as \eqref{eq:scalargaugetransf}
\begin{equation}\label{eq:goldstone}
\delta\phi = \lambda^\al{2} \cP_\al{2} \,,
\end{equation}
is expected to solve \eqref{eq:allordervariation}.
This variation describes one independent solution \(\lambda^{\alpha'_2}\) for each non-vanishing Killing vector \(\cP_{\alpha'_2}\).
Therefore there is one massless scalar field for each spontaneously broken generator of the gauge group \(G^g\).
Nonetheless, these fields cannot be counted as moduli.
As Goldstone bosons of a spontaneously broken gauge symmetry they describe the additional degrees of freedom of the massive gauge fields \(A_M^{\al{2}'}\) and get eaten by the St\"uckelberg mechanism.
In other words the scalar modes \eqref{eq:goldstone} are pure gauge and therefore non-physical.

Let us now explicitly show that \eqref{eq:goldstone} solves \eqref{eq:QPmoduli}.
Before we can compute the variations of \(\cQ^R_\al{2}\) and \(\cP_\al{2}\) with respect to \eqref{eq:goldstone},
we need to determine how the covariant derivative acts on the vielbein \(\cV_\al{2}^I\).
We denote the covariant derivative in a Killing direction by \(\cD_\al{2} = \cP^\al{1}_\al{2} \cD_\al{1}\) and
recall its definition in terms of connection form \(\theta\),
\begin{equation}
\cD_\al{2}\cV^I_\be{2}  = \cL_\al{2} \cV^I_\be{2} + \iota_\al{2} {\theta_\be{2}}^\ga{2} \cV^I_\ga{2} \,,
\end{equation}
where we used the fact that the Lie derivative acts on \(\cV_\al{2}^I\) as an ordinary derivative.
From \eqref{eq:generalvielbeinvariation} and the definition of the moment map \eqref{eq:genmomentmap} we obtain
\begin{equation}\label{eq:genvielbeincovderiv}
\cD_\al{2}\cV^I_\be{2} = \left[- {X_{\al{2}\be{2}}}^\ga{2} + {\bigl(\cQ_\al{2}\bigr)_\be{2}}^\ga{2} \right] \cV^I_\ga{2} \,.
\end{equation}
From this we can compute
\begin{equation}
\cD_\al{2} \cQ^R_\be{2} = \bigl(\cD_\al{2} \cV^I_\be{2}\bigr) \cQ^R_I + \cV^I_\be{2} \bigl(\cD_\al{2} \cQ^R_I\bigr) \,. \\
\end{equation}
Inserting \eqref{eq:genvielbeincovderiv} and the covariant derivative of the moment map \eqref{eq:Qderiv} gives
\begin{equation}\begin{aligned}\label{eq:Qgoldstone}
\cD_\al{2} \cQ^R_\be{2} &= - {X_{\al{2}\be{2}}}^\ga{2} \cQ^R_\ga{2} + {(\cQ_\al{2})_\be{2}}^\ga{2} \cQ^R_\ga{2} + \Omega(\cP_\al{2}, \cP_\be{2}) \\
&= {(\cQ_\al{2})_\be{2}}^\ga{2} \cQ^R_\ga{2} - \bigl[\cQ^R_\al{2}, \cQ^R_\be{2}\bigr] = 0 \,,
\end{aligned}\end{equation}
where we used the equivariance condition \eqref{eq:equivariance}.
In the last step we used that the \(\cQ^R_\al{2}\) span a subalgebra of \(H_R\) with generalized structure constants given by \({(\cQ^R_\al{2})_\be{2}}^\ga{2}\) (compare the discussion below \eqref{eq:HgRcomm}) and that \({(\cQ_\al{2})_\be{2}}^\ga{2} \cQ^R_\ga{2} 
 = {(\cQ^R_\al{2})_\be{2}}^\ga{2} \cQ^R_\ga{2}\).
In a similar fashion we can also compute the covariant derivative of \(\cP_\hal{2}\) from the covariant derivative of \(\cP_I\) given in \eqref{eq:Pcovderiv},
\begin{equation}\begin{aligned}\label{eq:Pgoldstone}
\cD_\al{2} \cP^\al{1}_\hbe{2} &= \bigl(\cD_\al{2} \cV^I_\hbe{2}\bigr) \cP^\al{1}_I + \cV^I_\hbe{2} \bigl(\cD_\al{2} \cP^\al{1}_I\bigr) \\
&= {(\cQ_\al{2})_\hbe{2}}^\hga{2} \cP^\al{1}_\hga{2} - {(\cQ_\al{2})_\be{1}}^\al{1} \cP^\be{1}_\hbe{2} = 0
 \,.
\end{aligned}\end{equation}
Together \eqref{eq:Qgoldstone} and \eqref{eq:Pgoldstone} show that the ansatz \eqref{eq:goldstone} indeed satisfies \eqref{eq:QPmoduli}.
By applying \eqref{eq:Qgoldstone} and \eqref{eq:Pgoldstone} recursively to themselves one can also show that all higher-order derivatives of \(\cQ^R_\al{2}\) and \(\cP_\hal{2}\) with respect to \eqref{eq:goldstone} vanish.
Note that we inserted the AdS conditions \eqref{eq:AdSconditionsQP} only in the very last step.

We have just seen that the Goldstone bosons appear generically as solutions of \eqref{eq:allordervariation}, however, they do not contribute to the moduli space.
Here, we do not attempt to find the remaining solutions of \eqref{eq:allordervariation}, which span the moduli space, in a similar general fashion.
This has been achieved explicitly for various theories in \cite{deAlwis:2013jaa,Louis:2014gxa,Louis:2015mka,Louis:2015dca,Louis:2016qca}.
Instead, we only consider theories where the scalar manifold is a symmetric homogeneous space, as introduced in chapter~\ref{sec:coset}.

If the scalar manifold is a symmetric homogeneous space \(\cM = G/H\),
it is most convenient to parametrize the scalar variation \(\delta\phi\) in terms of the corresponding \(\k\) valued quantity \(\cP_{\delta\phi}\), defined as
\begin{equation}
\cP_{\delta\phi} = \iota_{\delta\phi} \cP \in \k \,.
\end{equation}
To compute the (covariant) variations of the general AdS conditions \eqref{eq:AdSconditionsQP} it is necessary to determine the variations of the moment maps \(\cQ_I\) and Killing vectors \(\cP_I\) as well as of the vielbeins \(\cV^I_\al{2}\).
From \eqref{eq:dressedgenerator} we infer that in the coset case \(\cQ_I\) and \(\cP_I\) are given by the \(\h\)-components and the \(\k\)-components of the dressed gauge group generators \(\cT_I\).
Applying \eqref{eq:cosetreprderiv} to the definition \eqref{eq:dressedgenerator} of \(\cT_I\) yields
\begin{equation}
\cD_{\delta\phi} \cT_I = \bigl[\cT_I, \cP_{\delta\phi}\bigr] \,,
\end{equation}
and after splitting this into an \(\h\)-part and a \(\k\)-part one obtains
\begin{equation}\label{eq:PQvariation}
\cD_{\delta\phi} \cQ_I = \bigl[\cP_I, \cP_{\delta\phi}\bigr] \,,\qquad \cD_{\delta\phi} \cP_I = \bigl[\cQ_I, \cP_{\delta\phi}\bigr] \,.
\end{equation}
On the other hand, as discussed in the last paragraph of chapter~\ref{sec:coset}, the vielbeins \(\cV^\al{2}_I\) are given by the coset representative \(L\) expressed in the appropriate representations.
Analogously \(\cV_\al{2}^I\) is given by the inverse vielbein \(L^{-1}\).
Hence its covariant derivative takes the same form as the covariant derivative of \(L^{-1}\) and is according to \eqref{eq:cosetreprderiv} given by
\begin{equation}\label{eq:cVvariation}
\cD_{\delta\phi} \cV^I_\al{2} = - {\bigl(\cP_{\delta\phi}\bigr)_\al{2}}^\be{2} \cV^I_\be{2} \,,
\end{equation}
where \(\bigl({\cP_{\delta\phi}\bigr)_\al{2}}^\be{2}\) denotes \(\bigl(\cP_{\delta\phi}\bigr)\) expressed in the \(\h\)-representation of the dressed vector fields (i.e.~the representation which is labeled by the index \(\al{2}\)).

After this preparation we are in the position to analyze the general conditions \eqref{eq:QPmoduli}.
With \eqref{eq:PQvariation} and \eqref{eq:cVvariation} they read
\begin{equation}\begin{aligned}\label{eq:cosetmoduli}
\cD_{\delta\phi} \cQ^R_\al{2} &= - {(\cP_{\delta\phi})_\al{2}}^\be{2} \cQ^R_\be{2} + \bigl[\cP_\al{2}, \cP_{\delta\phi}\bigr]\!^R = 0 \,, \\
\cD_{\delta\phi} \cP_\hal{2} &= 
- {(\cP_{\delta\phi})_\hal{2}}^\be{2} \cP_\be{2} + 
\bigl[\cQ_\hal{2}, \cP_{\delta\phi}\bigr] = 0 \,,
\end{aligned}\end{equation}
where \((\,\cdot\,)^R\) denotes the projection of an \(\h\)-valued quantity onto \(\h_R\).
To proceed we recall that it follows from \eqref{eq:reductive} that \(\k\) transforms in some representation of \(\h\) with respect to the adjoint action.
We can therefore decompose \(\k\) intro irreducible representations \(\k_i\) of the subalgebra \(\h^g_R\) of \(\h\), i.e.
\begin{equation}\label{eq:kdecomp}
\k = \bigoplus_{i = 1, \dots, N} \k_i \,,\qquad [\h^g_R, \k_i] \subseteq \k_i \,.
\end{equation}
Let us denote the set of all solutions of \eqref{eq:cosetmoduli} by \(\f\), i.e.
\begin{equation}
\f = \left\{\cP_{\delta\phi} \in \k : \cD_{\delta\phi} \cQ^R_\al{2} = \cD_{\delta\phi} \cP_\hal{2} = 0 \right\} \,.
\end{equation}
It follows directly from \eqref{eq:cosetmoduli} that for \(\cP_{\delta\phi} \in \f\) also \([\cQ_\hal{2}, \cP_{\delta\phi}] \in \f\) and therefore
\begin{equation}\label{eq:fdecomp}
\f = \bigoplus_{i \in I} \k_i \,,\qquad I \subseteq {1,\dots,N} \,,
\end{equation}
i.e.~if \eqref{eq:cosetmoduli} is satisfied by one element of some irreducible \(\h^g_R\)-representation, it holds for all elements of this representation.

Let us furthermore introduce 
\begin{equation}\label{eq:kg}
\k^g = \mathrm{span}(\cP_\al{2}) \,,
\end{equation}
i.e.~the projection of the Lie algebra \(\g^g\) of \(G^g\) onto \(\k\).
According to our previous considerations \(\k^g\) corresponds to the Goldstone bosons of the spontaneous symmetry breaking \(G^g \rightarrow H^g\).
Therefore, \(\k^g\) must always be contained in the set of solutions \(\f\), which can be seen directly by inserting \(\cP_{\delta\phi} = \cP_\al{2}\) into \eqref{eq:cosetmoduli}.
Also \(\k^g\) is a \(\h^g_R\) representation (not necessarily an irreducible one) in the above sense and hence
\begin{equation}\label{eq:fsplit}
\f = \k^g \oplus \k_{AdS} \,,
\end{equation}
where \(\k_{AdS}\) spans the non-trivial solutions of \eqref{eq:cosetmoduli} and therefore the candidates for supersymmetric moduli.
The second condition of \eqref{eq:cosetmoduli} implies \(\bigl[\cQ_\hal{2}, \cP_{\delta\phi}\bigr] \subseteq \k^g
\) or equivalently
\begin{equation}
[\h^g_R, \f] \subseteq \k^g \,.
\end{equation}
According to \eqref{eq:kdecomp} this is only possible for two \(\h^g_R\)-representations: \(\k^g_R\) itself and the singlets which commute with \(\h^g_R\).
Hence, we deduce
\begin{equation}\label{eq:singlets}
[\h^g_R, \k_{AdS}] = 0 \,.
\end{equation}
Consequently, all moduli must necessarily commute with \(\h^g_R\) or in other words they must be singlets with respect to the adjoint action of \(\h^g_R\).
This is often a strong statement and can highly constrain the existence of moduli spaces.
Moreover, finding singlets in the branching of a Lie algebra representation into irreducible representations of a subalgebra is a very well understood problem.

Using this result the conditions on supersymmetric moduli \eqref{eq:cosetmoduli} can be simplified even further.
In terms of the generators \(\cQ_\hal{2}\) of \(\h^g_R\) equation \eqref{eq:singlets} reads
\begin{equation}
\bigl[\cQ_\hal{2}, \cP_{\delta\phi}] = 0 \,.
\end{equation}
Inserting this back into \eqref{eq:cosetmoduli} gives
\begin{equation}
{(\cP_{\delta\phi})_\hal{2}}^\be{2} \cP_\be{2} = 0 \,,
\end{equation}
and using the split of the index \(\tal{2}\) into \(\tal{2}'\) and \(\tal{2}''\) introduced in section~\ref{sec:adsgaugings} we obtain
\begin{equation}\label{eq:Pdeltaphi1}
{(\cP_{\delta\phi})_\hal{2}}^{\tbe{2}'} = 0 \,.
\end{equation}
On the other hand, we infer from the first equation in \eqref{eq:cosetmoduli} that
\begin{equation}\label{eq:Pdeltaphi2}
{(\cP_{\delta\phi})_{\tal{2}''}}^\be{2} \cQ^R_\be{2} = 0 \,.
\end{equation}
We show in appendix~\ref{app:symP} that \({(\cP_{\delta\phi})_\hal{2}}^\tbe{2}\) is symmetric in its indices, i.e.
\begin{equation}
{(\cP_{\delta\phi})_\hal{2}}^\tbe{2} = \delta_{\hal{2}\hde{2}} \delta^{\tbe{2}\tga{2}} {(\cP_{\delta\phi})_\tga{2}}^\hde{2} \,.
\end{equation}
Applying this relation to \eqref{eq:Pdeltaphi1} and \eqref{eq:Pdeltaphi2} we find
\begin{equation}
{(\cP_{\delta\phi})_{\tal{2}}}^\be{2} \cQ^R_\be{2} = 0 \,.
\end{equation}
Therefore, we find the following set of conditions on the supersymmetric moduli \(\k_{AdS}\),
\begin{equation}\begin{gathered}\label{eq:cosetmoduli2}
{(\cP_{\delta\phi})_{\al{2}}}^\be{2} \cQ^R_\be{2} = {(\cP_{\delta\phi})_\hal{2}}^\be{2} \cP_\be{2} = 0 \,, \\
\bigl[\cQ_\hal{2}, \cP_{\delta\phi}] = \bigl[\cP_\al{2}, \cP_{\delta\phi}]^R = 0 \,.
\end{gathered}\end{equation}
These conditions are usually simpler to analyze than the original conditions \eqref{eq:cosetmoduli2} and will serve as the starting point for most of our further discussions.


However, a priori it is not clear that \(\k_{AdS}\) really describes the moduli space of the AdS solution, since we only checked for the vanishing of the first derivatives.
A simple sufficient condition for a solution \(\cP_{\delta\phi} \in \k_{AdS}\) of \eqref{eq:cosetmoduli} or \eqref{eq:cosetmoduli2} to be a moduli is that it keeps all generators \(\cT_\al{2}\) of the gauge group \(G^g\) invariant, i.e.
\begin{equation}\label{eq:Tmoduli}
\cD_{\delta\phi} \cT_\al{2} = - {(\cP_{\delta\phi})_\al{2}}^\be{2} \cT_\be{2} + \bigl[\cT_\al{2}, \cP_{\delta\phi}\bigr] = 0 \,,
\end{equation}
and not only \(\cQ^R_\al{2}\) and \(\cP_\hal{2}\) as in \eqref{eq:cosetmoduli}.
Due to the linear action of the covariant derivative \(\cD_{\delta\phi}\) all higher-order covariant derivatives of \(\cT_\al{2}\) vanish if the first derivative \eqref{eq:Tmoduli} vanishes.
Moreover, we show in appendix~\ref{app:Tmoduli} that if all elements of \(\k_{AdS}\) satisfy \eqref{eq:Tmoduli} the moduli space is a symmetric homogeneous space as well.
This means, that we can find a subalgebra \(\h_{AdS}\) of \(\h\) such that \(\g_{AdS} = \h_{AdS} \oplus \k_{AdS}\) is a subalgebra of \(\g\).
\(\g_{AdS}\) and \(\h_{AdS}\) in turn generate subgroups \(G_{AdS} \subseteq G\) and \(H_{AdS} \subseteq H\) and the moduli space is given by
\begin{equation}\label{eq:symmetricmodulispace}
\cM_{AdS} = \frac{G_{AdS}}{H_{AdS}} \,,
\end{equation}
which is symmetric because \(\g_{AdS}\) inherits the properties \eqref{eq:reductive} and \eqref{eq:symmetric} from \(\g\).

Let us discuss the implications of the general conditions \eqref{eq:cosetmoduli2} for different theories with specific numbers of supersymmetries.
We begin with four and five-dimensional theories with \(q = 8\) real supercharges (i.e.~$\cN =2$ supergravities).
A general discussion of their AdS vacua and the corresponding moduli spaces can be found in \cite{deAlwis:2013jaa, Louis:2016qca}.
The scalar field manifold \(\cM\) of such theories factorizes into the product
\begin{equation}\label{eq:N2scalarmanifold}
\cM = \cM_V \times \cM_H \,,
\end{equation}
where \(\cM_V\) is spanned by the scalar fields in vector multiplets and \(\cM_H\) is spanned by the scalar fields in hyper multiplets.
We denote the former by \(\phi_V\) and the latter by \(\phi_H\).
The geometry of \(\cM_V\) depends on the space-time dimension, \(\cM_H\) on the other hand is in both cases a quaternionic K\"ahler manifold.
Generically \(\cM_V\) and \(\cM_H\) are not necessarily symmetric but there exist many symmetric manifolds of the form \(G/H\) which describe viable scalar geometries for such theories.
In these cases it is possible to use our previous results to determine the moduli space of an AdS solution.

Note that for \(\cN = 2\) theories the gauge fields \(A^\al{2}\) are non-trivial sections only over the first factor \(\cM_V\) in \eqref{eq:N2scalarmanifold} and do not depend on \(\cM_H\).
Therefore, also the variation matrix \({(\cP_{\delta\phi})_\al{2}}^\be{2}\) acting on \(\cV^\al{2}_I\) depends only on the variation of the vector multiplet scalars \(\delta\phi_V \in T_{\left<\phi\right>} \cM_V\).
This implies that the first line of \eqref{eq:cosetmoduli2} is completely independent of \(\cM_H\) and only restricts \(\delta\phi_V\).
In the following we analyze the condition \({(\cP_{\delta\phi})_{\al{2}}}^\be{2} \cQ^R_\be{2} = 0\) separately for the two cases \(D=4\) and \(D=5\) and show that it determines \(\delta\phi_V\) completely, irrespective of the specific choice of \(\cM_V\) or the gauge group \(G^g\).

In five dimensions there is one (real) graviphoton field \(A^{\hal{2} = 0}\).
According to \eqref{eq:AdSconditionsQP} the corresponding moment map \(\cQ^R_{\hal{2} = 0}\) needs to be non-vanishing and generates the gauged R-symmetry group \(H^g_R = \U(1)\), see also the discussion in the following chapter.
Therefore, \eqref{eq:cosetmoduli2} implies that
\begin{equation}
{(\cP_{\delta\phi})_{\al{2}}}^0 = 0 \,.
\end{equation}
Moreover, we compute in appendix~\ref{app:symP} that \({(\cP_{\delta\phi})_{\tal{2}}}^0\) can be expressed directly in terms of the variation \(\delta \phi^{\al{1}}_V\) of the scalar fields on \(\cM_V\), see \eqref{eq:D5Pdeltaphi},
\begin{equation}
{(\cP_{\delta\phi})_{\tal{2} = \al{1}}}^0 = - \sqrt{\tfrac23} \delta_{\al{1}\be{1}} \delta\phi^\be{1}_V \,,
\end{equation} 
and hence
\begin{equation}
\delta \phi^{\al{1}}_V = 0 \,.
\end{equation}

In four dimensions the situation is similar, however, here the graviphoton \(A^0\) is complex.
We denote its complex conjugate by \(A^{\bar 0}\) and let the index \(\hal{2}\) take the values \(0\) and \(\bar 0\).
Therefore, we only have
\begin{equation}\label{eq:D4variationV}
{(\cP_{\delta\phi})_{\al{2}}}^0 \cQ^R_0 + {(\cP_{\delta\phi})_{\al{2}}}^{\bar 0} \cQ^R_{\bar 0} = 0 \,,
\end{equation}
where \(\cQ^R_{\bar 0}\) denotes the complex conjugate of \(\cQ^R_0\), which -- as in five dimensions -- has to be non-vanishing.
Moreover, \(\cM_V\) is a complex Manifold (to be precise a special K\"ahler manifold), so it is possible to describe the variation \(\delta\phi_V\) by a complex vector \(\delta\phi_V^\al{1}\) and its complex conjugate \(\delta\bar\phi_V^{\bar\alpha_1}\).
Inserting the explicit expressions \eqref{eq:D4Pdeltaphi} for \({(\cP_{\delta\phi})_{\al{2}}}^0\) and \({(\cP_{\delta\phi})_{\al{2}}}^{\bar 0}\) into \eqref{eq:D4variationV} gives
\begin{equation}
\delta\phi^\al{1}_V \cQ^R_0 = \delta\bar\phi^{\bar\alpha_1}_V \cQ^R_{\bar 0} = 0 \,,
\end{equation}
which in turn implies the vanishing of \(\delta\phi_V\).

As well in four as in five dimensions the variations of the vector multiplet scalars \(\delta\phi_V\) must vanish.
Therefore the geometry of \(\cM_V\) is not directly relevant for the structure of the moduli space.
It only restricts the possible gauge groups  to be contained in the isometry group of \(\cM_V\).
Consequently, a non-trivial moduli space \(\cM_{AdS}\) can be spanned only by scalar fields in hyper multiplets, i.e.
\begin{equation}
\cM_{AdS} \subseteq \cM_H \,,
\end{equation}
and is determined by the conditions in the second line of \eqref{eq:cosetmoduli2}.
The details of this computation will depend on the choice of a symmetric quaternionic K\"ahler manifold \(\cM_H\) and the gauge group \(G^g\).

For half-maximal supergravities (\(q = 16\)) the scalar manifold is given by the symmetric coset space%
\footnote{This is not true for the chiral theories in six and ten dimensions. However, these theories do not allow for supersymmetric AdS solutions.}
\begin{equation}\label{eq:halfmaximalcoset}
\cM = \frac{G^\ast}{H^\ast} \times \frac{\SO(10-D, n_V)}{\SO(10-D) \times \SO(n_V)} \,,
\end{equation}
where \(n_V\) denotes the number of vector multiplets.
In most cases \(G^\ast\) is given by \(\SO(1,1)\), only in four dimensions it is given by \(\SU(1,1)\). 
\(H^\ast\) is the maximal compact subgroup of \(G^\ast\), so in four dimensions \(H^\ast = \U(1)\) and in all other cases it is trivial.
The gauge fields transform in the vector representation of \(\SO(10-D, n_V)\) and also non-trivially with respect to \(G^\ast\).
Only in five dimensions there is an additional gauge field transforming as a singlet with respect to \(\SO(10-D, n_V)\).
Moreover, all scalar fields are either part of the gravity multiplet or of vector multiplets, therefore \({(P_{\delta\phi})_\al{2}}^\be{2}\) depends on the variation of all scalar fields in \(\cM\), in contrast to theories with \(q = 8\) supercharges.
For this reason the first condition in \eqref{eq:cosetmoduli2},
\begin{equation}\label{eq:halfmaximalcondition}
{(\cP_{\delta\phi})_{\al{2}}}^\be{2} \cQ^R_\be{2} = 0 \,,
\end{equation}
is particularly strong and often constraints the existence of supersymmetric moduli considerably.
The group \(G^\ast\) does not mix fields from different multiplets, therefore variations in the first factor \(G^\ast / H^\ast\) of \eqref{eq:halfmaximalcoset} contribute only to \({(\cP_{\delta\phi})_\hal{2}}^\hbe{2}\) and \({(\cP_{\delta\phi})_\tal{2}}^\tbe{2}\).
On the other hand, variations in the second factor of \eqref{eq:halfmaximalcoset} give rise only to \({(\cP_{\delta\phi})_\hal{2}}^\tbe{2}\) and \({(\cP_{\delta\phi})_\tal{2}}^\hbe{2}\).
For this reason the condition \({(\cP_{\delta\phi})_{\hal{2}}}^\hbe{2} \cQ^R_\hbe{2}\) enforces all variations in \(G^\ast / H^\ast\) to vanish, as we will illustrate in the next chapter for a concrete example.
Consequently a possible moduli space can only be a submanifold of the second factor of \eqref{eq:halfmaximalcoset}.


In four dimensions there is also a supergravity theory with \(q = 12\) real supercharges.
The scalar manifold of this theory is given by
\begin{equation}
\cM = \frac{\SU(3,n_V)}{\mathrm{S}[\U(3) \times \U(n_V)]} \,,
\end{equation}
where \(n_V\) again denotes the number of vector multiplets.
The gauge fields arrange themself into the complex vector representation of \(\SU(3,n_V)\).
The analysis of the moduli space is very similar to the half-maximal case.
In the next chapter we show explicitly that \eqref{eq:halfmaximalcondition} enforces the moduli space to be trivial.

Let us finally draw our attention to theories with more than 16 real supercharges, which thus have the gravitational multiplet as their only supermultiplet.
For these theories the conditions \eqref{eq:cosetmoduli} simplify considerably and become
\begin{equation}\begin{aligned}\label{eq:maximalmoduli}
\cD_{\delta\phi} \cQ^R_\hal{2} &= - {(\cP_{\delta\phi})_\hal{2}}^\hbe{2} \cQ^R_\hbe{2} = 0 \,, \\
\cD_{\delta\phi} \cP_\hal{2} &= 
\bigl[\cQ^R_\hal{2}, \cP_{\delta\phi}\bigr] = 0 \,.
\end{aligned}\end{equation}
Moreover, here the only generators of the gauge group are \(\cT_\hal{2} = \cQ^R_\hal{2} + \cP_\hal{2}\), see \eqref{eq:maximalgenerators}.
Therefore, \eqref{eq:maximalmoduli} is equivalent to \eqref{eq:Tmoduli} which shows that all solutions of \eqref{eq:maximalmoduli} are moduli and that the moduli space is a symmetric homogeneous space of the form \eqref{eq:symmetricmodulispace}.

To make this a bit more specific we note that there is no spontaneous symmetric breaking due to the vanishing of all Killing vectors \(\cP_{\hal{2}}\) in the background.
This is consistent with the observation \eqref{eq:maximalGg} that the entire gauge group is only given by \(H^g_R\).
Therefore we do not have to worry about possible Goldstone bosons and \(\k_{AdS}\) comprises all solutions of \eqref{eq:maximalmoduli}, i.e.
\begin{equation}\label{eq:kads}
\k_{AdS} = \left\{\cP \in \k : \bigl[ \cP, \cQ^R_\hal{2}\bigr] = {\cP_\hal{2}}^\hbe{2} \cQ^R_\hbe{2} = 0 \right\} \,.
\end{equation}
To extend this to a proper subalgebra of \(\g\) we define
\begin{equation}\label{eq:hads}
\h_{AdS} = \left\{\cQ \in \h : \bigl[ \cQ, \cQ^R_\hal{2}\bigr] = {\cQ_\hal{2}}^\hbe{2} \cQ^R_\hbe{2} = 0 \right\} \,,
\end{equation}
and \(\g_{AdS} = \k_{AdS} \oplus \h_{AdS}\).
It is straightforward to show that \(\h_{AdS}\) and \(\g_{AdS}\) are subalgebras of \(\h\) and \(\g\), respectively, i.e.~they are closed with respect to the Lie bracket.
Consequently \(\k_{AdS}\) corresponds to the tangent space of the coset manifold \(\cM_{AdS} = G_{AdS}/H_{AdS}\).
We illustrate our techniques in the next chapter and compute the AdS moduli spaces for all theories with more than 16 supercharges explicitly.

\chapter{Examples for Maximally Supersymmetric AdS Solutions}\label{chap:adsmoduli}

In this chapter we apply our previous general results to specific supergravity theories with
 a symmetric scalar field space
 and discuss their maximally supersymmetric AdS backgrounds with group theoretical methods.

In the first section~\ref{sec:greater16} we discuss the gauge groups and moduli spaces for all maximally supersymmetric AdS solutions with more than 16 supercharges.
These cases are particularly constrained due to the absence of any other multiplets than the gravity multiplet.
In section~\ref{sec:equal16} we turn to theories with exactly 16 supercharges, i.e.~half-maximal supergravities.
Here, there can be additional vector multiplets which makes the analysis slightly more involved.
As an example we explicitly discuss the seven-dimensional case.
In section~\ref{sec:N=3} we study a second example with vector multiplets, namely four-dimensional \(\cN = 3\) supergravity, which has 12 supercharges.

\section{AdS solutions with \texorpdfstring{$q > 16$}{q > 16} supercharges}\label{sec:greater16}

At first we need to determine which theories allow for maximally supersymmetric AdS solutions at all.
It is well-known from \cite{Nahm:1977tg} that the corresponding AdS superalgebras exist only in certain dimensions and also not for arbitrary numbers of supercharges.
Consequently, one expects that only those theories where an AdS superalgebra exists can be gauged in such a way that a (maximally supersymmetric) AdS solution is possible.
Maximally supersymmetric AdS backgrounds are characterized by the general condition \eqref{eq:adsconditions}, this means that we have to seek for theories which allow for \(A_0 \neq 0\) but \(A_1 = 0\). Note that \(A_1 = 0\) is already enough to ensure \((A_0)^2 \sim \id\), which is necessary for unbroken supersymmetry.

This task simplifies a lot if the scalar manifold is a symmetric homogeneous space of the form \(\cM = G/H\).
As we describe in chapter~\ref{sec:coset}, here the gaugings can be conveniently described in terms of the T-tensor \(\cT\) \eqref{eq:Ttensor}, which is a scalar field dependent object with a well-defined transformation behavior with respect to \(H\).
Moreover, the shift matrices \(A_0\) and \(A_1\) are built from the appropriate \(H\)-irreducible components of \(\cT\).
In table~\ref{tab:A0A1} we explicitly list which irreducible components of \(\cT\) correspond to \(A_0\) and \(A_1\). 
Due to its \(H\)-invariance the condition \(A_1 = 0\) implies that every irreducible component of \(\cT\) which is present in \(A_1\) must vanish identically.
Therefore, \(A_0 \neq 0\) is only possible if there is an irreducible component of \(\cT\) which is part of \(A_0\) but not of \(A_1\).%
\footnote{The situation is slightly more subtle if there are two independent components of \(\cT\) which are both transforming in the same representation.
If both of them are part of \(A_1\) it is possible that only a certain linear combination of them is set equal to zero.
A second linear combination that might be part of \(A_0\) might still be non-vanishing.
However, if we consider these two different linear combinations as independent irreducible representations our argumentation is still valid.}
Inspection of table~\ref{tab:A0A1} shows that for more than 16 real supercharges this is only possible in dimensions \(D = 4, 5\) and \(7\).\footnote{Note that we restrict the discussion to \(D \geq 4\).}

\begin{table}[htb]
\centering
\begin{tabular}{|c|c|c|c|c|c|c|}
\hline
$D$ & $q$ & $H = H_R$ & $A_0$ & $A_1$ & Ref. &  AdS$_D$ \\
\hline
11 & 32 & - & - & - & & \\
\hline
10 & $(32,0)$ & $\U(1)$ & - & - & & \\
& $(16,16)$ & -  & $\mathbf{1}_m$ & $\mathbf{1}_m$ & \cite{Romans:1985tz} & \\
\hline
9 & 32 & $\U(1)$ & $\mathbf{0} \oplus \mathbf{1}_a$ & $\mathbf{0} \oplus \mathbf{1}_a \oplus \mathbf{1}_b$ & \cite{Bergshoeff:2002nv, FernandezMelgarejo:2011wx} &  \\
\hline 
8 & 32 & $\U(2)$ & $\mathbf{1_{+1}}$ & $\mathbf{1_{+1}} \oplus \mathbf{3_{+1}} \oplus \mathbf{5_{+1}}$ & \cite{Bergshoeff:2003ri, deRoo:2011fa} & \\ 
\hline
7 & 32 & $\USp(4)$ & $\mathbf 1 \oplus \mathbf 5$ & $\mathbf{5} \oplus \mathbf{14} \oplus \mathbf{35}$ & \cite{Samtleben:2005bp} & \textbullet \\
\hline
6 & (16,16) & $\USp(4) \times \USp(4)$ & $(\mathbf{4},\mathbf{4})$ & $(\mathbf{4},\mathbf{4}) \oplus (\mathbf{4},\mathbf{16}) \oplus (\mathbf{16},\mathbf{4})$ & \cite{Bergshoeff:2007ef} & \\
& (16,8) & $\USp(4) \times \USp(2)$ & $(\mathbf{4},\mathbf{2})_a$ & $(\mathbf{4},\mathbf{2})_a \oplus (\mathbf{4},\mathbf{2})_b \oplus (\mathbf{16},\mathbf{2})$ & \cite{Roest:2009sn} & \\
\hline
5 & 32 & $\USp(8)$ & $\mathbf{36}$ & $\mathbf{315}$ & \cite{deWit:2004nw} & \textbullet \\
& 24 & $\USp(6)$ & $\mathbf{21}_a$ & $\mathbf{14} \oplus \mathbf{21}_b \oplus \mathbf{70}$ & &  \textbullet \\
\hline
4 & 32 & $\SU(8)$ & $\mathbf{36}$ & $\mathbf{420}$ & \cite{deWit:2007kvg} & \textbullet \\
& 24 & $\U(6)$ & $\mathbf{21_{+1}}$ & $\mathbf{15_{+1}} \oplus \mathbf{35_{-3}} \oplus \mathbf{105_{+1}}$ & \cite{Andrianopoli:2008ea, Roest:2009sn} &  \textbullet \\
& 20 & $\U(5)$ & $\mathbf{15_{+1}}$ & $\mathbf{\overline{5}_{-3}} \oplus \mathbf{10_{+1}} \oplus \mathbf{\overline{40}_{+1}}$ & \cite{Trigiante:2016mnt} & \textbullet \\
\hline
\end{tabular}
\caption{Deformations of supergravities with \(q >16\) and \(D \geq 4\).
The last column indicates whether a maximally supersymmetric AdS solution is possible.
A subscript ``$m$'' denotes a massive deformation.
If there are multiple deformations transforming in the same \(H_R\) representation they are distinguished by subscripts ``$a$'', ``$b$'', ... .
For $D=9$ and $D=6$, $q = (16,8)$ all components of \(A_1\) have to vanish independently.}
\label{tab:A0A1}
\end{table}

In the following we want to analyze the gaugings which can lead to AdS solutions and the respective moduli spaces for the allowed theories from table~\ref{tab:A0A1} explicitly.
The first step consists in finding the subgroup \(H^g_R \subset H_R\) which is generated by the moment maps \(\cQ_\hal{2}\).
\(H^g_R\) is a subgroup of \(H_R\) under which \(A_0\) does not transform and which is gaugeable by the graviphotons, we will see in the examples that it is always the maximal such subgroup of \(H_R\).
We determine \(H^g_R\) in a case-by-case analysis for the dimensions \(D = 4,5\) and \(7\) separately und verify the results using the explicit formula \eqref{eq:AdSconditionsQP} for the moment maps \(\cQ_\hal{2}\).
We want to stress that the results for \(H^g_R\) are universal and not restricted to theories with \(q > 16\).
However, if \(q > 16\) the only possible multiplet is the gravitational multiplet and there can be no other gauge fields than the graviphotons.
Therefore, as explained in chapter~\ref{sec:adsgaugings}, the gauge group \(G^g\) must be reductive and is uniquely fixed by \(G^g = H^g_R\).
 
The knowledge of the gauge group \(H^g_R\) finally allows us to determine the moduli spaces of the AdS solutions.
The key result of chapter~\ref{sec:adsmoduli} is that moduli must necessarily be uncharged with respect to \(H^g_R\).
As explained in chapter~\ref{sec:coset} the Lie algebra \(\g\) of \(G\) splits into the Lie algebra \(\h\) of \(H\) and its orthogonal complement \(\k\).
It is \(\k\) which corresponds to the non-compact directions of \(G\) and therefore to the physical scalar fields.
Moreover, \(\h\) and \(\k\) satisfy \(\left[\h,\k\right] \subseteq \k\) so \(\k\) transforms in an \(\h\)-representation with respect to the adjoint action.
As \(\h^g_R\) is a subalgebra of \(\h\) we can decompose \(\k\)
 into irreducible representations of \(\h^g_R\).
We have seen that only the singlets in this decomposition can be candidates for moduli.

We summarize the results for \(H^g_R\) and the relevant decompositions in table~\ref{tab:adsdecomp}.
It shows that the only theory with \(\h^g_R\)-singlets in the decomposition of \(\k\) is the five-dimensional maximal (i.e. \(\cN = 8\) or \(q = 32\)) supergravity.
We argue in due course that the corresponding scalar fields are indeed moduli.
The absence of singlets shows that all the other theories cannot have a non-trivial moduli space.

\begin{table}[htb]
\centering
{\tabulinesep=0.7mm
\begin{tabu}{|c|c|c|c|c|c|}
\hline
$D$ & $q$ & $G/H$ & $H^g_R$ & $\g \rightarrow \h \oplus \k$ & $\k \rightarrow \bigoplus \k_i$  \\
\hline
7 & 32 & 
$\frac{\SL(5)}{\SO(5)}$ 
& $\SO(5)$ &$\mathbf{24} \rightarrow \mathbf{10} \oplus \mathbf{14}$ & $\mathbf{14} \rightarrow 
\mathbf{14}$ \\
\hline
5 & 32 & $\frac{\E_{(6,6)}}{\USp(8)}$ & $\SU(4)$ & $\mathbf{78} \rightarrow \mathbf{36} \oplus \mathbf{42}$ & $\mathbf{42} \rightarrow 
2 \cdot \mathbf{1} \oplus \mathbf{10} \oplus \mathbf{\overline{10}} \oplus \mathbf{20'}$\\
& 24 & $\frac{\SU^*(6)}{\USp(6)}$ & $\U(3)$ & $\mathbf{35} \rightarrow \mathbf{21} \oplus \mathbf{14}$ & $\mathbf{14} \rightarrow
\mathbf{3_{-1}} \oplus \mathbf{\overline{3}_{+1}} \oplus \mathbf{8_0}$ \\
\hline
4 & 32 & $\frac{\E_{(7,7)}}{\SU(8)}$ & $\SO(8)$ & $\mathbf{133} \rightarrow \mathbf{63} \oplus \mathbf{70}$ & $\mathbf{70} \rightarrow \mathbf{70}$ \\
& 24  & $\frac{\SO^*(12)}{\U(6)}$ & $\SO(6)$ & $\mathbf{66} \rightarrow \mathbf{1_0} \oplus \mathbf{35_0} \oplus \mathbf{15_1} \oplus \mathbf{\overline{15}_{-1}}$ & $\mathbf{15_1} \oplus \mathbf{\overline{15}_{-1}} \rightarrow 
2 \cdot \mathbf{15}$ \\
& 20 & $\frac{\SU(5,1)}{\U(5)}$ & $\SO(5)$ & $\mathbf{35} \rightarrow \mathbf{1_0} \oplus \mathbf{24_0} \oplus \mathbf{5_1} \oplus \mathbf{\overline{5}_{-1}}$ & $\mathbf{5_1} \oplus \mathbf{\overline{5}_{-1}}\rightarrow 
2 \cdot \mathbf{5}$ \\
\hline
\end{tabu}}
\caption{Relevant representation theoretical decompositions for the determination of AdS moduli spaces.
Firstly, the branching of the adjoint representation of \(\g\) into \(\h\)-representations and secondly the branching of the \(\h\)-representation corresponding to \(\k\) into representations of \(\h^g_R\).}
\label{tab:adsdecomp}
\end{table}

In the following we discuss each of the three dimensions \(D=4,5,\) and \(7\) independently.
For each case we demonstrate how to explicitly compute the gauge group \(G^g = H^g_R\) using the general formula \eqref{eq:AdSconditionsQP}.
Moreover, for the maximal five-dimensional theory we show that the two singlets in the decomposition of \(\k\) are indeed moduli and compute the corresponding moduli space.

Let us shortly outline our strategy:
\begin{enumerate}
\item Find the maximal subalgebra \(\x \subseteq \h_R\) such that \(\bigl[\x, A_0\bigr] = 0\), i.e.~\(\x\) is the stabilizer of \(A_0\) in \(\h_R\), and decompose the graviphotons \(A^\hal{2}\) into irreducible representations with respect to \(\x\).
\item The adjoint representation of the gauge algebra \(\h^g_R\) must be contained in this decomposition.
The result can be confirmed explicitly using \eqref{eq:AdSconditionsQP}.
\item Decompose the scalar fields \(\k\) into representations of \(\h^g_R\) (see table~\ref{tab:adsdecomp}). The singlets are candidates for moduli.
\end{enumerate}

\subsection*{Four-dimensional AdS solutions}

The R-symmetry group of a four-dimensional supergravity with \(q = 4 \cN\) real supercharges is
\begin{equation}\label{eq:d4ralgebra}
H_R = \begin{cases}
\U(\cN) & \text{if}\; \cN \neq 8 \\
\SU(\cN) & \text{if}\; \cN = 8 
\end{cases} \,,
\end{equation}
where \(\cN\) is the number of chiral supersymmetry parameters \(\epsilon^i_+ = \Gamma_\ast \epsilon^i_+\).
Their charge conjugates \(\epsilon_{-i} = (\epsilon^i_+)^C\) have opposite chirality, i.e.~\(\epsilon_{-i} = -\Gamma_\ast \epsilon_{-i}\).\footnote{Our spinor conventions are outlined in appendix~\ref{app:conventions}.}
\(\mathrm{(S)}\U(\cN)\) indices are raised and lowered by complex conjugation.
We summarize some properties of four-dimensional supergravities in appendix~\ref{app:sugras}.


As outlined there, the shift matrix \((A_0)_{ij} = \bigl((A_0)^{ij}\bigr)^\ast\) is a symmetric matrix. 
The condition \eqref{eq:adsconditions} on maximally supersymmetric AdS vacua reads
\begin{equation}
(A_0)_{ik} (A_0)^{kj} = - \frac{\Lambda}{12} \delta_i^j \,.
\end{equation}
It implies that all eigenvalues \(\lambda_i\) of \(A_{ij}\) satisfy \(\left|\lambda_i\right| = \sqrt{\frac{\Lambda}{12}}\) , but they can in principle differ by a complex phase.
As outlined above we need to find the stabilizer algebra of \(A_0\) in \(\h^g_R\), i.e.~the maximal subalgebra \(\x \in \h^g_R\) commuting with \(A_0\).
As explained for example in \cite{ORaifeartaigh:1986agb}, there is always an element \(U \in \SU(\cN)\) such that
\begin{equation}\label{eq:orthogonalA}
(A_0)_{kl} U^k_i U^l_j = e^{i\omega} \sqrt{\frac{\Lambda}{12}} \delta_{ij} \,,
\end{equation}
i.e.~it is possible to align the phases of all eigenvalues of \(A_0\) by a special unitary transformation.
If \(H_R = \U(\cN)\) we can perform an additional \(\U(1)\) rotation to remove the overall phase factor \(e^{i\omega}\) as well.
However, this is not possible if \(H_R\) is only \(\SU(\cN)\).
\eqref{eq:orthogonalA} is invariant with respect to orthogonal transformations and therefore
\begin{equation}
\x = \so(\cN) \,.
\end{equation}

Next we decompose the dressed graviphotons \(A^\hal{2}\) into irreducible representations of \(\x\).
They are given by \(A^{[ij]}_M\) and their complex conjugates \(A_{M[ij]} = (A^{[ij]}_M)^\ast\).
Both transform in the same way with respect to \(\x = \so(\cN)\), namely in the antisymmetric tensor representation.
This is at the same time also the adjoint representation of \(\so(\cN)\),
so we expect the gauged R-symmetry algebra to be given by \(\h^g_R = \so(\cN)\).
For the \(\cN = 6\) theory there is an additional graviphoton \(A^0_M\), transforming as an R-symmetry singlet.
However, there is no generator of \(\x\) left which could be gauged by \(A^0_M\).


To compute the generators \(\cQ_\hal{2}\) of \(\h^g_R\) explicitly, using the general formula \eqref{eq:AdSconditionsQP}, it is necessary to combine the fundamental and anti-fundamental representation of \(\su(\cN)\) into a column vector, e.g.~\(\epsilon^i = (\epsilon^i_+, \epsilon_{-i})^T\), see also appendix~\ref{app:sugras}.
Analogously we arrange \((A_0)_{ij}\) and \((A^0)^{ij}\) into a \((2\cN) \times (2\cN)\) matrix as
\begin{equation}\label{eq:D4A0}
A_0 = \begin{pmatrix} 0 & (A_0)^{ij} \\ (A_0)_{ij} & 0 \end{pmatrix} \,.
\end{equation}
Inserting \eqref{eq:D4A0} together with the explicit expression for \(B_\hal{2}\) given in \eqref{eq:appD4B} and \eqref{eq:appD4Bb}
into \eqref{eq:AdSconditionsQP} yields
\begin{equation}\label{eq:D4Q}
\cQ_{ij} \equiv \begin{pmatrix}{\bigl(\cQ_{ij}\bigr)^k}_l & \bigl(\cQ_{ij}\bigr)^{kl} \\ \bigl(\cQ_{ij}\bigr)_{kl} & {\bigl(\cQ_{ij}\bigr)_k}^l \end{pmatrix} 
= \frac{1}{\sqrt{2}}\begin{pmatrix}\delta^k_{[i} (A_0)_{j]l} & 0 \\ 0 & - \delta^k_{[i} (A_0)_{j]l} \end{pmatrix} \,,
\end{equation}
and an analogous result for \(\cQ^{ij}\).
After diagonalizing \((A_0)_{ij}\) and \((A_0)^{ij}\) by an \(\SU(\cN)\) transformation \eqref{eq:orthogonalA}
we find from \eqref{eq:D4Q} the following generators of the gauged R-symmetry \(\h^g_R\),
\begin{equation}
{(\cQ_{ij})^k}_l = -{(\cQ_{ij})_l}^k = e^{i \omega} \sqrt{\frac{|\Lambda|}{24}} \delta^k_{[i} \delta_{j]l} \,,\qquad \cQ^{ij} = \bigl(\cQ_{ij}\bigr)^\ast \,.
\end{equation} 
We recognize the generators of \(\so(\cN)\). Therefore, for all four-dimensional theories the gauged R-symmetry is indeed given by
\begin{equation}
H^g_R = \SO(\cN) \,.
\end{equation}
We want to point out again that for \(\cN \neq 8\) we can use the left-over \(\U(1)\) freedom to annihilate the complex phase \(e^{i\omega}\).
For \(\cN = 8\), however, this is not possible and \(\omega\) parametrizes a family of inequivalent \(\SO(8)\)-gaugings, known as \(\omega\)-deformations \cite{DallAgata:2012mfj}.%
\footnote{See \cite{Borghese:2014gfa} for a discussion of \(\omega\)-deformations in \(\cN = 6\) supergravity.}

Let us finally discuss the role of the additional gauge field \(A^\tal{2}_M = A^0_M\) in the \(\cN = 6\) theory, which could in principle gauge another isometry
generated by
\begin{equation}
\cT_0 = \cQ_0 + \cP_0 \,.
\end{equation}
Since \(B_0 = 0\) \eqref{eq:appD4B0} it follows directly from \eqref{eq:AdSconditionsQP} that
\begin{equation}
\cQ_0 = 0 \,.
\end{equation}
However, for the same reason \eqref{eq:AdSconditionsQP} a priori does not require \(\cP_0 = 0\),
but if we evaluate the commutator between a generator \(\cQ_\hal{2}\) of \(H^g_R\) and \(\cP_0\) we find
\begin{equation}\label{eq:N6QP}
\bigl[\cQ_\hal{2}, \cP_0\bigr] = \bigl[\cT_\hal{2}, \cT_0\bigr] = {\bigl(\cT_\hal{2}\bigr)_0}^\al{2} \cT_\al{2} = 0 \,,
\end{equation}
since \(A^0\) is uncharged with respect to \(G\).
Moreover, we can read of from table~\ref{tab:adsdecomp} that there are no \(\h^g_R\) singlets in \(\k\).
Therefore, \eqref{eq:N6QP} implies
\begin{equation}
\cP_0 = 0 \,,
\end{equation}
and \(A^0_M\) cannot gauge an isometry of \(\cM\)
Nonetheless, \(A^0_M\) can still generate an independent \(U(1)\)-gauge symmetry which does not correspond to an isometry \cite{Borghese:2014gfa}.

As mentioned above and summarized in table~\ref{tab:adsdecomp} none of the four-dimensional solutions with \(q \geq 16\) admits for \(\h^g_R\) singlets in the decomposition of \(\k\).
Therefore for all three cases the moduli space is trivial.

\subsection*{Five-dimensional AdS solutions}

The R-symmetry group of a five-dimensional supergravity with \(q = 4 \cN\) real supercharges is given by
\begin{equation}
H_R = \USp(\cN) \,,
\end{equation}
where \(\cN\) is the number of supersymmetry parameters \(\epsilon^i\) satisfying the symplectic Majorana condition \eqref{eq:symplmajorana}.
The relevant properties of five-dimensional supergravities are summarized in appendix~\ref{app:sugras}.

Firstly, note that in five dimensions the shift matrix \(\left(A_0\right)_{ij} =  \Omega_{ki} A^k_{0\,j}\) is symmetric \eqref{eq:appD5A09}.
\(\Omega_{ij}\) is the \(\usp(\cN)\) invariant tensor introduced in \eqref{eq:symplmajorana} which can be used to raise and lower indices.
Moreover, we require \(A_0\) to satisfy the condition \eqref{eq:adsconditions} on maximally supersymmetric AdS vacua which reads
\begin{equation}\label{eq:D5AdS}
A^i_{0\,k} A^k_{0\,j} = \lambda^2 \delta^i_j \,,\qquad \lambda^2 = \frac{\left|\Lambda\right|}{24} \,.
\end{equation}
Let us determine the maximal subalgebra \(\x \subseteq \usp(\cN)\) which commutes with \(A_0\).
For this purpose we note that \eqref{eq:D5AdS} implies together with \(A^i_{0\, i} = A_{0\,ij} \Omega^{ij} = 0\) that
the eigenvalues of \(A^i_{0\,j}\) are given by \(\pm \lambda\), with multiplicity \(\cN/2\) each.
We denote the respective eigenvectors by \(e^i_\alpha\) and \(e^i_{\bar\alpha}\) and introduce \(A_{\alpha\beta} =  e^i_\alpha A_{0\, ij} e^j_\beta\), \(\Omega_{\alpha\beta} = e^i_\alpha \Omega_{ij} e^j_\beta\), et cetera.
The symmetry of \(A_0\) requires that
\begin{equation}\begin{aligned}
A_{\alpha\beta} &= - \lambda \Omega_{\alpha\beta} = 0 \,, \\
A_{\bar\alpha\bar\beta} &= \lambda \Omega_{\bar\alpha\bar\beta} = 0 \,, \\
A_{\alpha\bar\beta} &= A_{\bar\beta\alpha} = \lambda \Omega_{\alpha\bar\beta} \,.
\end{aligned}\end{equation}
Expressed in this basis \(A_0\) has the form of a hermitian metric which is invariant with respect to unitary transformations and therefore
\begin{equation}\label{eq:D5x}
\x = \u(\cN/2) = \u(1) \oplus \su(\cN/2) \,.
\end{equation}
Working in the eigenbasis of \(A_0\) corresponds to splitting the fundamental representation of \(\usp(\cN)\) labeled by \(i\) into the fundamental and anti-fundamental representation of \(\x\), labeled by \(\alpha\) and \(\bar\alpha\).
Note that it is moreover possible to choose a convenient basis of eigenvectors in which
\begin{equation}
\Omega_{\alpha\bar\beta} = \delta_{\alpha\bar\beta} \,.
\end{equation}

In the next step we have to look at the dressed graviphoton fields \(A^\hal{2}\) \eqref{eq:appD5graviphotons}.
In five-dimensional supergravities there generically exist the graviphoton fields \(A^{[ij]}_M\) constrained by the condition \(A^{ij}_M \Omega_{[ij]} = 0\), i.e.~transforming in the traceless antisymmetric tensor representation of \(\usp(\cN)\).
Moreover, for theories with \(\cN \neq 8\) there is an additional graviphoton \(A^0_M\), transforming under \(\usp(\cN)\) as a singlet. 
To understand how these representations branch into representations of \(\x\) we express them in the eigenbasis of \(A_0\).
The \(\usp(\cN)\) singlet \(A^0_M\) stays of course inert under \(\u(\cN)\) and therefore transforms in the adjoint representation of \(\u(1)\).
On the other hand, the vector fields \(A^{[ij]}_M\) decompose as
\begin{equation}\label{eq:ialphabaralpha}
A^{[ij]}_M \rightarrow A^{[\alpha\beta]}_M \oplus A^{[\bar\alpha\bar\beta]}_M \oplus A^{\alpha\bar\beta}_M \,,
\end{equation}
where the last summand satisfies \(A_M^{\alpha\bar\beta} \delta_{\alpha\bar\beta} = 0\).
Therefore, the \(A_M^{\alpha\bar\beta}\) transform in the adjoint representation of \(\su(\cN)\).
Consequently, we expect that the gauged R-symmetry algebra \(\h^g_R\) is given by \(\h^g_R = \u(\cN/2)\) if the singlet \(A^0_M\) is present (i.e.~for \(\cN \neq 8\)) and otherwise by \(\h^g_R = \su(\cN/2)\) (i.e.~for \(\cN = 8\)).

Let us now give an explicit verification of this result.
Inserting the expression \eqref{eq:appD5B} for the matrices \(B_\hal{2}\)  into \eqref{eq:AdSconditionsQP} yields
\begin{equation}\begin{aligned}\label{eq:D5Rgen}
(\cQ_0)^k_l &=  2i \sqrt{\tfrac{8-\cN}{2\cN}} (A_0)^k_l \,,\\
 (\cQ_{ij})^k_l &= 2i \left((A_0)^k_{[i} \Omega_{j]l} - \delta^k_{[i} (A_0)_{j]l} + \tfrac2\cN \Omega_{ij} (A_0)^k_l \right) \,.
\end{aligned}\end{equation}
These are the generators of \(\h^g_R\).
We want to express them in the basis of eigenvectors of \(A_0\).
The result reads
\begin{equation}\begin{aligned}
(\cQ_0)^\gamma_\delta &= 2 i \lambda \sqrt{\tfrac{8-\cN}{2\cN}} \delta^\gamma_\delta \,, \\
(\cQ_{\alpha\bar\beta})^\gamma_\delta &= - 2 i \lambda \left(\delta^\gamma_\alpha \delta_{\bar\beta\delta} - \tfrac2\cN \delta_{\alpha\bar\beta} \delta^\gamma_\delta\right) \,,
\end{aligned}\end{equation}
and similarly for \((\cQ_0)^{\bar\gamma}_{\bar\delta}\) and \((\cQ_{\alpha\bar\beta})^{\bar\gamma}_{\bar\delta}\).
All other components are either determined by antisymmetry or vanish identically.
We recognize that \(\cQ_0\) commutes with all other generators and thus spans the abelian algebra \(\u(1)\).
The \(\cQ_{\alpha\bar\beta}\) on the other hand are hermitian and traceless and therefore are
 the generators of \(\su(\cN/2)\).
This confirms that the gauged R-symmetry is given by
\begin{equation}
H^g_R = \begin{cases}
\U(\cN/2) & \text{if}\; \cN \neq 8 \\
\SU(\cN/2) & \text{if}\; \cN = 8 
\end{cases} \,.
\end{equation}

The next step is the determination of the moduli space.
The relevant decompositions of the representation \(\k\) of the scalar fields into irreducible representations  of \(\h^g_R\) are summarized in table~\ref{tab:adsdecomp}.
Only for the maximal theory with \(\h^g_R = \su(4)\) there are singlets in the decomposition, which thus is the only theory where a non-trivial moduli space can exist.

Let us check that these singlets are indeed moduli and determine the geometry of the manifold they span.
From table~\ref{tab:adsdecomp} we read off that the scalar manifold of the maximal theory is given by 
\begin{equation}
\cM = \frac{\E_{(6,6)}}{\USp(8)} \,,
\end{equation}
and that the decomposition of the adjoint representation of \(\mathfrak{e}_{(6,6)}\) into representations of \(\usp(8)\) reads \(\mathbf{78} \rightarrow \mathbf{36} \oplus \mathbf{42}\).
The \(\mathbf{36}\) is the adjoint representation of \(\h = \usp(8)\) and the \(\mathbf{42}\) corresponds to \(\k\).
To determine the geometry of the moduli space \(\cM_{AdS}\) (which is a subspace of \(\cM\)) we decompose both into representations of \(\h^g_R = \su(4)\) and find
\begin{equation}\begin{aligned}\label{eq:d5n8modulidecomp}
\h &\colon\quad \mathbf{36} \rightarrow \mathbf{1} + \mathbf{10} + \mathbf{\overline{10}} + \mathbf{15}\,, \\
\k &\colon\quad \mathbf{42} \rightarrow 2 \cdot \mathbf{1} + \mathbf{10} + \mathbf{\overline{10}} + \mathbf{20'} \,.
\end{aligned}\end{equation}
Next we determine the algebra \(\g_{AdS}\) spanned by the three singlets in \eqref{eq:d5n8modulidecomp}.
For this purpose we note that the \(\mathbf{36}\) corresponds to a symmetric \(\usp(8)\)-tensor \(\Lambda_{(ij)}\), and that the \(\mathbf{42}\) is given by a completely antisymmetric \(\usp(8)\) 4-tensor \(\Sigma_{[ijkl]}\), constrained by the tracelessness condition \(\Omega^{ij}\Sigma_{ijkl} = 0\) \cite{deWit:2004nw}.
Together \(\Lambda_{ij}\) and \(\Sigma_{ijkl}\) span the adjoint representation of \(\mathfrak{e}_{(6,6)}\) and satisfy the commutator relations
\begin{equation}\begin{aligned}\label{eq:e6}
\bigl[\Lambda_{ij}, \Lambda_{kl}\bigr] &= \Omega_{ik} \Lambda_{jl} + \dots \,, \\
\bigl[\Lambda_{ij}, \Sigma_{klmn}\bigr] &= \Omega_{ik} \Sigma_{jlmn} + \dots \,, \\
\bigl[\Sigma_{ijkl}, \Sigma_{mnop}\bigr] &= \Omega_{im} \Omega_{jn} \Omega_{ko} \Lambda_{lp} + \dots \,, 
\end{aligned}\end{equation} 
where the ellipses stand for all terms which need to be added to obtain the correct (anti-) symmetry on the right-hand side.
Moreover, a generator \(T = \lambda^{ij} \Lambda_{ij} + \sigma^{ijkl} \Sigma_{ijkl}\) of \(\mathfrak{e}_{(6,6)}\) acts on a tensor \(X_{[ij]}\) in the antisymmetric traceless representation (i.e.~the \(\mathbf{27}\)) of \(\usp(8)\) as \cite{deWit:2004nw}
\begin{equation}\label{eq:usp8action}
(T X)_{ij} = - 2 {\lambda_{[i}}^k X_{j]k} + {\sigma_{ij}}^{kl} X_{kl} \,.
\end{equation}
To reproduce the decomposition \eqref{eq:d5n8modulidecomp} we express \(\Lambda_{ij}\) and \(\Sigma_{ijkl}\) in the eigenbasis of \(A_0\) which was constructed above.
The three singlets are given by
\begin{equation}
\Lambda^0 = \tfrac{1}{4}\delta^{\alpha\bar\beta} \Lambda_{\alpha\bar\beta} \,,\quad
\Sigma^- = \tfrac{1}{4!}\epsilon^{\alpha\beta\gamma\delta} \Sigma_{\alpha\beta\gamma\delta} \,,\quad
\Sigma^+ = \tfrac{1}{4!}\epsilon^{\bar\alpha\bar\beta\bar\gamma\bar\delta} \Sigma_{\bar\alpha\bar\beta\bar\gamma\bar\delta} \,. 
\end{equation}
From \eqref{eq:e6} we find
\begin{equation}
\left[\Lambda^0, \Sigma^\pm\right] = \pm \Sigma^\pm \,,\qquad \left[\Sigma^-, \Sigma^+\right] = \Lambda_0 \,,
\end{equation}
These are the well-known commutator relations of \(\su(1,1)\).
Moreover, from \eqref{eq:usp8action} it follows that \(\Lambda^0\) and \(\Sigma^{\pm}\) indeed satisfy the conditions \eqref{eq:kads} and \eqref{eq:hads} on supersymmetric moduli and therefore
\begin{equation}
\h_{AdS} = \mathrm{span}\bigl(\{\Lambda^0\}\bigr) \,,\qquad \k_{AdS} = \mathrm{span}\bigl(\{\Sigma^-, \Sigma^+\}\bigr) \,,
\end{equation}
and \(\g_{AdS} = \h_{AdS} \oplus \k_{AdS} = \su(1,1)\).
Consequently, the moduli space is given by the coset space
\begin{equation}
\cM_\mathrm{AdS} = \frac{\SU(1,1)}{\U(1)} \,.
\end{equation}

\subsection*{Seven-dimensional AdS solutions}

The R-symmetry group of a seven-dimensional supergravity theory with \(q = 8 \cN\) real supercharges is given by
\begin{equation}
H_R = \USp(\cN) \,,
\end{equation}
where \(\cN\) is the number of supersymmetry parameters \(\epsilon^i\) satisfying the symplectic Majorana condition \eqref{eq:symplmajorana}.
We summarize the essential properties of seven-dimensional supergravities in appendix~\ref{app:sugras}.

In seven dimensions the shift matrix \(\left(A_0\right)_{ij}\) is antisymmetric \eqref{eq:appD7A0}.
Hence, in general it decomposes into two irreducible \(\usp_\cN\) representations: The singlet representation (proportional to \(\Omega_{ij}\)) and the antisymmetric traceless representation.
However, as we see from table~\ref{tab:A0A1} the second AdS condition \(A_1 = 0\) enforces the antisymmetric traceless part to vanish\footnote{In \(D=7\) there only exist the \(\cN=4\) and the \(\cN = 2\) theories.
The case \(\cN = 2\) is not contained in table~\ref{tab:A0A1} but an antisymmetric traceless representation of \(\USp(2)\) does not exist.} and therefore
\begin{equation}\label{eq:D7A0AdS}
(A_0)_{ij} = \pm \sqrt\frac{|\Lambda|}{60}\Omega_{ij} \,.
\end{equation}
Consequently, the maximal subalgebra \(\x\) of \(\h_R = \usp(\cN)\) commuting with \(A_0\) is \(\usp(\cN)\) itself, i.e.
\begin{equation}
\x = \usp(\cN)\,.
\end{equation}
Therefore the decomposition of the dressed graviphotons \(A^\hal{2}_M\) into representations of \(\x\) is trivial.
As stated in \eqref{eq:appD7graviphotons} the graviphotons are given by \(A^{(ij)}_M\), i.e.~they transform in the symmetric tensor representation of \(\usp(\cN)\).
This is also its adjoint representation,
so we expect the gauged R-symmetry algebra to be given by \(\h^g_R = \usp(\cN)\).
Let us verify this explicitly.
The matrices \(B_\hal{2}\) are given in \eqref{eq:appD7B}.
Inserting the expression stated there as well as \eqref{eq:D7A0AdS} into \eqref{eq:AdSconditionsQP} gives
\begin{equation}
(\cQ_{ij})^k_l = 6 \sqrt\frac{\left|\Lambda\right|}{30} \delta^k_{(i} \Omega_{j)l} \,.
\end{equation}
These indeed are the generators of \(\usp(\cN)\) in the fundamental representation, which
confirms our above result and hence
\begin{equation}
H^g_R = \USp(\cN) \,.
\end{equation}

In seven-dimensions the only supergravity with \(q > 16\) is the maximal \(\cN = 4\) theory.
Also here the decomposition of \(\k\) into irreducible representations of \(\h^g_R\) does not contain any singlets, see table~\ref{tab:adsdecomp}.
This shows that the AdS moduli space is trivial.

\section{AdS solutions in seven-dimensional \texorpdfstring{$\cN = 2$}{N = 2} supergravity}\label{sec:equal16}

In the previous section we discussed the AdS solutions and the corresponding moduli spaces of all gauged supergravities with more than 16 real supercharges by group theoretical arguments.
For less supersymmetric theories, however, the analysis is more complicated due to the appearance of multiplets other than the gravity multiplet.
For example, half-maximal theories (i.e.~theories with \(q=16\) real supercharges) can be coupled to an arbitrary number of vector multiplets.
The vector fields in these multiplets can be used to gauge additional symmetries and thus the gauge group \(G^g\) can be larger than \(H^g_R\) and possibly non-compact.
Nonetheless, the scalar geometry of all half-maximal theories is still described by a symmetric space of the form \(\cM = G/H\), so we expect our results from chapter~\ref{sec:adsmoduli} to be applicable.

In the following we illustrate the analysis of less supersymmetric AdS solutions with the example of half-maximal supergravity in seven dimensions.
A more detailed and explicit discussion can be found in \cite{Louis:2015mka}.
The AdS solutions of half-maximal supergravity in four, five and six dimensions have been analyzed in \cite{Louis:2014gxa, Louis:2015dca, Karndumri:2016ruc}.

The global symmetry group \(G\) of seven-dimensional half-maximal (i.e.~\(\cN = 2\)) supergravity coupled to \(n\) vector multiplets is given by \cite{Townsend:1983kk, Bergshoeff:1985mr}
\begin{equation}
G = \SO(1,1)
\times \SO(3,n) \,.
\end{equation}
Let us denote its generators by \(t_0\) and \(t_{[IJ]}\), where \(I,J = 1, \dots, 3+n\).
In the vector representation the generators \(t_{[IJ]}\) read
\begin{equation}\label{eq:tIJ}
{(t_{IJ})_K}^L = \delta^L_{[I}\, \eta_{J]K} \,,
\end{equation}
where \(\eta_{IJ} = \mathrm{diag}(-1,-1,-1;+1, \dots, +1)\) is the canonical \(\SO(3,n)\) metric.
The maximal compact subgroup \(H = H_R \times H_\mathrm{mat}\) of \(G\) is
\begin{equation}
H = \SO(3)_R \times \SO(n) \,.
\end{equation} 
This is consistent with the above statement that the R-symmetry group of all seven-dimensional supergravities is given by \(\USp(\cN)\) since locally \(\SO(3) \cong \SU(2) \cong \USp(2)\).
The collective vector fields \(A^I_M\) transform in the vector representation of \(\SO(3,n)\) and carry an \(\SO(1,1)\)-charge of \(+ 1/2\), i.e.
\begin{equation}\label{eq:t0}
t_0 A^I = + \tfrac{1}{2} A^I \,.
\end{equation}
Consequently the embedding tensor \(\Theta\) sits in the product representation \(\ydiagram{1}_{-\frac12} \otimes \left(1_0 \oplus \ydiagram{1,1}_0\right)\).
After imposing the linear constraint \(\Theta\) reads \cite{Bergshoeff:2007vb, Dibitetto:2015bia}
\begin{equation}
\Theta \colon \ydiagram{1}_{-\frac12} \oplus \ydiagram{1,1,1}_{-\frac12} \,. \\
\end{equation}
Moreover, there exists a massive deformation which cannot be described as the gauging of a global symmetry \cite{Bergshoeff:2005pq, Bergshoeff:2007vb}.
Altogether, the deformations can be explicitly parametrized by the two independent embedding tensor components \(f_{[IJK]}\), \(\xi_I\), and by a mass parameter \(h\).
According to \eqref{eq:gaugegenerators} the generators \(X_I\) of the gauge group \(G^g\) read in terms of the embedding tensor
\begin{equation}\label{eq:D7X}
X_I = {\Theta_I}^0 t_0 + {\Theta_I}^{JK} t_{JK} \,,
\end{equation}
where \(\Theta^0_I\) and \({\Theta_I}^{JK}\) are given by \cite{Bergshoeff:2007vb}
\begin{equation}\begin{aligned}\label{eq:D7Theta}
{\Theta_I}^{JK} &= {f_I}^{JK} + \delta_I^{[J}\xi^{K]} \,, \qquad\quad
{\Theta_I}^0 &= \xi_I \,.
\end{aligned}\end{equation}
Inserting \eqref{eq:D7Theta} as well as \eqref{eq:tIJ} and \eqref{eq:t0} back into \eqref{eq:D7X} yields the generators \(X_I\) of the gauge group,
\begin{equation}\label{eq:D7XIJK}
{\bigl(X_I\bigr){}_J}^K = {X_{IJ}}^K = - {f_{IJ}}^K - \frac{1}{2} \eta_{IJ} \xi^K + \delta^K_{[I} \xi_{J]} \,,
\end{equation}
Note that this is precisely the same expression as found in \cite{Schon:2006kz} for the half-maximal five-dimensional gauged supergravities.

The transition from \(G\) to \(H\) is performed via the coset representative \(L = (L^\alpha_I) = (L^\hal{}_I, L^\tal{}_I)\) of \(\cM = G/H\), where \(\hal{} = 1,2,3\) and \(\tal{} = 1,\dots, n\).
So \(\hal{}\) labels the vector representation of \(SO(3)\) and \(\tal{}\) labels the vector representation of \(SO(n)\).
The contractions of \(f_{[IJK]}\) and \(\xi_I\) with \(L\) yields the T-tensor.
The distribution of its irreducible components on the shift matrices reads
\begin{equation}\begin{aligned}
A_0 &\colon (\mathbf{1}, \mathbf{1}) \oplus (\mathbf{1}, \mathbf{1})_m \,, \\
A_1 &\colon (\mathbf{1}, \mathbf{1}) \oplus (\mathbf{1}, \mathbf{1})_m \oplus (\mathbf{3}, \mathbf{1}) \oplus (\mathbf{3}, \mathbf{n}) \oplus (\mathbf{1}, \mathbf{n}) \,,
\end{aligned}\end{equation}
where \((\mathbf{1}, \mathbf{1})_m\) denotes the massive deformation parametrized by h.
Even though not directly visible from this schematic decomposition, it is crucial that different linear combinations of the two \(H\)-singlet deformations enter \(A_0\) and \(A_1\).
Therefore a maximally supersymmetric solution is only possible if the theory is not only gauged but also deformed by the mass parameter \(h\).
Only then it is possible to have \(A_0 \neq 0\) and \(A_1 = 0\).
Explicitly, the condition \(A_1 = 0\) imposes
\begin{equation}\begin{aligned}\label{eq:D7strcond}
f_{\hal{}\hbe{}\hga{}} &\sim h \epsilon_{\hal{}\hbe{}\hga{}} \,, \\
f_{\hal{}\hbe{}\tga{}} &= \xi_\hal{} = \xi_\tal{} = 0 \,.
\end{aligned}\end{equation}
As explained in chapter~\ref{sec:coset} the contraction of \({X_{IJ}}^K\) with the coset representatives yields the T-tensor and as its components the moment maps \(\cQ_\alpha\) and Killing vectors \(\cP_\alpha\). 
Consequently, we infer from \eqref{eq:D7XIJK} that the conditions \eqref{eq:D7strcond} translate into conditions on \(\cQ_\alpha\) and \(\cP_\alpha\) as
\begin{equation}\begin{gathered}\label{eq:QPD7half}
{\left(\cQ_\hal{}\right)_\hbe{}}^\hga{} \sim \epsilon_{\hal{}\hbe{}\hga{}} \,,\qquad
{\left(\cQ_\tal{}\right)_\hbe{}}^\hga{} = \cP_\hal{} = 0 \,.
\end{gathered}\end{equation}
We see that the \(\left(\cQ_\hal{}\right)^\hga{}_\hbe{}\) generate the group \(H^g_R = SO(3)\).
This is also what we expect from the analysis in the previous section
and \eqref{eq:QPD7half} is fully consistent with the general conditions \eqref{eq:adsconditions}.
The remaining unconstrained generators of the gauge group are spanned by
\begin{equation}\begin{aligned}\label{eq:D7generators}
{\left(\cQ_\hal{}\right)_\tbe{}}^\tga{} &= - {\bigl(\cP_\tbe{}\bigr)_\hal{}}^\tga{} = - {\bigl(\cP_\tbe{}\bigr)_\tga{}}^\hal{} = - {f_{\hal{}\tbe{}}}^\tga{} \,, \\
{\left(\cQ_\tal{}\right)_\tbe{}}^\tga{} &= - {f_{\tal{}\tbe{}}}^\tga{} \,. \\
\end{aligned}\end{equation} 
Therefore, the gauge group \(G^g\) can be larger than just \(H^g_R\) and in particular also non-compact.

Let us now go over to the discussion of the moduli space.
As usual we begin with the split of the Lie algebra \(\g\) of \(G\) into \(\h\) and \(\k\).
Therefore, we decompose the generators \(t_{[IJ]}\) of \(\g\) according to
\begin{equation}
t_{[IJ]} \rightarrow t_{[\hal{}\hbe{}]} \oplus t_{[\tal{}\tbe{}]} \oplus t_{\hal{}\tbe{}} \,,
\end{equation}
hence \(\h\) is spanned by \(t_{[\hal{}\hbe{}]}\) and \(t_{[\tal{}\tbe{}]}\) whereas \(\k\) is spanned by \(t_0\) and \(t_{\hal{}\tbe{}}\).
Consequently, we can expand every \(\cP_{\delta\phi} \in \k\) representing a scalar variation \(\delta\phi\) as
\begin{equation}
\cP_{\delta\phi} = \delta\phi^0 t_0 + \delta\phi^{\hal{}\tbe{}} t_{\hal{}\tbe{}} \,.
\end{equation}
To constrain the supersymmetric moduli space we use the condition \eqref{eq:halfmaximalcondition}.
Together with \eqref{eq:t0} we find
\begin{equation}
0 = {\bigl(\cP_{\delta\phi}\bigr)_\hal{}}^\hbe{} \cQ^R_\hbe{} = \delta\phi^0 {\bigl(t_0\bigr)_\hal{}}^\hbe{} \cQ^R_\hbe{} = - \tfrac{1}{2} \delta\phi^0 \cQ^R_\hal{}\,,
\end{equation}
and therefore \(\delta\phi^0 = 0\).
On the other hand, evaluating \eqref{eq:halfmaximalcondition} for \(\alpha = \tal{}\) together with \eqref{eq:tIJ} yields
\begin{equation}
0 = {\bigl(\cP_{\delta\phi}\bigr)_\tal{}}^\hbe{} \cQ^R_\hbe{} = \delta\phi^{\hga{}\tde{}} {\bigl(t_{\hga{}\tde{}}\bigr)_\tal{}}^\hbe{} \cQ^R_\hbe{} = \tfrac{1}{2} \delta\phi^{\hbe{}\tal{}} \cQ^R_\hbe{}\,,
\end{equation}
and consequently that \(\delta\phi^{\hal{}\tbe{}} = 0\).
Therefore the moduli space is trivial.

\section{AdS solutions in four-dimensional \texorpdfstring{$\cN = 3$}{N = 3} supergravity}\label{sec:N=3}

In this section we discuss the maximally supersymmetric AdS solutions of four-dimensional $\cN = 3$ supergravity as a second example for AdS solutions of a supergravity theory admitting vector multiplets.%
\footnote{Aspects of AdS solutions and gaugings of four-dimensional $\cN = 3$ supergravities have also been discussed in \cite{Karndumri:2016tpf}.}

The scalar manifold of this theory is given by the coset space \cite{Castellani:1985ka}
\begin{equation}
\cM = \frac{\SU(3,n)}{\mathrm{S}[\U(3) \times \U(n)]} \,,
\end{equation}
where \(n\) denotes the number of vector multiplets.
The gauge fields \(A^I_M\) transform in the \((\mathbf{3 + n}) \oplus \overline{(\mathbf{3 + n})}\) representation of \(G = \SU(3,n_V)\).
Consequently, the dressed gauge fields \(A^\al{2}_M\) transform in the \((\mathbf{3}, \mathbf{1})_{-1} \oplus (\mathbf{1},\overline{\mathbf{n}})_{-3/n}\) representation of \(H = \U(3) \times \SU(n)\), where the subscripts denote the \(\U(1)\) charge.
Moreover, we denote the complex conjugate of \(A^\al{2}_M\) by \(A^{\bar\alpha_2}_M\).

To determine if maximally supersymmetric AdS solutions exist we need to know how the irreducible \(H\)-representations of the T-tensor distribute on the shift matrices \(A_0\) and \(A_1\).
We can read them of from \cite{Trigiante:2016mnt},
\begin{equation}\begin{aligned}
A_0 &\colon (\mathbf{6}, \mathbf{1})_{+1} \,, \\
A_1 &\colon (\overline{\mathbf{3}}, \mathbf{1})_{+1} \oplus (\mathbf{1},\mathbf{n})_{3/n} \oplus (\mathbf{3},\mathbf{n})_{2+3/n} \oplus (\mathbf{8},\mathbf{n})_{3/n} \,.
\end{aligned}\end{equation}
We observe that the \(H\)-representation of \(A_0\) does not appear in \(A_1\), therefore \(A_0 \neq 0\) and \(A_1 = 0\) is possible.
Consequently, the conditions \eqref{eq:adsconditions} can be solved and maximally supersymmetric AdS solutions exist.
Moreover, the \((\mathbf{6}, \mathbf{1})_{+1}\) representation of \(A_0\) is the symmetric tensor representation of \(\U(3)\), which agrees precisely with our general considerations in chapter~\ref{sec:greater16}.
There, we have furthermore determined that this form of \(A_0\) implies that the three moment maps \(\cQ^R_\hal{2}\) generate the gauged R-symmetry group
\begin{equation}
H^g_R = \SO(3) \,.
\end{equation}
Of course, the general gauge group \(G^g \subset \SU(3,n)\) can be much more complicated, however, its precise form is not relevant for our further analysis.

Let us now discuss the moduli spaces of such solutions.
We denote the generators of \(\SU(3,n)\) by \(t_{I \bar J}\).
In the fundamental representation they read
\begin{equation}
{(t_{I \bar J})_K}^L = \delta_{\bar J K} \delta^L_I - \tfrac1{3+n} \delta_{I \bar J} \delta^L_K \,.
\end{equation}
According to the splitting \(\su(3,n) \rightarrow \u(3) \oplus \su(n)\) they decompose as
\begin{equation}
t_{I \bar J} \rightarrow t_{\hal{2} \bar{\hat \beta}_2} \oplus t_{\tal{2} \bar{\tilde \beta}_2} \oplus t_{\hal{2} \bar{\tilde \beta}_2} \oplus t_{\tal{2} \bar{\hat \beta}_2} \,.
\end{equation}
The first two terms span the maximally compact subalgebra \(\h =  \u(3) \oplus \su(n)\) and the second two terms span the non-compact part \(\k\) which corresponds to the tangent space of \(\cM\).
Therefore, we can expand the variation matrix \(\cP_{\delta\phi} \in \k\) as
\begin{equation}
\cP_{\delta\phi} = \delta\phi^{\hal{2} \bar{\tilde \beta}_2} t_{\hal{2} \bar{\tilde \beta}_2} +  \delta\phi^{\tal{2} \bar{\hat \beta}_2} t_{\tal{2} \bar{\hat \beta}_2} \,.
\end{equation}
In particular \( \delta\phi^{\tal{2} \bar{\hat \beta}_2}\) is the complex conjugate of \(\delta\phi^{\hal{2} \bar{\tilde \beta}_2}\).
Inserting this parametrization into \eqref{eq:halfmaximalcondition} yields
\begin{equation}
{(\cP_{\delta\phi})_{\tal{2}}}^\hbe{2} \cQ^R_\hbe{2} = \delta_{\tal{2} \bar{\tilde\gamma}_2 } \delta\phi^{\hbe{2} \bar{\tilde\gamma}_2} \cQ^R_\hbe{2} = 0 \,,
\end{equation}
and thus
\begin{equation}
\delta\phi^{\hal{2} \bar{\tilde\beta}_2} = 0 \,.
\end{equation}
In the same way we infer from \({(\cP_{\delta\phi})_{\bar{\tilde\alpha}_2}}^{\bar{\hat\beta}_2}\cQ^R_{\bar{\hat\beta}_2} = 0\) the vanishing of \(\delta\phi^{\tal{2} \bar{\hat \beta}_2}\).
This shows that the moduli space is trivial.

\chapter{Marginal Deformations of \texorpdfstring{$(1,0)$}{(1,0)} SCFTs}\label{chap:SCFT}

In this chapter we study possible marginal deformations of six-dimensional \(\cN=(1,0)\) superconformal field theories.
These theories can serve as holographic duals of the seven-dimensional AdS backgrounds with \(\cN = 2\) supersymmetry which we analyzed in the previous chapter.

As explained in the \hyperref[chap:introduction]{introduction}, 
we can deform an (S)CFT by adding local operators \(\cO_i\) to its Lagrangian\footnote{As mentioned before this prescription is only symbolic for non-Lagrangian theories. Nonetheless, it can be still equipped with a sensible meaning via conformal perturbation theory.}
\begin{equation}\label{eq:SCFTdeformations}
\cL \rightarrow \cL + \lambda^i \cO_i \,,
\end{equation}
with coupling constants \(\lambda^i\) parametrizing the deformation.
Depending on their scaling or conformal dimensions \(\Delta_{\cO_i}\) we characterize the operators \(\cO_i\) as irrelevant, relevant or marginal deformations, as these three cases lead to a qualitatively different behavior.
While irrelevant and relevant deformations necessarily destroy the conformal invariance of the theory, marginal deformations preserve it at leading order in the coupling constants \(\lambda^i\).
By definition the conformal dimension \(\Delta_{\cO_i}\) of a marginal operator agrees with the space-time dimension \(d\) such that the corresponding couplings are dimensionless.
This is a necessary condition for the preservation of conformal invariance.
However, if we take higher-order corrections in \(\lambda^i\) into account, the conformal dimension of a marginal operator can get renormalized and we distinguish again between marginally irrelevant, marginally relevant and exactly marginal deformations.
Only for exactly marginal deformations \(\Delta_{\cO_i}\) does not get renormalized.
Therefore, these deformations do not break the conformal symmetry at all orders in perturbation theory and a deformation of the form \eqref{eq:SCFTdeformations} gives another (S)CFT.
The space spanned by the corresponding exactly marginal couplings \(\lambda^i\) is called the conformal manifold \(\cC\).
As already explained in detail, the conformal manifold corresponds to the moduli space of an holographically dual AdS solution.

Since we are dealing with SCFTs, we are furthermore interested in deformations that preserve supersymmetry as well, i.e.~supersymmetric exactly marginal deformations.
The action of the deformed theory stays invariant under supersymmetry transformations if the deformations \(\cO_i\) satisfy
\begin{equation}
\bigl[Q, \cO_i\bigr] = \partial_\mu (\dots) \,,
\end{equation}
where \(Q\) schematically  stands for all supercharges and \(\partial_\mu (\dots)\) is the total derivative of a well-defined operator.
In an SCFT all operators -- and hence also the supersymmetric marginal deformations --  have to arrange into unitary representations of the underlying superconformal symmetry algebra, i.e.~into supermultiplets.
However, unitarity in combination with superconformal invariance imposes strong bounds on the conformal dimensions of the components of the allowed supermultiplets.
Schematically, these bounds take the form
\begin{equation}
\Delta \geq f(s, r) \,,
\end{equation}
for some function \(f\) depending on the spin \(s\) and the R-symmetry charge \(r\) of the respective operator.
These unitarity bounds are often so strong that they forbid the existence of supersymmetric marginal deformations completely.

The goal of this chapter is to show that for six-dimensional \(\cN = (1,0)\) SCFTs indeed all supersymmetric exactly marginal deformations are excluded by unitarity bounds.
To this end, we first review in section~\ref{sec:superconfalgebra} the representation theory of the corresponding superconformal algebra \(\mathfrak{osp}(6,2|2)\).
Afterwards we present the main part of our analysis in section~\ref{sec:marginaloperators}.

\section{Unitary representations of \texorpdfstring{$\OSp(6,2|2)$}{OSp(6,2|2)}}\label{sec:superconfalgebra}

Let us start with a brief review of the representation theory of the conformal algebra \(\so(6,2)\), which is a subalgebra of \(\mathfrak{osp}(6,2|2)\).\footnote{Our presentation follows \cite{Cordova:2016emh}. For a more detailed discussion of the representation theory of superconformal algebras see e.g.~\cite{Minwalla:1997ka, Bhattacharya:2008zy, Rychkov:2016iqz} and references therein.}
The generators of \(\mathfrak{osp}(6,2|2)\) are the
 Lorentz transformations \(M_{[\mu\nu]}\), the dilatation operator \(D\), the momenta \(P_\mu\) as well as the special conformal transformations \(K_\mu\).
Together \(M_{[\mu\nu]}\) and \(D\) are the generators of the maximal compact subalgebra \(\so(6) \oplus \so(2)\) of \(\so(6,2)\), while its non-compact part is spanned by \(P_\mu\) and \(K_\mu\).
Every element of a conformal multiplet, i.e.~of an irreducible representation of \(\so(6,2)\), is a local operator with a distinct transformation behaviour with respect to Lorentz transformations and dilatations, or in other words it is part of an irreducible representation of \(\so(6) \oplus \so(2)\).
Consequently, we label each operator \(\cO\) by three half-integer \(\so(6)\) weights \((h_1,h_2,h_3)\) and an \(\so(2)\) weight \(\Delta_\cO\) which is called the operator's conformal dimension.%
\footnote{It is sometimes convenient to translate the \(\so(6)\)  weights  \((h_1,h_2,h_3)\) into \(\su(4)\) Dynkin labels \([a_1 a_2 a_3]\) via
$
a_1 = h_2 - h_3 \,,\
a_2 = h_1 + h_2 \,,\
a_3 = h_2 + h_3 \,.
$
This implies, in particular, that they are not completely arbitrary but that they need to satisfy the constraint \(h_1 \geq h_2 \geq \left|h_3\right|\).
For example, \((\frac{1}{2}, \frac{1}{2}, \pm\frac{1}{2})\) denotes the
(anti-)chiral spinor representation, while \((1,0,0)\) is the
\(\SO(6)\) vector representation.}
The representation theory of the compact Lie algebra \(\so(6) \oplus \so(2)\) is well-known, we can therefore focus on the role of the momenta \(P_\mu\) and special conformal transformations \(K_\mu\).
Note that often when we are talking about a single operator \(\cO\), we actually mean the whole  \(\so(6) \oplus \so(2)\)-representation of which \(\cO\) is a member.

One infers from the commutation relations of \(P_\mu\) and \(K_\mu\) with \(D\) that \(P_\mu\) and \(K_\mu\) carry conformal dimension \(\Delta_P = + 1\) and \(\Delta_K = -1\).
This means that the action of \(P_\mu\) or \(K_\mu\) on an operator \(\cO\) raises or lowers the conformal dimension of \(\cO\) accordingly.
In each irreducible conformal multiplet there is a distinct operator of lowest conformal dimension, it is called the conformal primary \(\cP\).
It can be equivalently characterized by the requirement that it is annihilated by all special conformal transformations \(K_\mu\), i.e.
\begin{equation}
\bigl[K_\mu, \cP\bigr] = 0 \,.
\end{equation}
The rest of the multiplet is obtained by acting on \(\cP\) with the momenta \(P_\mu\).
The resulting operators \(\cO\) are called conformal descendants.
Therefore, we can label the conformal multiplet by the \(\so(6) \oplus \so(2)\) weights of its primary \(\cP\).
Since \(P_\mu\) acts on local operators as a spacetime derivative \(\partial_\mu\), the descendant operators are nothing but the derivatives of the conformal primary \(\cP\).
Consequently, the only operator which can not be written as the total derivative \(\partial_\mu (...)\) of some other operator is the conformal primary \(\cP\).

We can define an inner product \((\cdot, \cdot)\) of local operators by using their two-point function, i.e.
\begin{equation}\label{eq:operatorproduct}
\bigl(\cO_1, \cO_2\bigr) = \bigl< \cO_1^\dagger \cO_2 \bigr> \,.
\end{equation}
Equivalently, we can use the correspondence between local operators and states in radial quantization.
It assigns to each operator \(\cO\) the state \(\bigr|\cO\bigl>\) that is obtained by acting with \(\cO\) on the vacuum state \(\bigl|\Omega\bigr>\), i.e. \(\bigr|\cO\bigl> = \cO \bigl|\Omega\bigr>\).
The scalar product \eqref{eq:operatorproduct} is then simply given by the scalar product of the corresponding states.
This allows us to introduce the notion of a unitary representation.
In a unitary conformal multiplet we require all operators to be non-negative with respect to the norm
\begin{equation}\label{eq:norm}
\left\|\cO \right\|^2 = \left(\cO, \cO\right) \,, 
\end{equation}
induced by \eqref{eq:operatorproduct}.
In particular, it is possible to compute the norm of every descendant operator in terms of the norm, the \(\so(6) \oplus \so(2)\) weights \((h_1,h_2,h_3)\) and \(\Delta_\cP\) of the primary operator by using the conformal algebra and \(P^\dagger_\mu = K_\mu\).
The outcome of these computations shows that for arbitrary values of \((h_1,h_2,h_3)\) and \(\Delta_\cP\), not necessarily all descendants have a positive norm.
This implies unitarity bounds on the conformal dimension \(\Delta_\cP\) of the form
\begin{equation}\label{eq:CFTunitarity}
\Delta_\cP \geq f(h_1, h_2, h_3) \,,
\end{equation}
where the function \(f\) is explicitly determined in \cite{Minwalla:1997ka}.
We have to distinguish the following two situations.
Generically, the bound \eqref{eq:CFTunitarity} is not saturated and all descendants have a strictly positive norm.
In this case the multiplet is called a long multiplet.
For certain values of \(\Delta_\cP\) and \((h_1,h_2,h_3)\) it is however possible that the bound is saturated.
In this case some descendants must have vanishing norm.
All operators of vanishing norm form a representation of the conformal algebra as well, thus the original multiplet is reducible.
It is therefore possible to consistently project out all operators of vanishing norm, the resulting irreducible representation is called a short multiplet.

We are now in the position to review the representation theory of the full superconformal algebra \(\mathfrak{osp}(6,2|2)\).
The bosonic subalgebra of \(\mathfrak{osp}(6,2|2)\) is \(\mathfrak{osp}(6,2) \times \su(2)_R\), where \(\su(2)_R\) is the R-symmetry algebra.
Analogous to the previous discussion we characterize every member of a superconformal multiplet by the three half-integer \(\so(6)\) weights \((h_1,h_2,h_3)\), its conformal dimension \(\Delta\), as well as an half-integer \(\su(2)_R\) weight \(k\).
The fermionic part of \(\mathfrak{osp}(6,2|2)\) is generated by an R-doublet of supercharges \(Q^i_\alpha\) and an R-doublet of superconformal charges \(S_i^\alpha\).
Here \(\alpha = 1,\ldots,4,\) denotes
the fundamental representation of \(\SU(4) = \mathrm{Spin}(6)\) and \(i=1,2\) 
labels the fundamental representation of the \(\SU(2)_R\).
They satisfy the schematic anticommutator relations \(\left\{Q,Q\right\} \sim P\) and  \(\left\{S,S\right\} \sim K\) and therefore carry the conformal dimensions \(\Delta_Q = + \frac12\) and \(\Delta_S = - \tfrac12\).

As in the conformal case every superconformal multiplet contains one unique operator \(\cS\) of lowest conformal dimension \(\Delta_\cS\).
It is called the superconformal primary operator and satisfies
\begin{equation}
\bigl[S^\alpha_i, \cS\bigr] = 0 \,.
\end{equation}
Note that every superconformal primary is automatically a conformal primary as well, but not necessarily the other way around.
In fact, a generic superconformal multiplet contains multiple conformal primary operators but only one superconformal primary.
The remaining operators in the superconformal multiplet, called superconformal descendants, are obtained by acting with the supercharges \(Q^i_\alpha\) on \(\cS\).
Moreover, an operator \(\cO\) obtained by the action of \(l\) supercharges is often called a level-\(l\) descendant and has conformal dimension \(\Delta_\cO = \Delta_\cS + \frac{l}{2}\).
As a consequence every superconformal multiplet decomposes into a direct sum of finitely many conformal multiplets, i.e.~irreducible representations of \(\mathfrak{osp}(6,2) \times \su(2)_R\).
Let us explain this in a bit more detail:
If we act with a certain number of supercharges on the superconformal primary it is often possible to trade two of the supercharges for a momentum \(P_\mu\) using \(\left\{Q, Q\right\} \sim P\).
In this case the resulting operator is a conformal descendant as well.
However, if this is not possible the resulting operator is a conformal primary.
On top of each of these conformal primaries we can build a conformal multiplet as we have explained above. 
If we are only interested in the supermultiplet consisting of these conformal primary operators, we might consequently use the effective anticommutator relation \(\left\{Q,Q\right\} \sim 0\).
Therefore each supermultiplet can exist of only finitely many conformal primaries.
Moreover, we call a conformal primary operator from which we cannot obtain another conformal primary by the action of \(Q\) a top component of the corresponding supermultiplet.
In other words, such a top component is annihilated by all supercharges \(Q\) up to a total derivative.

The unitarity condition that all operators must have non-negative norm \eqref{eq:norm} again imposes bounds on the conformal dimension \(\Delta_\cS\) of the superconformal primary operators \(\cS\).
Due to the existence of more superconformal descendants than just conformal descendants these bounds are stronger compared to the conformal case.
These bounds can be computed using \(Q^\dagger = S\) and the (anti-)commutator relations listed in appendix~\ref{app:superconfalgebra}.
They take the generic form
\begin{equation}\label{eq:genericbound}
\Delta_\cS \geq f(h_1, h_2, h_3; k) \,,
\end{equation}
where the function \(f\) is explicitly determined in \cite{Minwalla:1997ka, Dobrev:2002dt}.
We recall it in the following section.
We again distinguish between the situation where the inequality is strict, corresponding to a long multiplet, and the situation in which the bound is saturated, in which case some descendants have vanishing norm and the corresponding superconformal multiplet is short.
Moreover, for special values of \((h_1,h_2,h_3)\) there can be isolated short representations at particular isolated values of \(\Delta_\cS\) which are smaller than allowed by the generic bound \eqref{eq:genericbound}.

\section{Classification of marginal operators}\label{sec:marginaloperators}

After these preliminaries we are now in the position to show that six-dimensional \(\cN=(1,0)\) SCFTs do not allow for supersymmetric marginal deformations.
A supersymmetric marginal deformation is an operator \(\cO\) with the following properties.
At first, it has to be a Lorentz scalar operator with conformal dimension \(\Delta_{\cO} = 6\). Otherwise it would break conformal invariance.
Note that the Lorentz algebra is a subalgebra of the conformal algebra.
For similar reasons we demand \(\cO\) to be a singlet with respect to the R-symmetry, since the R-symmetry algebra is a subalgebra of the full superconformal algebra.
Moreover, adding a total derivative to the Lagrangian \eqref{eq:SCFTdeformations} does not deform the theory, hence \(\cO\) has to be a conformal primary.
Finally, we want \(\cO\) to preserve supersymmetry, so it has to be the top-component of a supermultiplet, which means that it is annihilated by all supercharges up to a total derivative.

From our previous discussion it is in principle possible to determine all unitary representations of \(\mathfrak{osp}(6,2|2)\)
and to scan the resulting catalogue for multiplets which contain a scalar top-component of the correct conformal dimension.\footnote{This approach has been followed recently in \cite{Cordova:2016xhm, Cordova:2016emh}.}
However, we follow a slightly different strategy and write down all scalar operators of conformal dimension \(\Delta = 6\) which are in principle compatible with the unitarity bound.
This gives a (short) finite list of operators for which we can check explicitly whether they fulfill the above requirements.   
As we discussed, they must be part of a unitary representation of the 
superconformal algebra and therefore  are either superconformal primary operators or
descendant operators
that are obtained by acting with \(l\) supercharges \(Q^i_{\alpha}\) on
a superconformal primary operator. 
However, the primary operators that are invariant under Lorentz symmetry, R-symmetry and
supersymmetry have been shown to be proportional to the identity
operator \cite{Green:2010da}. 
Therefore, we can restrict our further analysis to descendant operators.
Note that the order of  supercharges in a descendant operator does not matter for our analysis as supercharges anticommute up to a moment operator, i.e.~a total derivative.

If we start with a primary operator with \(\so(6)\) weights
\((h_1,h_2,h_3)\), we can only find Lorentz invariant descendant
operators at level
\begin{equation}\label{eq:invlevel}
l=2(h_1+h_2+h_3) +4n \,,
\end{equation}
with $n$ being an arbitrary non-negative integer.
In appendix~\ref{app:minlevel} we give a proof of this statement.
Thus the conformal dimension of the primary operator needs to be
\begin{equation}\label{eq:confdimdesc}
\Delta_\cS = 6 - \frac{l}{2} = 6 - h_1 - h_2 - h_3 - 2 n \,. 
\end{equation}
Moreover, we will use in the following that
\(k=0\) is only possible if \(l\) is even as 
descendants with an odd number of supercharges cannot be R-singlets.
The general bound from \cite{Minwalla:1997ka, Dobrev:2002dt} for a unitary representation reads 
\begin{equation}\label{eq:generalbound}
\Delta_\cS \geq h_1 + h_2 - h_3 + 4k + 6\,,
\end{equation}
which is not compatible with \eqref{eq:confdimdesc}, since \(h_1\) and
\(h_2\) are necessarily non-negative. Therefore, all descendants
of primary operators in long representations are excluded.

For special choices of the weights \((h_1, h_2, h_3)\) there
exist isolated short representations which we now turn to.
The following cases can be distinguished.

a) If \(h_1 - h_2 > 0\) and \(h_2 = h_3\), there is a short representation with
\begin{equation}
\Delta_\cS = h_1 + 4k + 4 \,.
\end{equation}
Together with \eqref{eq:confdimdesc} the only possible solution is
\begin{equation}
(h_1, h_2, h_3) = (1, 0, 0)\,,\quad k=0\,,\quad\Delta_\cS = 5 \,.
\end{equation}
A primary operator with these properties carries no R-symmetry indices and 
has to be an antisymmetric $\SU(4)$-tensor (which is isomorphic to
the six-dimensional vector representation of $\SO(6)$).
Thus the corresponding candidate descendant operator has to take the form
\begin{equation}
\cO_2 = \epsilon^{\alpha\beta\gamma\delta}\left\{Q_{i\alpha},[Q^i_\beta,\cS_{[\gamma\delta]}]\right\} \,,
\end{equation}
where $\cS_{[\gamma\delta]}$ is the associated primary operator with $\Delta_\cS=5$.
The norm of this operator can be computed straightforwardly
by using 
the superconformal algebra given in
appendix~\ref{app:superconfalgebra}
with the result \(\left\|\cO_2\right\|^2 \sim \Delta_0-5=0\).
As zero-norm states are not allowed in a unitary theory,
the operator \(\cO_2\) has to vanish.\footnote{Note
that this operators is a total derivative for any $\Delta_\cS$.
This is the case because the contraction of the R-symmetry indices is performed with an $\epsilon$-symbol, so \(\cO_2\) is symmetric under the exchange of the two supercharges and using \eqref{eq:qanticom} we see that \(\cO_2 \sim \left[P^{\alpha\beta}, \cS_{\alpha\beta}\right]\).}

b) For \(h_1 = h_2 = h_3 = h \neq 0\) there are additional short representations if
\begin{subequations}
\begin{align}
\Delta_\cS &= 2 + h + 4k\ ,\qquad \text{or} \label{eq:boundc1}\\
\Delta_\cS &= 4 + h + 4k\ . \label{eq:boundc2}
\end{align}
\end{subequations}
While \eqref{eq:boundc2} is not compatible with \eqref{eq:confdimdesc}, there are two solutions for \eqref{eq:boundc1}, namely 
\begin{equation}\label{sol1}
h=\frac{1}{2}\,,\quad k=\frac{1}{2}\,,\quad \Delta_\cS = \frac{9}{2} \,,
\end{equation} 
and
\begin{equation}\label{sol2}
h=1\,,\quad k=0\,,\quad \Delta_\cS = 3 \,.
\end{equation}
Denoting the primary operator for the first solution \eqref{sol1}
 by \(U^i_\alpha\), 
it is indeed possible to identify a Lorentz and R-symmetry invariant
descendant operator \(\cO_3\) at level \(l=3\) 
\begin{equation}
\cO_3 = \epsilon^{\alpha\beta\gamma\delta}\left\{Q_{i\alpha},\left[Q^i_\beta,\{Q_{j\gamma}, \cS^j_\delta\}\right]\right\} \,.
\end{equation}
Computing the norm yields
\(
\left\|\cO_3\right\|^2 \sim \left(\Delta_\cS - \frac{9}{2}\right)\left(\Delta_\cS + \frac{7}{2}\right)
\)
which vanishes at the critical value \(\Delta_\cS = 6-\frac{l}{2}
=\frac{9}{2}\). Consequently, \(\cO_3\) itself vanishes and cannot be
considered as a possible marginal operator.
Notice that it is in principle possible to contract the R-symmetry
indices in a different fashion but all such operators 
differ from \(\cO_3\) only by a total derivative.
Moreover, we have checked that all these other combinations have
vanishing norm as well. 

For the second solution \eqref{sol2} the primary operator has the
index structure \(U_{(\alpha\beta)}\) (with \(h=1\) and \(k=0\)) and we can build a Lorentz and R-symmetry invariant descendant operator \(\cO_6\) at level \(l=6\),
\begin{equation}
\cO_6 = \epsilon^{\alpha\beta\gamma\delta} \epsilon^{\epsilon\zeta\eta\theta} \left\{Q_{i\alpha},\left[Q^i_\epsilon,\left\{Q_{j\beta},\left[Q^j_\zeta,\left\{Q_{k\gamma},[Q^k_\eta, \cS_{(\delta\theta)}]\right\}\right]\right\}\right]\right\} \,.
\end{equation}
There are also other possibilities to contract the indices within \(\cO_6\), which would however lead to total derivatives.
In any case all these \(l=6\) operators are descendants of the operator \([Q_{i[\alpha},\cS_{(\beta]\gamma)}]\), whose norm is \((\Delta_\cS - 3)\) and hence vanishes.

c)
Finally for \(h_1 = h_2 = h_3 = 0\) there are short representations for
\begin{equation}\label{noname}
\Delta_\cS = 4k\,,\qquad \Delta_\cS = 4k+2\,,\qquad \Delta_\cS = 4k+4 \,.
\end{equation}
Since we have eight distinct supercharges, a descendant operator at
level \(l > 8\) is always zero by means of \eqref{eq:qanticom},
so according to \eqref{eq:invlevel} the only levels at
which we should look for suitable candidate operators are \(l = 4,8\).

At level \(l = 4\) we need $\Delta_\cS=4$ and there is one operator with \(k = 0\),
\begin{equation}\label{eq:O4}
\cO_4 = \epsilon^{\alpha\beta\gamma\delta} \left\{Q_{i\alpha},\left[Q^i_\beta,\left\{Q_{j\gamma},\left[Q^j_\delta,\cS\right]\right\}\right]\right\} \,,
\end{equation}
which has norm \(\left\|\cO_4\right\|^2 \sim \Delta_\cS(\Delta_\cS-2)\). 
It does not vanish for \(\Delta_\cS = 4\), but we find that the norm of \(\left[Q^i_\alpha, \cO_4\right]\) is proportional to \(\Delta_\cS(\Delta_\cS - 2)(\Delta_\cS + 1)\), so \(\left[Q^i_\alpha, \cO_4\right]\) vanishes only if \(\cO_4\) itself vanishes.
This means that \(\cO_4\)  breaks supersymmetry and thus cannot be a
supersymmetric marginal operator. 
Moreover, \(\cO_4\) is also a total derivative. 

The only possibility for non-vanishing \(k\) is \(k=1\) as
\eqref{noname}
implies for $k>1$ that $\Delta_\cS>4$ while for $k=\frac12$ the level
$l$ cannot be even. 
The  operator with $k=1$ reads
\begin{equation}
\cO'_4 = \epsilon^{\alpha\beta\gamma\delta} \left\{Q_{i\alpha},\left[Q^i_\beta,\left\{Q_{j\gamma},\left[Q_{k\delta},\cS^{(jk)}\right]\right\}\right]\right\} \,.
\end{equation}
We can compute \(\left\|\cO'_4\right\|^2 \sim (\Delta_\cS-4)(\Delta_\cS+6)(\Delta_\cS+8)\), and thus this operator is ruled out as well.
Clearly, it is again also a total derivative.

At level \(l=8\) we need $\Delta_\cS=2$. Using the same argument as
above there is no operator with
\(k \neq 0\).
Hence, a Lorentz invariant level \(l=8\) operator is (up to total
derivatives) always a descendant of the $l=2$ operator 
\begin{equation}
\cO^{ij}_{\alpha\beta} = \left\{Q^i_{[\alpha},\left[Q^j_{\beta]},\cS\right]\right\} \,.
\end{equation}
If we antisymmetrize also in the R-symmetry indices \(i\) and \(j\),
we find  \(\bigl\|\cO^{[ij]}_{\alpha\beta}\bigr\|^2 \sim \Delta_\cS \),
but this operator is symmetric under the exchange of the two supercharges
and we end up with a total derivative.
On the other hand we find for the symmetric component that
\(\bigl\|\cO^{(ij)}_{\alpha\beta}\bigr\|^2 \sim \Delta_\cS (\Delta_\cS - 2)\),
so it vanishes at the dimension we are interested in.
Let us show for the sake of completeness that also all the level \(l=8\)
descendants of \(\cO^{[ij]}_{\alpha\beta}\) have vanishing or negative
norm at \(\Delta_\cS=2\).
They are in turn descendants of the \(l=4\) operator
\begin{equation}
\cO^{ij} = \epsilon^{\alpha\beta\gamma\delta} \left\{Q^i_\alpha,\left[Q^j_\beta, \epsilon_{kl}\cO^{kl}_{\gamma\delta}\right]\right\}
= \epsilon^{\alpha\beta\gamma\delta} \left\{Q^i_\alpha,\left[Q^j_\beta,\left\{Q_{k\gamma},\left[Q^k_\delta,\cS\right]\right\}\right]\right\} \,.
\end{equation}
While the antisymmetric part \(\cO^{[ij]}\) of this operator is nothing else
than \(\cO_4\) from \eqref{eq:O4} with norm \(\Delta_\cS(\Delta_\cS-2)\),
the symmetric part \(\cO^{(ij)}\) has norm \(\Delta_\cS(\Delta_\cS-2)(\Delta_\cS-4)\),
and so both operators have vanishing norm for $\Delta_\cS=2$.

To conclude, we have thus shown that all candidates for marginal
operators either have zero norm or are not supersymmetric.
Notice that most of the operators are also total derivatives but we
did not have to use this fact in our argument. 
Let us close with the observation that 
the above analysis can be easily extended to relevant operators with conformal dimension \(\Delta < 6\).
In this case the dimension of the primary operator needs to satisfy
\begin{equation}
\Delta_\cS = \Delta - \frac{l}{2} < 6 - h_1 - h_2 - h_3 - 2 n \,,\qquad n \in \bbN \,,
\end{equation}
which is clearly also not compatible with the general bound \eqref{eq:generalbound}.
Moreover, for generic \(\Delta < 6\) all isolated short representations are ruled out as well.
Only for \(\Delta = 4\) the operators from c) with \(k=0\) remain possible candidate operators,
but we have shown that their norms are negative at the appropriate dimensions.

\chapter{Conclusion and Outlook}\label{chap:conclusions}

In this thesis we investigated maximally supersymmetric solutions of gauged supergravity theories, with a special focus on anti-de Sitter solutions.
Supersymmetric AdS backgrounds are especially relevant in the context of the AdS/CFT correspondence; in particular we studied their moduli spaces which are related to the conformal manifolds of the dual SCFTs.

In chapter~\ref{chap:classification} we gave an exhaustive classification of maximally supersymmetric solutions in gauged or otherwise deformed supergravities in \(3 \leq D \leq 11\) dimensions.
These solutions split in two different classes.
Firstly, if there are no background fluxes, the background space-time has to be maximally symmetric and is therefore either Minkowskian or AdS\(_D\).
While Minkowskian backgrounds exist in both ungauged as well as gauged theories, AdS$_D$ backgrounds require a non-trivial potential and therefore are restricted to gauged or derformed theories.
However, there are certain conditions on the fermionic shift matrices \(A_0\) and \(A_1\) which constrain the possible gaugings or deformations.
The second class of solutions has non-trivial background fluxes.
This implies that the shift matrices \(A_0\) and \(A_1\) as well as the R-symmetry connection \eqref{eq:gaugedQ} have no background value.
Consequently, the fermionic supersymmetry variations take the same form as in the ungauged case and therefore the possible maximally supersymmetric solutions agree with the solutions of the corresponding ungauged theory.
Moreover, solutions with background flux can only exist if there are no spin-1/2 fermions in the gravity multiplet at all or if the theory is chiral.
This restricts such solutions to only a small class of supergravity theories, listed in table~\ref{tab:pformfluxes} together with the allowed fluxes.
From the correspondence with the ungauged theories we infer that for all these theories all maximally supersymmetric solutions are known and classified; they are listed in table~\ref{tab:adsbackgrounds}.

In chapter~\ref{chap:ads} we exclusively focused on AdS$_D$ solutions in gauged supergravities in dimensions \(D \geq 4\).
In this case unbroken supersymmetry imposes algebraic conditions \eqref{eq:adsconditions} on the shift matrices \(A_0\) and \(A_1\) which in turn restrict the admissible gauge groups.
We found that the gauge group -- after a possible spontaneous symmetry breaking -- is always of the form \(H^g_R \times H^g_\mathrm{mat}\), where \(H^g_R\) is unambiguously determined by the conditions on \(A_0\) and \(A_1\).
This resembles the structure of the global symmetry groups of SCFTs, where \(H^g_R\) corresponds to the R-symmetry group and \(H^g_\mathrm{mat}\) to a possible flavor symmetry.
Moreover, the conditions on \(A_0\) and \(A_1\) determine at which points of the scalar field space AdS$_D$ solutions exist.
A continuous family of such points corresponds to a non-trivial moduli space of solutions.
However, some points in the space of solutions can be related by a gauge transformation and therefore are physically equivalent.
The corresponding directions in the scalar field space are Goldstone bosons arrising from the spontaneous breaking of a gauge symmetry and must be modded out of the moduli space.
To obtain explicit results, we focused on theories where the scalar field space is a symmetric homogeneous space \(\cM = G/H\).
We found explicit conditions on potential supersymmetric moduli, in particular they have to be uncharged with respect to the gauged R-symmetry \(H^g_R\).
Consequently, the determination of the admissible gauge groups is an important first step in the discussion of the moduli spaces of AdS solutions.
A particularly simple class of such theories is given by the supergravities with more than 16 real supercharges.
They mostly behave like maximal supergravities and have the gravitational multiplet as their only supermultiplet.
The moduli spaces of the AdS solutions of these theories are symmetric spaces as well and can be determined by purely group theoretical arguments.

Using these results, we discussed the maximally supersymmetric AdS solutions of all gauged supergravities with more than 16 real supercharges in chapter~\ref{chap:adsmoduli}.
Restricting to \(D \geq 4\), they only exist in dimensions \(D = 4, 5\) and \(7\).
We explicitly determined their gauge groups and showed that almost all of them do not allow for non-trivial moduli spaces.
The only exception occurs for maximal supergravity in five dimension where the moduli space is given by \(\SU(1,1)/\U(1)\).
These results are in one-to-one agreement with predictions from the AdS/CFT correspondence.
It has been shown in \cite{Cordova:2016xhm} that the dual SCFTs do not admit for supersymmetric marginal deformations as well and thus do not have conformal manifolds.
Moreover, the \(\SU(1,1)/\U(1)\) moduli space in five dimensions corresponds to the complex gauge coupling of the dual four-dimensional \(\cN = 4\) super Yang-Mills theory.

We also considered two less-supersymmetric examples and studied the AdS solutions of half-maximal supergravity in seven dimensions and of \(\cN = 3\) supergravity in four dimensions.
In the first case, the theory does not only have to be gauged but unlike most other cases also needs to be deformed by a mass parameter.
For both theories we found that there is no moduli space, as for most of the previously discussed examples.
We confirmed the holographic interpretation of the seven-dimensional result and explicitly showed in chapter~\ref{chap:SCFT} that the dual six-dimensional \(\cN = (1,0)\) SCFTs cannot be deformed by supersymmetric marginal operators.
This follows purely from the representation theory of the underlying superconformal algebra \(\mathfrak{osp}(6,2|2)\);
any candidate for a supersymmetric marginal deformation violates the unitarity bounds and is therefore forbidden.

Using our general results and following the examples presented in this thesis it should be possible to find the moduli spaces of maximally supersymmetric AdS solutions for all theories with a symmetric scalar field space \(\cM = G/H\).
In particular not only the theories with more than 16 supercharges but also all supergravities with \(8 < q \leq 16\) supercharges are of this type.
They are characterized by the existence of only two different types of supermultiplets, the gravity multiplet and an arbitrary number of vector multiplets.
Our last examples of half-maximal supergravity in seven dimensions, discussed in chapter~\ref{sec:equal16}, and \(\cN=3\) supergravity in four dimensions, discussed in chapter~\ref{sec:N=3}, are of this type.
The AdS solutions and moduli spaces of other half-maximal theories are determined in \cite{Louis:2014gxa, Louis:2015dca, Karndumri:2016ruc}.
Following a similar reasoning as in the seven-dimensional case it should be straightforward to check the agreement with our findings.
Moreover, also supergravities with \(q \leq 8\) can have symmetric scalar field spaces, even though here more general geometries are allowed.
It would be interesting to apply our methods to these examples and to compare the results with \cite{deAlwis:2013jaa,Louis:2016qca}.

As mentioned in the introduction the AdS/CFT correspondence in its usual formulation involves ten- or eleven-dimensional backgrounds of string theory or M-theory.
In this context the gauged supergravity approach corresponds to a (hopefully consistent) truncation of the full higher-dimensional spectrum.
However, it is not clear if every consistently gauged supergravity possesses a higher-dimensional origin.
The same question therefore arises for the AdS backgrounds studied in this thesis.
Do they all allow for an interpretation as the consistent truncation of a higher-dimensional string theory or M-theory background, and if not what is their role in the vicinity of the AdS/CFT correspondence?
As we found, the existence of a maximally supersymmetric AdS solution constrains the allowed gaugings, we are therefore only dealing with a subclass of all gauged supergravities.
So even if there were some gaugings without higher-dimensional origin, one could still ask the same question restricted to the gaugings leading to maximally supersymmetric AdS solutions.
Of course, for many AdS solutions there exists a straightforward higher-dimensional origin. 
For example, the solutions of maximal supergravity (see chapter~\ref{chap:adsmoduli}) directly correspond to the \(AdS \times S\) solutions of ten- and eleven-dimensional supergravity discussed in chapter~\ref{chap:classification}.
This question is thus most relevant for the less supersymmetric cases with not as strongly restricted gaugings.

On the other hand, even if a higher-dimensional origin exists, one might still wonder to which extend the gauged supergravity approach reproduces the relevant behavior of the full solution, especially whether they share the same moduli spaces. 
A priori, it is conceivable that there exist truncations which do not contain all moduli of the higher-dimensional solution.
Nonetheless, it could still be possible to find at least one other truncation to a gauged supergravity which contains all moduli.%
\footnote{See \cite{Hoxha:2000jf, Louis:2016msm} for an example where this does not seem to be the case.}
Only in this case the constraints on the moduli space derived from the gauged supergravity approach would be  valid also for the higher-dimensional backgrounds.
For string theory backgrounds which do not allow for such a truncation, it could be in principle possible that their moduli spaces do not agree with the form of the moduli spaces allowed in a gauged supergravity. 
It would be very interesting to find a general solution to this problem.

\chapter*{Acknowledgments}

First of all, I am deeply grateful to my supervisor Jan Louis.
I benefited a lot from his advise, numerous discussions and his continuous support.

I am also very thankful to my second supervisor Volker Schomerus as well as to Caren Hagner, Ingo Runkel and Marco Zagermann for agreeing to take part in the disputation committee.

I would like to thank Markus Dierigl and Constantin Muranaka for reading the manuscript of this thesis. I would also like to thank Constantin Muranaka for our countless discussions.

I am grateful to all my colleagues and friends at DESY, the II. Institute for Theoretical Physics and the GRK 1670
for the pleasant and enjoyable time as a PhD student.
In particular, I would like to thank
Constantin Muranaka,
Jonny Frazer,
Mafalda Dias, 
Markus Dierigl
and
Markus Ebert
as well as 
Alessandra Cagnazzo,
Ander Retolaza,
Benedict Broy,
David Ciupke,
Fabian R\"uhle,
Giovanni Rabuffo,
Jakob Moritz,
Jan Hesse,
L\'or\'ant Szegedy,
Lucila Z\'arate,
Marco Scalisi,
Paul Oehlmann,
Peter-Simon Dieterich,
Piotr Pietrulewicz,
Rob Klabbers,
Shruti Patel,
So Young Shim,
Stefan Liebler, 
Stefano Di Vita
and
Yannick Linke
for the fun last three years in Hamburg.

I am particularly thankful to my parents and family for their constant encouragement and support during my studies.

Moreover, I would like to thank the German Science Foundation (DFG) for financial support under the Research Training Group (RTG) 1670 ``Mathematics inspired by String Theory and Quantum Field Theory''.

\begin{appendix}

\chapter{Conventions and Notations}\label{app:conventions}

In this appendix we summarize the conventions and notations used in this thesis.
We mostly follow the sign and spinor conventions of \cite{Freedman:2012zz}. 

\subsection*{Metric}
The space-time metric is mostly positive, i.e.
\(
\eta_{MN} = \mathrm{diag}(-, +, \dots, +)
\).

\subsection*{Indices}
In our description of supergravities we use the following indices:
\vspace{0.3em}
\begin{itemize}
\setlength{\itemsep}{0.3em}
\item space-time: \(M, N,  \ldots \in \{0,1, \dots, D-1\} \)
\item gravitini (R-symmetry): \(i,j, \ldots \in \{1,\dots, \cN\} \)
\item spin-1/2 fermions: \(a,b,\ldots\)
\item scalars: \(r, s, \ldots\)
\item gauge fields: \(I, J, \ldots\)
\item $p$-form field strengths: \(I_p, J_p, \ldots\)
\item dressed $p$-form field strengths: \(\al{p}, \be{p}, \ldots\)
\end{itemize}
\vspace{0.3em}
Moreover, we use a hat or a tilde over an index to indicate whether a field belongs to the gravity multiplet or any other multiplet, i.e.~the fermions \(\chi^{\hat a}\) belong to the gravity multiplet and \(\chi^{\tilde a}\) to matter multiplets.

%

\subsection*{$\Gamma$-matrices}

The \(D\)-dimensional gamma matrices \(\Gamma^M\) span a Clifford algebra and are defined via their anti-commutation relation
\begin{equation}\label{eq:gammadef}
\Gamma^M \Gamma^N + \Gamma^N \Gamma^M = 2 g^{MN} \id \,.
\end{equation}
In the main text their antisymmetric products 
appear frequently and we abbreviate
\begin{equation}\label{eq:gammaprod}
\Gamma^{M_1\dots M_p} = \Gamma^{[M_1} \dots \Gamma^{M_p]} \,,
\end{equation}
where the antisymmetrization \([\dots]\) is with total weight 1, i.e. \(\Gamma^{MN} = \frac{1}{2}\left(\Gamma^M\Gamma^N - \Gamma^N \Gamma^M\right)\).
In even dimensions \(D = 2m\) we additionally have the chirality operator 
\(\Gamma_\ast\) defined by 
\begin{equation}\label{eq:gamma5}
\Gamma_\ast = (-i)^{m+1} \Gamma_0 \Gamma_1 \dots \Gamma_{D-1} \,,
\end{equation}
which allows to define projection operators 
\begin{equation}\label{eq:Ppm}
P_{\pm} = \tfrac{1}{2} (\id \pm \Gamma_{\ast}) \,.
\end{equation}
From the definition of \(\Gamma_\ast\) one infers the relations \cite{Freedman:2012zz}
\begin{equation}\label{eq:gammahodgeeven}
\Gamma^{M_1 \dots M_p}\, \Gamma_\ast = -(-i)^{m+1} \frac{1}{(D-p)!}\, {\epsilon^{M_p \dots M_1}}_{N_1 \dots N_{D-p}}\, \Gamma^{N_1 \dots N_{D-p}} \,,
\end{equation}
while in odd dimensions \(D = 2m +1\) one instead has
\begin{equation}\label{eq:gammahodgeodd}
\Gamma^{M_1 \dots M_p} = i^{m+1} \frac{1}{(D-p)!} {\epsilon^{M_1 \dots M_p}}_{N_{D-p} \dots N_1} \Gamma^{N_1 \dots N_{D-p}} \,.
\end{equation}
In even dimensions all anti-symmetric products $\Gamma^{M_1\dots M_p}$
are linearly independent whereas in odd dimensions this only holds for \(p \leq m\) due to \eqref{eq:gammahodgeodd}. 
We denote the contraction with $\Gamma$-matrices by a dot ``\(\cdot\)", i.e. for a \(p\)-form~\(F\) we define
\begin{equation}
F \cdot \Gamma = F_{M_1 \dots M_p}\, \Gamma^{M_1 \dots M_p} \,.
\end{equation}
Moreover, one introduces the charge conjugation matrix \(C\) which is defined by the properties
\begin{equation}\label{eq:C}
C^T = - t_{0} C \,,\qquad (\Gamma^M)^T = t_0 t_1 C \Gamma^M C^{-1} \,,
\end{equation}
where \(t_0\) and \(t_1\) are sign factors collected in Table~\ref{tab:spinors}.

\subsection*{Spinors}

For a set of complex spinors \(\epsilon^i\) transforming as a vector in the fundamental representation of the R-symmetry group \(H_R\) we denote the (Dirac) conjugates by \(\bar\epsilon_i\) with a lowered index, i.e.
\begin{equation}
\bar\epsilon_i \equiv (\epsilon^i)^\dagger i \Gamma^0 \,.
\end{equation}
It is convenient to introduce the spinor \(\epsilon_i\) with lowered index as the charge conjugate of \(\epsilon^i\), i.e. \(\epsilon_i = (\epsilon^i)^C\), defined by the relation
\begin{equation}\label{eq:chargeconjugate}
\bar\epsilon_i = (-t_0 t_1) \epsilon_i^T C \,,
\end{equation}
where \(\epsilon_i^T C\) is called the Majorana conjugate of \(\epsilon_i\).
With this notation bilinears of spinors \(\epsilon^i\) and \(\eta^j\) satisfy \cite{Freedman:2012zz}
\begin{equation}\label{eq:bilinearswap}
\bar\epsilon_i \Gamma^{M_1\dots M_p} \eta^j = t_p\, \bar\eta^j \Gamma^{M_1\dots M_p} \epsilon_i \,,
\end{equation}
where \(t_2 = -t_0\), \(t_3 = -t_1\), \(t_{p+4} = t_p\) and \(\bar\eta^j \equiv (-t_1) \eta_i^\dagger i \Gamma^0 = (-t_0 t_1) (\eta^i)^T C\).
This relation is particularly useful if there is a relation between \(\epsilon^i\) and \(\epsilon_i\), i.e.~if the spinors satisfy a (symplectic) Majorana condition.

Applying charge conjugation twice yields \(\bigl((\epsilon^i)^C\bigr){}^C = (- t_1) \epsilon^i\) and according to the sign of \((- t_1)\) we can introduce Majorana or symplectic Majorana spinors.
If \(t_1 = - 1\), the charge conjugation is a strict involution and it is consistent to impose the reality constraint
\begin{equation}\label{eq:majorana}
\epsilon^i = \delta^{ij} \epsilon_j \,,
\end{equation}
with \(\delta^{ij}\) the identity matrix.
A spinor satisfying \eqref{eq:majorana} is called a Majorana spinor and has half as many real degrees of freedom compared with an unconstrained spinor.

If \(t_1 = + 1\), the above Majorana condition would be inconsistent but we can instead impose the symplectic Majorana condition,
\begin{equation}\label{eq:symplmajorana}
\epsilon^i = \Omega^{ij} \epsilon_j \,,
\end{equation}
where \(\Omega^{ij} = (\Omega_{ij})^\ast\) is a non-degenerate antisymmetric matrix satisfying \(\Omega_{ik}\Omega^{jk} = \delta_i^j\).
Note that this condition is only consistent for an even number of spinors \(\epsilon^i\) because otherwise a matrix \(\Omega_{ij}\) with the required properties does not exist.

If not denoted otherwise we always assume spinors to fulfill the (symplectic) Majorana conditions \eqref{eq:majorana} or \eqref{eq:symplmajorana}, respectively.
The benefit of this choice is that it gives spinor bilinears well-defined reality properties.
For example, symplectic Majorana spinors \(\epsilon^i\) and \(\eta^i\) satisfy
\begin{equation}\label{eq:reality}
\left(\bar \epsilon_i \Gamma^{M_1\dots M_p} \eta^j\right)^\ast = (-t_0 t_1)^{(p+1)} \Omega^{ik} \Omega_{jl} \bar \epsilon_k \Gamma^{M_1\dots M_p} \eta^l \,.
\end{equation}
By replacing \(\Omega_{ij}\) with \(\delta_{ij}\) one obtains the analogous relation for Majorana spinors.
This allows us to easily construct real Lagrangians.
We illustrate this with the example of the gravitino mass term \eqref{eq:appfermionmass}.
Up to a prefactor it is given by
\begin{equation}\label{eq:majoranagravmass}
{(A_0)^i}_j \bar\psi_{iM} \Gamma^{MN} \psi^j_N \,,
\end{equation}
and is real if
\begin{equation}
\left({(A_0)^i}_j \right)^\ast = (-t_0 t_1) {(A_0)_i}^j = (-t_0 t_1) \Omega^{ik} \Omega_{jl} {(A_0)^l}_k \,.
\end{equation}
Consequently, we assume all objects with indices \(i,j,\dots\) to be pseudo real or pseudo imaginary, which means that indices can be raised or lowered by complex conjugation (up to a sign factor).

However, using (symplectic) Majorana conditions can sometimes obscure the action of the R-symmetry, especially in even dimensions where we furthermore can distinguish between left- and right-handed spinors.

If \(D\) is odd (symplectic) Majorana spinors are the only minimal spinor representations.
Note that the Majorana condition \eqref{eq:majorana} is invariant under \(H_R = \SO(\cN)\) transformations, where \(\cN\) denotes the number of spinors \(\epsilon^i\).
The symplectic Majorana condition \eqref{eq:symplmajorana}, on the other hand, is invariant under \(H_R = \USp(\cN)\).

In even dimensions \(D\) the situation is slightly more complicated since here the projectors \(P_{\pm}\) \eqref{eq:Ppm} can be used to define chiral or Weyl spinors.
We need to distinguish between two different cases.
Let us first consider the situation where \((\Gamma_\ast \epsilon^i)^C = - \Gamma_\ast (\epsilon^i)^C\) (as well as \(t_1 = - 1\)), which implies that the charge conjugate of a left-handed spinor is right-handed and vice versa.
Therefore, a Majorana spinor cannot have a definite chirality.
Nonetheless, we can decompose \(\epsilon^i\) into its left and right handed component, i.e.
\begin{equation}
\epsilon^i = \epsilon^i_+ + \epsilon_{-i} \,,
\end{equation}
with 
\begin{equation}
\epsilon^i_+ \equiv P_+ \epsilon^i \,,\qquad \epsilon_{-i} \equiv P_- \epsilon^i \,.
\end{equation}
Note that \((\epsilon^i_+)^C = \epsilon_{-i}\), i.e.~the positioning of the indices is consistent with \eqref{eq:chargeconjugate}.
On the other hand this also implies that \(\epsilon^i_+\) and \(\epsilon_{-i}\) do not satisfy the Majorana condition \eqref{eq:majorana} individually.
Consequently, we loose the ability to raise and lower indices with \(\delta_{ij}\).
Moreover, Weyl spinors \(\epsilon^i_+\) do not satisfy the reality property \eqref{eq:reality} anymore.
However, we can still write down a relation similar to the Majorana condition \eqref{eq:majorana} if we replace \(\epsilon^i\) by a column vector \(\epsilon^I\) consisting of \(\epsilon^i_+\) and \(\epsilon_{-i}\), i.e.
\begin{equation}
\epsilon^i \rightarrow \epsilon^I \equiv \begin{pmatrix} \epsilon^i_+ \\ \epsilon_{-i} \end{pmatrix} \,,\qquad \text{and} \qquad \epsilon_i \rightarrow \epsilon_I \equiv (\epsilon^I)^\cc = \begin{pmatrix} \epsilon_{-i} \\ \epsilon^i_+ \end{pmatrix} \,.
\end{equation}
With this notation we have
\begin{equation}
\epsilon^I = \Delta^{IJ} \epsilon_J \,,\qquad\text{where}\qquad \Delta^{IJ} = \begin{pmatrix} 0 & \delta^j_i \\
\delta^i_j & 0 \end{pmatrix} \,,
\end{equation}
which formally resembles \eqref{eq:majorana} or \eqref{eq:symplmajorana}.
The formal replacement of \(\epsilon^i\) by \(\epsilon^I\) (and analogously for all other involved spinors) enables us to convert our general formulae (collected in appendix~\ref{app:susy}) from (symplectic) Majorana spinors to Weyl spinors.

Let us illustrate this with the gravitino mass term.
Using chiral spinors \(\psi^i_{+M}\) and \(\psi_{M-i}\) it reads
\begin{equation}\label{eq:weylgravmass}
(A_0)_{ij} \, \bar\psi^i_{+M} \Gamma^{MN} \psi^j_{+N} + \mathrm{h.c.} =  (A_0)_{ij} \, \bar\psi^i_{M+} \Gamma^{MN} \psi^j_{N+} + (A_0)^{ij} \, \bar\psi_{M-i} \Gamma^{MN} \psi_{N-j} \,,
\end{equation}
where \((A_0)^{ij} = \left((A_0)_{ij}\right)^\ast\).
Note that we stick to our convention that raising and lowering indices is related to complex conjugation.
\eqref{eq:weylgravmass} can be cast into a form equivalent to \eqref{eq:majoranagravmass} by combining \(\psi^i_{M+}\) and \(\psi_{M-i}\) into a column vector, i.e.~\(\psi^I_M = \left(\psi^i_{M+}, \psi_{M-i}\right)^T\) and by introducing
\begin{equation}
 A^I_{0\,J} = \begin{pmatrix} 0 & (A_0)^{ij} \\ (A_0)_{ij} & 0 \end{pmatrix} \,.
\end{equation}
With this notation \eqref{eq:weylgravmass} reads
\begin{equation}
{(A_0)^J}_J \bar\psi_{IM} \Gamma^{MN} \psi^J_N \,,
\end{equation}
which is (after the replacements \(\psi^i_M \rightarrow \psi^I_M\) and \(A^i_{0\,j} \rightarrow A^I_{0\,J}\)) of the same form as \eqref{eq:majoranagravmass}.

We finally want to mention that the Weyl condition \(\epsilon^i_+ = P_+ \epsilon^i_+\) is invariant with respect to \(H_R = (\mathrm{S})\U(\cN)\).

Now we turn to the second case where \((\Gamma_\ast \epsilon^i)^C = \Gamma_\ast (\epsilon^i)^C\). Here one can consistently define (symplectic) Majorana-Weyl spinors.
This means we can have two independent sets of spinors \(\epsilon^i_+\) and \(\epsilon^{i'}_{+}\),
\begin{equation}
P_{\pm} \epsilon^i_\pm = \epsilon^{i'}_\pm \,,
\end{equation}
which individually satisfy \eqref{eq:majorana} or \eqref{eq:symplmajorana}, respectively.
Analogously to the odd-dimensional case we find \(H_R = \SO(\cN_+) \times \SO(\cN_-)\) or \(H_R = \USp(\cN_+) \times \USp(\cN_-)\), where \(\cN_+\) denotes the number of chiral spinors \(\epsilon^i_+\) and \(\cN_-\) the number of anti-chiral spinors \(\epsilon^{i'}_-\).%
\footnote{The notation \(\cN = (\cN_+, \cN_-)\) is also common.}
In this case a sum over the index \(i\) in a general formula is implicitly understood to run over \(i'\) as well, unless stated otherwise.%
\footnote{This prescription can be formalized by replacing \(\epsilon^i\) with \(\epsilon^I = (\epsilon^i_+, \epsilon^{i'}_+)^T\), similarly as in our previous discussion.}
We summarize the irreducible spinor representations together with the compatible R-symmetry groups \(H_R\) for various dimensions in table~\ref{tab:spinors}.

\begin{table}[htb]
\centering
\begin{tabular}{|c|cc|c|c|}
\hline
$D$  (mod 8) & $t_0$ & $t_1$ & irrep. & $H_R$ \\
\hline
3 & $+$ & $-$ & M & $\SO(\cN)$ \\
4 & $+$ & $-$ & M / W & $(\mathrm{S})\U(\cN)$ \\
5 & $+$ & $+$ & S & $\USp(\cN)$ \\
6 & $-$ & $+$ & SW & $\USp(\cN_+) \times \USp(\cN_-)$ \\
7 & $-$ & $+$ & S & $\USp(\cN)$ \\
8 & $-$ & $-$ & M / W & $(\mathrm{S})\U(\cN)$ \\
9 & $-$ & $-$ & M & $\SO(\cN)$ \\
10 & $+$ & $-$ & MW & $\SO(\cN_+) \times \SO(\cN_-)$ \\
\hline
\end{tabular}
\caption{Spinor conventions in various dimensions \cite{Freedman:2012zz}. \(t_0\) and \(t_1\) are the sign factors introduced in \eqref{eq:C}. ``M'' stands for Majorana spinors, ``S'' for symplectic Majorana spinors and ``W'' for Weyl spinors.
In four and eight dimensions one can have either Majorana or Weyl spinors (but not both), while in six and ten dimensions (symplectic) Majorana-Weyl spinors are possible.}
\label{tab:spinors}
\end{table}

\chapter{Supersymmetry Variations}\label{app:susy}

In this appendix we summarize the general form of the supergravity Lagrangian and supersymmetry variations and derive some important relations between the fermionic shift matrices and the Killing vectors and moment maps.

In appendix~\ref{app:susyvariations} we summarize the supersymmetry variations of the fermions and bosons in a general supergravity theory and comment on some of the properties of the involved objects.
In appendix~\ref{app:lagrangian} we review the general form of a supergravity Lagrangian.
In appendix~\ref{app:susycalculations} we compute the Killing vectors \(\cP_I\) and their moment maps \(\cQ^R_I\) in terms of the shift matrices \(A_0\) and \(A_1\).
In appendix~\ref{app:sugras} we give explicit expressions for some of the previously introduced objects in dimensions \(D = 4,5,7\).

\section{Supersymmetry variations}\label{app:susyvariations}

In this appendix we collectively present the general form of the supersymmetry variations of the fields present in a (gauged) supergravity theory.
These expressions are universal and not restricted to a specific dimension or number of supercharges.
Moreover, we assume all spinors to satisfy to be (symplectic) Majorana.
See appendix~\ref{app:conventions} for our spinor conventions and for the conversion from Majorana to chiral spinors.

The supersymmetry variations of the bosonic fields read
\begin{subequations}\begin{align}
\delta e^A_M &= \tfrac12 \bar\epsilon_i \Gamma^A \psi^i_M \,,  \label{eq:vielbeinvariation} \\
\delta A^{I_p}_{N_1\dots N_{p-1}} &= \tfrac{p!}{2} \cV^{I_p}_{\alpha_p} 
\left[\bigl(B^{\alpha_p}\bigr)^i_j \bar \psi_{i[N_1} \Gamma_{N_2 \dots N_{p-1}]} \epsilon^j + \bigl(C^\al{p}\bigr)^a_i \bar\chi_a \Gamma_{N_1 \cdots N_{p-1}} \epsilon^i\right] + \dots \,, \label{eq:Avariation}
\end{align}\end{subequations}
where we have omitted possible terms that depend on the other $p$-form fields and their supersymmetry variations.
The supersymmetry variations of the fermionic fields up to terms of higher order in the fermionic fields are given by
\begin{subequations}\begin{align}
\delta \psi^i_M &= \hat\cD_M  \epsilon^i  + \left(\cF_M\right)^i_j \epsilon^j + A^i_{0\,j} \Gamma_M \epsilon^j + \dots \,, \label{eq:appgravitinovariation} \\
\delta \chi^a &= \cF^a_i \epsilon^i + A^a_{1\,i} \epsilon^i + \dots \,, \label{eq:appspin12variation}
\end{align}\end{subequations}
where \(\hat\cD\) is the covariant derivative introduced in \eqref{eq:gaugedQ}.
The shift matrices \(A_0\) and \(A_1\) generically depend on the scalar fields.
Moreover, we have defined the abbreviations
\begin{equation}\label{eq:appFM}
\big(\cF_M\big)^i_j  =  \tfrac{1}{2(D-2)} \sum_{p \geq 2} 
\big(B^{(p)}_{ \hat\alpha_p}\big)^i_j\,
F^{\hat\alpha_p}_{{N_1}\dots {N_p}} {T_{(p)}^{{N_1}\dots {N_p}}}{}_M \,,
\end{equation}
with
\begin{equation}\label{eq:T}
{T_{(p)}^{{N_1}\dots {N_p}}}{}_M = {\Gamma^{{N_1}\dots {N_p}}}{}_M + p\tfrac{D-p-1}{p-1} \Gamma^{[N_1\dots N_{p-1}} \delta^{N_p]}_M \,,
\end{equation}
as well as
\begin{equation}
\cF^{a}_i = \tfrac12 \sum_{p \geq 1} \big(C^{(p)}_{ \hat\alpha_p}\big)^{a}_i \, F^{\alpha_p}_{N_1\dots N_p} \Gamma^{N_1\dots N_p} \epsilon^i \,.
\end{equation}
The matrices \(B_\al{p}\) and \(C_\al{p}\) are constant and mediate between the different representations of \(H\) that occur in the theory.
To be more specific, we denote the generators of \(H\) in the respective representations by \({(J_A)_i}^j\), \({(J_A)_a}^b\) and \({(J_A)_\al{p}}^\be{p}\) and demand
\begin{equation}\begin{aligned}\label{eq:appBCproperty}
{\bigl(J_A\bigr)_\al{p}}^\be{p} B_\be{p} &= \bigl[J_A, B_\al{p}\bigr]  \,, \\
{\bigl(J_A\bigr)_\al{p}}^\be{p} \bigl(C_\be{p}\bigr)_i^a &= \bigl(J_A\bigr)_i^j \bigl(C_\al{p}\bigr)^a_j -  \bigl(C_\al{p}\bigr)^b_i {\bigl(J_A\bigr)_b}^a \,.
\end{aligned}\end{equation}
To keep the notation compact we defined
\begin{equation}
B_{\tilde \alpha_p} = B_{\alpha_1} = 0 \,.
\end{equation}
The closure of the supersymmetry algebra imposes a Clifford algebra like condition on \(B_\al{p}\) and \(C_\al{p}\),
\begin{equation}\label{eq:BCanticom}
\frac{(p!)^2}{D-2}\frac{D-p-1}{p-1} \left(B^\dagger_{\alpha_p} B_{\beta_p} + B^\dagger_{\beta_p} B_{\alpha_p}\right) +  (p!)^2 \left(C^\dagger_{\alpha_p} C_{\beta_p} + C^\dagger_{\beta_p} C_{\alpha_p}\right)  = 2 \delta_{\alpha_p \beta_p} \id \,.
\end{equation}

\section{The general Lagrangian}\label{app:lagrangian}

In this appendix we state the general Lagrangian of a (gauged) supergravity theory at the two derivative level.

The Lagrangian can be split into a purely bosonic part and a part that also depends on the fermionic fields, i.e.
\begin{equation}
\cL = \cL_B + \cL_F \,,
\end{equation}
The bosonic Lagrangian is already given in \eqref{eq:bosonicaction}, we restate it here for the sake of completeness
\begin{equation}\begin{aligned}\label{eq:appbosoniclagrangian}
e^{-1} \cL_B &= -\frac{R}{2} 
- \frac{1}{2} \sum_{p \geq 1} M^{(p)}_{I_{p} J_{p}}\!\left(\phi\right)\, F^{I_p} \wedge \ast F^{J_p} - V + e^{-1} \cL_\mathrm{top} \\
&= -\frac{R}{2} 
- \frac{1}{2} \sum_{p \geq 1} \delta_{\al{p}\be{p}} \, F^\al{p} \wedge \ast F^\be{p}  - V+ e^{-1} \cL_\mathrm{top} \,.
\end{aligned}\end{equation}
Note that we often denote the dressed scalar field strengths \(\hat\cP^\al{1} \equiv F^\al{1}\). 
The scalar potential reads \eqref{eq:generalpotential}
\begin{equation}\label{eq:appgeneralpotential}
V = - \tfrac{2(D-1)(D-2)}{\cN} \tr(A_0^\dagger A_0) + \tfrac{2}{\cN}\tr(A_1^\dagger A_1) \,,
\end{equation}
where \(A_0\) and \(A_1\) are the fermionic shift matrices from \eqref{eq:appgravitinovariation} and \eqref{eq:appspin12variation}.
Moreover, there can be a topological term \(\cL_\mathrm{top}\) which does not depend on the space-time metric.

The fermionic Lagrangian (which despite its name in general also depends on the bosonic fields) is of the general form
\begin{equation}
\cL_F = \cL_\mathrm{kin,f} + \cL_\mathrm{pauli} + \cL_\mathrm{mass} + \cO(f^4) \,.
\end{equation}
The kinetic terms of the fermions read
\begin{equation}\label{eq:appkinferm}
e^{-1} \cL_\mathrm{kin,f} = - \frac{1}{2} \bar \psi_{iM} \Gamma^{MNP} \hat\cD_N \psi^i_P - \frac{1}{2} \bar \chi_{a} \Gamma^M \hat\cD_M \chi^a \,,
\end{equation}
where \(\hat\cD\) denotes the gauge covariant derivative given in \eqref{eq:gaugedQ} and \eqref{eq:chigaugedcovderiv}.
Local supersymmetry requires the existence of Pauli-like interaction terms between the $p$-form field strengths \(F^\al{2}\) and the fermions.
They are of the form
\begin{equation}\label{eq:pauli}
\cL_\mathrm{pauli} = \sum_{p \geq 1} \left(\cL^{(p)}_{F\bar\psi\psi} + \cL^{(p)}_{F\bar\chi\psi} + \cL^{(p)}_{F\bar\chi\chi} \right) \,,
\end{equation}
where
\begin{subequations}\begin{align}
e^{-1}\cL^{(p)}_{F\bar\psi\psi} &= - \frac{1}{4(p-1)} F^{\alpha_p}_{M_1 \dots M_p}   \left(B_{\alpha_p}\right)^i_j \bar\psi^N_i \Gamma_{[N} \Gamma^{M_1 \dots M_p} \Gamma_{P]} \psi^{P\,j} \,, \label{eq:apppaulia} \\
e^{-1}\cL^{(p)}_{F\bar\chi\psi} &= \frac12 F^{\alpha_p}_{M_1 \dots M_p} \left(C_{\alpha_p}\right)^a_i \bar\chi_a \Gamma^N \Gamma^{M_1\dots M_p} \psi^i_N \,, \label{eq:apppaulib} \\
e^{-1} \cL^{(p)}_{F\bar\chi\chi} &= \frac12 F^{\alpha_p}_{M_1 \dots M_p} \left(D_{\alpha_p}\right)^a_b \bar\chi_a \Gamma^{M_1\dots M_p} \chi^b \,.
\end{align}\end{subequations}
\(B_\al{p}\) and \(C_\al{p}\) are the same matrices as in the supersymmetry variations \eqref{eq:appgravitinovariation} and \eqref{eq:appspin12variation}.
The matrices \(D_\al{p}\) have similar properties. Their precise form, however, is not relevant for our discussion.
If the theory is gauged (or otherwise deformed) the Langrangian also includes mass terms for the fermions which read
\begin{equation}\label{eq:appfermionmass}
e^{-1} \cL_\mathrm{mass} = \frac{D-2}{2} A^i_{0\,j} \bar\psi_{iM} \Gamma^{MN} \psi^j_N + A^a_{1\,i} \bar\chi_a \Gamma^M \psi^i_M + M^a_b \bar\chi_a \chi^b \,,
\end{equation}
where \(A_0\) and \(A_1\) are the same matrices as in \eqref{eq:appgravitinovariation} and \eqref{eq:appspin12variation}.
The third mass matrix \(M^a_b\) also depends on the scalar fields and the gaugings/deformations, but it is not relevant for our discussion.
Moreover, the supersymmetric completion of the Lagrangian requires terms of higher order in the fermions which we do not give here.

\section{Killing vectors and moment maps}\label{app:susycalculations}

In a supergravity theory the variation of the vielbein \(e^A_M\) of the space-time metric \eqref{eq:vielbeinvariation} induces additional terms in the variation of the sigma model kinetic term in \eqref{eq:appbosoniclagrangian} which are not present in global supersymmetry.
They read
\begin{equation}\label{eq:deltaPP}
\delta \cL_{\hat\cP\hat\cP} = -\frac{e}{2} \delta_{\alpha_1\beta_1} \hat\cP_M^{\alpha_1} \hat\cP_N^{\beta_1} \bar\epsilon_i \left(\tfrac12 g^{MN} \Gamma^P - \Gamma^{(M} g^{N)P} \right) \psi^i_P+ \dots \,.
\end{equation}
These terms are canceled by the Pauli term \eqref{eq:apppaulib} for \(p = 1\).
Indeed, plugging the variation \eqref{eq:appspin12variation} of the spin-$\frac12$ fermions \(\chi^a\) into \eqref{eq:apppaulib} yields
\begin{equation}\label{eq:deltaPchipsi}
e^{-1} \delta\cL^{(1)}_{\hat \cP\bar\chi\psi} = - \frac{1}{2} \hat\cP^{\alpha_1}_M \hat\cP_N^{\beta_1} \bigl(C_{\alpha_1}\bigr)^a_i \bigl(C^\dagger_{\beta_1}\bigr)_a^j \bar \epsilon_j \left(\tfrac12\Gamma^{MNP} - \tfrac12g^{MN} \Gamma^P + \Gamma^{(M} g^{N)P}\right) \psi^i_P  + \dots \,,
\end{equation}
which cancels \eqref{eq:deltaPP} due to \eqref{eq:BCanticom}.
Only the term cubic in the \(\Gamma\)-matrices does not have a counterpart in \eqref{eq:deltaPP}.
This term, however, is canceled by the kinetic term of the gravitini in \eqref{eq:appkinferm}.
From the gravitino variation \eqref{eq:appgravitinovariation} and \eqref{eq:Dcommgauged} we see that its variation contains
\begin{equation}\begin{aligned}\label{eq:psikinvariation}
e^{-1} \delta \cL_{\bar\psi\hat\cD\psi} &= \frac{1}{2}\bigl(\hat\cH^R_{MN}\bigr)^i_j \bar \psi_{iP} \Gamma^{MNP} \epsilon^j +\ldots \,, \\
&= \frac{1}{2}\bigl(\cH^R_{MN} + F^\al{2}_{MN} Q^R_\al{2}\bigr)^i_j \bar \psi_{iP} \Gamma^{MNP} \epsilon^j +\ldots \,,
\end{aligned}\end{equation}
where \(\cH^R_{MN}\) is the field strength of the R-connection \(\cQ^R_M\) \eqref{eq:Dcomm} and \(\cQ^R_\al{2}\) are the generalized moment maps defined in \eqref{eq:genmomentmap} and \eqref{eq:Qderiv}.
Comparing \eqref{eq:psikinvariation} with \eqref{eq:deltaPchipsi} requires
\begin{equation}
\cH^R_{MN} = -\tfrac12 C^\dagger_{\alpha_1} C_{\beta_1} \hat\cP^{\alpha_1}_{[M} \hat\cP^{\beta_1}_{N]} \,.
\end{equation}
However, in a gauged theory we still need to take care of the second term in \eqref{eq:psikinvariation} which contains the 2-form field strengths \(F^\al{2}_{MN}\).
For this purpose we vary the \(p=2\) Pauli terms \eqref{eq:apppaulia} and \eqref{eq:apppaulib} as well as the fermionic mass terms \eqref{eq:appfermionmass}.
The relevant terms in their variations are given by
\begin{equation}\begin{aligned}
e^{-1}\delta \cL_{\bar\psi\psi} &= \tfrac12 F^{\hat\alpha_2}_{MN} A^i_{0\,k}
\left(B_{\hat\alpha_2}\right)^k_j \bar\psi^P_i \left(-(D-3)
   {\Gamma^{MN}}_P + 2 \delta_P^{[M}\Gamma^{N]}\right) \epsilon^j
+\ldots \,, \\
e^{-1} \delta \cL^{(2)}_{F\bar\psi\psi} &= \tfrac12 F^{\hat\alpha_2}_{MN}  \left(B_{\hat\alpha_2}\right)^i_k A^k_{0\,j} \bar\psi^P_i \left( -(D-3) {\Gamma^{MN}}_P - 2 \delta_P^{[M}\Gamma^{N]}\right) \epsilon^j +\ldots\,, \\
e^{-1} \delta \cL_{\bar\chi\psi} &= \tfrac12 F^\al{2}_{MN} \bigl(A^\dagger_1\bigr)^i_a \left(C_\al{2}\right)^a_j  \bar\psi_j^P \left(- {\Gamma^{MN}}_P - 2 \delta_P^{[M}\Gamma^{N]}\right) \epsilon^j + \dots \,, \\
e^{-1} \delta \cL^{(2)}_{F\bar\chi\psi} &= \tfrac12 F^\al{2}_{MN}  \bigl(C^\dagger_\al{1} \bigr)^i_a A^a_{1\,j} \bar\psi_j^P \left({\Gamma^{MN}}_P - 2 \delta_P^{[M}\Gamma^{N]}\right) \epsilon^j + \dots \,.
\end{aligned}\end{equation}
The terms cubic in the \(\Gamma\)-matrices have to cancel \eqref{eq:psikinvariation} so we determine that the moment maps \(\cQ^R_\al{2}\) are given by
\begin{equation}\label{eq:QRA0A1}
\cQ^R_{\alpha_2} = (D-3) \bigl\{A_0, B_{\alpha_2}\bigr\} + \bigl(A^\dagger_1 C_{\alpha_2} - C^\dagger_{\alpha_2} A_1 \bigr) \,. \\
\end{equation}
Let us also derive a similar condition on the Killing vectors \(\cP_\al{2}\) \eqref{eq:gaugedP}.
Similar relations have first been obtained for \(D=4\) in \cite{DAuria:2001rlt}.
For this purpose we compute another term in the supersymmetry variation of the kinetic term of the scalar fields.
The gauged \(\hat\cP^\al{1}\) depend on the gauge fields \(A_M^\al{2}\) via \(\hat\cP^\al{1}_M = \cP_M^\al{1} + \cP^\al{1}_\al{2} A_M^\al{2}\),
therefore plugging the variation of \(A_M^\al{2}\) \eqref{eq:Avariation} into \eqref{eq:appbosoniclagrangian} gives
\begin{equation}\label{eq:scalarkinvar}
e^{-1} \delta \cL_{\hat \cP \hat \cP} = \delta_{\al{1}\be{1}}\hat \cP^\al{1}_M \cP^\be{1}_\al{2} \bigl(B^\al{2}\bigr)^i_j \bar\psi^M_i \epsilon^j + \dots \,.
\end{equation}
Similar to the above analysis we compute the relevant terms in the variations of the \(p=1\) Pauli terms \eqref{eq:apppaulib} and of the fermionic mass terms \eqref{eq:appfermionmass}.
The result reads
\begin{equation}\begin{aligned}\label{eq:appfermpaulip1var}
e^{-1} \delta \cL_{\bar\chi_\psi} &= \tfrac12 \hat \cP^\al{1}_M \bigl(A^\dagger_1\bigr)^i_a \left(C_\al{1}\right)^a_j \bar\psi^N_i \left({\Gamma^M}_N - \delta^M_N \right) \epsilon^j + \dots \,, \\
e^{-1} \delta \cL^{(1)}_{\hat \cP \bar\chi_\psi} &= \tfrac12 \hat \cP^\al{1}_M \bigl(C^\dagger_\al{1} \bigr)^i_a A^a_{1\,j}  \bar\psi^N_i \left({\Gamma^M}_N + \delta^M_N \right) \epsilon^j + \dots \,.
\end{aligned}\end{equation}
Comparing this with \eqref{eq:scalarkinvar} yields
\begin{equation}\label{eq:PA1}
\delta_{\al{1}\be{1}}\cP^\be{1}_\al{2} B^\al{2} = \tfrac12 \bigl(A^\dagger_1 C_{\alpha_1} - C^\dagger_{\alpha_1} A_1 \bigr) \,.
\end{equation}
We finally want to cancel also the terms quadratic in \(\Gamma\) in \eqref{eq:appfermpaulip1var}.
This gives rise to a gradient flow equation for \(A_0\) \cite{DAuria:2001rlt}.
Inserting \eqref{eq:appgravitinovariation} into the kinetic term of the gravitini \eqref{eq:appkinferm} gives
\begin{equation}\begin{aligned}
e^{-1} \delta \cL_{\bar\psi\hat\cD\psi} &= - (D-2)\bigl(\cD_M A_0\bigr)^i_j \bar\psi_{iN} \Gamma^{MN} \epsilon^j + \dots \\
&= - (D-2) \cP^\al{1}_M \bigl(\cD_\al{1} A_0\bigr)^i_j \bar\psi_{iN} \Gamma^{MN} \epsilon^j + \dots \,,
\end{aligned}\end{equation}
The comparison with 
\eqref{eq:appfermpaulip1var} yields
\begin{equation}\label{eq:appgradientflow}
\cD_{\al{1}} A_0 = \tfrac{1}{2(D-2)} \bigl(A^\dagger_1 C_{\alpha_1} + C^\dagger_{\alpha_1} A_1 \bigr) \,.
\end{equation}
Note that it is possible to derive a similar relation expressing \(\cD_\al{1} A_1\) in terms of \(A_0\) and the third fermion mass matrix \(M^a_b\) \cite{DAuria:2001rlt}.

\section{Supersymmetry variations in various dimensions}\label{app:sugras}

In this appendix we give some explicit expressions for the general formulae collected above.
In particular we state the properties of \(A_0\) and give expressions for \(B_\hal{2}\).
We only consider the dimensions \(D=4,5,7\) which are the relevant cases for chapter~\ref{chap:adsmoduli}.

\subsection*{$D=4$}

In four dimensions we have, according to Table~\ref{tab:spinors}, the choice between Majorana or Weyl spinors.
However, as explained in appendix~\ref{app:conventions} the R-symmetry is manifest only if we select the latter.
Accordingly, we choose the gravitini \(\psi^i_{M+}\) to be chiral, i.e. \(\Gamma_\ast \psi^i_{M+} = \psi^i_{M+}\).
Therefore, their charge conjugates \(\psi_{M-i} = (\psi^i_{M+})^C\) are antichiral.
The gravitini transform in the fundamental (or antifundamental representation, respectively) with respect to the R-symmetry group \(H_R\), given by
\begin{equation}\label{eq:appd4ralgebra}
H_R = \begin{cases}
\U(\cN) & \text{if}\; \cN \neq 8 \\
\SU(\cN) & \text{if}\; \cN = 8 
\end{cases} \,,
\end{equation}
where \(i = 1, \dots, \cN\).

To apply the results of the previous section we arrange \(\psi^i_{M+}\) and \(\psi_{M-i}\) in a combined column vector, and similarly for the supersymmetry parameters \(\epsilon^i_+\) and \(\epsilon_{-i}\),
\begin{equation}
\psi^i_M \rightarrow \begin{pmatrix} \psi^i_{M+} \\ \psi_{M- i} \end{pmatrix} \qquad\text{and}\qquad \epsilon^i \rightarrow \begin{pmatrix}\epsilon^i_+ \\ \epsilon_{-i} \end{pmatrix} \,,
\end{equation}
see also the discussion in appendix~\ref{app:conventions}.
Sticking to this notation, the gravitino shift matrix \(A_0\) in \eqref{eq:appgravitinovariation} reads
\begin{equation}
A_0 = \begin{pmatrix} 0 & (A_0)^{ij} \\ (A_0)_{ij} & 0 \end{pmatrix} \,,
\end{equation}
where \((A_0)^{ij} = \left((A_0)_{ij}\right)^\ast\).
This form is due to the fact that the multiplication with one \(\Gamma\)-matrix inverts the chirality of a spinor.
Moreover, the formula \eqref{eq:bilinearswap} applied to the gravitino mass term in \eqref{eq:appfermionmass} shows that \((A_0)_{ij}\) is symmetric.
In combination these properties imply that \(A_0\) is a hermitian matrix.

The dressed vector fields \(A^{\hal{2}}\) from the gravity multiplet  (i.e.~the graviphotons) are given by
\begin{equation}\label{eq:appD4graviphotons}
A^\hal{2}_M = 
\bigl(A^{[ij]}_M,A_{M[ij]}\bigr) \,,
\end{equation} 
where \(A_{M[ij]} = \bigl(A^{[ij]}_M\bigr)^\ast\).
Only for the \(\cN = 6\) there is an additional gauge field in the gravity multiplet which is a singlet with respect to the global symmetry group of the theory and hence also with respect to \(H_R\).
We denote it by
\begin{equation}
A_M^\tal{2} = \bigl(A^0_M\bigr) \,,
\end{equation}
as if it would belong to an additional vector multiplet.
This is consistent since -- as we will see below -- the corresponding field strength \(F^0\) does not enter the supersymmetry variation of the gravitini.
This field content is constructed easiest by starting with the maximal \(\cN = 8\) theory \cite{Cremmer:1978ds,Cremmer:1979up,deWit:1982bul} and then decomposing the R-symmetry according to \(\SU(8) \rightarrow \U(\cN) \times \SU(8 - \cN)\) (see e.g.~\cite{Andrianopoli:2008ea}).
The spectrum of a theory with \(\cN\) supersymmetries is obtained by keeping only those fields which transform as singlets with respect to the second factor \(\SU(8 - \cN)\).
This also explains the appearance of the additional vector field \(A^0_M\) in the \(\cN = 6\) theory, where \(A^0_M = A^{[78]}_M\) is indeed invariant under \(\SU(2)\).
Note that \eqref{eq:appD4graviphotons} implies that there are no graviphotons for \(\cN = 1\).

In the same spirit one can determine the general form of the matrices \((B_\hal{2})\) in the supersymmetry variations of the gravitini \eqref{eq:appFM}.
For the \(\cN = 8\) theory they can be read off from \cite{deWit:1982bul} and are given by
\begin{equation}\label{eq:appD4B}
B_{ij} = \begin{pmatrix} 0 & (B_{ij})^{kl} \ \\ 0 & 0 \end{pmatrix} \qquad \text{and} \qquad
B^{ij} = \begin{pmatrix} 0 & 0 \ \\ -(B^{ij})_{kl} & 0 \end{pmatrix} \,,
\end{equation}
with 
\begin{equation}\label{eq:appD4Bb}
(B_{ij})^{kl} = \tfrac{1}{\sqrt{2}} \delta^{kl}_{ij} \,.
\end{equation}
Following the above argument, these expressions also hold for all other theories with \(\cN \neq 8\).
For \(\cN = 6\) there could in principle also be a matrix \(B_0\), but 
\begin{equation}\label{eq:appD4B0}
B_0 = 0 \,,
\end{equation}
since \((B_0)^{ij} = (B_{[78]})^{ij} = 0\) for \(i,j = 1, \dots 6\) and analogously for \((B_0)_{ij}\).
This justifies to treat \(A^0\) formally not as a graviphoton \(A^\hal{2}\).
The general structure of \eqref{eq:appD4B} and \eqref{eq:appD4Bb} is determined by the requirement that they transform invariantly with respect to \(H_R\).
Moreover, we can use the \(\cN = 2\) case to fix the numerical prefactor in \eqref{eq:appD4Bb}.
For \(\cN = 2\) theories there are no spin-1/2 fermions \(\chi^{\hat a}\) in the gravity multiplet and thus the matrices \(C_\hal{2}\) do not exist.
Therefore \eqref{eq:BCanticom} uniquely fixes the factor in \(B_\hal{2}\).

\subsection*{$D=5$}

In five dimensions we are using symplectic Majorana spinors, accordingly the R-symmetry group is given by
\begin{equation}
H_R = \USp(\cN) \,,
\end{equation}
where \(\cN\) denotes the number of gravitini \(\psi^i_M\), \(i = 1, \dots \cN\), satisfying the symplectic Majorana constraint \eqref{eq:symplmajorana}.
Every pair of symplectic Majorana spinors has 8 indepedent real components, hence the admissible values for \(\cN\) are \(2, 4, 6\) and \(8\).
In particular, we use the \(\USp(\cN)\)-invariant tensor \(\Omega_{ij} = (\Omega^{ij})^\ast\) to raise or lower indices.

Applying \eqref{eq:bilinearswap} and \eqref{eq:reality} to the gravitino mass term in \eqref{eq:appfermionmass} shows that the shift matrix \((A_0)_{ij}\) is symmetric and that
\begin{equation}\label{eq:appD5A09}
(A_0)_{ij} = (A_0)_{(ij)} = -(A_0^{ij})^\ast \,.
\end{equation}
In combination with the symmetry of \(A_0\), \eqref{eq:appD5A09} implies that \(A^i_{0\,j} = \Omega^{ik} (A_0)_{kj}\) is a hermitian matrix.

For the graviphotons \(A^\hal{2}_M\) we follow a similar strategy as in four dimensions and start with the maximal theory with \(\cN = 8\), where \cite{Gunaydin:1984qu,Pernici:1985ju,Gunaydin:1985cu}
\begin{equation}\label{eq:D5N8graviphotons}
A^\hal{2}_M = A^{[ij]}_M \,,\qquad A^{ij}_M \Omega_{ij} = 0 \,.
\end{equation}
To obtain the theories with \(\cN < 8\) we decompose \(\USp(8) \rightarrow \USp(\cN) \times \USp(8-\cN)\) and keep only those fields in \eqref{eq:D5N8graviphotons} which are singlets with respect to the second factor \(\USp(8-\cN)\).
This yields
\begin{equation}\label{eq:appD5graviphotons}
A^\hal{2}_M = \bigl(A^{[ij]}_M, A^0_M\bigr) \,,\qquad A^{ij}_M \Omega_{ij} = 0 \,,
\end{equation}
so for \(\cN \neq 8\) there is an additional vector field \(A^0_M\) in the gravity multiplet which is a singlet with respect to \(H_R\).
Note that for \(\cN = 2\) there is only \(A^0_M\). 
Analogously we obtain \(B_0\) and \(B_{[ij]}\) for all \(\cN\) from starting with the expression for \(B_{[ij]}\) for the \(\cN = 8\) case.
The result reads
\begin{equation}\label{eq:appD5B}
\bigl(B_0\bigr)^k_l = \tfrac i2\sqrt{\tfrac{8-\cN}{2\cN}}\delta^k_l \,,\qquad \bigl(B_{ij}\bigr)^k_l = i \delta^k_{[i} \Omega_{j]l} + \tfrac i\cN \Omega_{ij} \delta^k_l \,.
\end{equation}
The general structure of these matrices is determined by \(\USp(\cN)\) invariance, and
as in four dimensions we can use \eqref{eq:BCanticom} to fix the numerical prefactor in \(B_0\) for \(\cN = 2\), which in turn determines the prefactors in \(B_0\) as well as in \(B_{[ij]}\) for all \(\cN\).

\subsection*{$D=7$}

In seven dimensions we are using symplectic Majorana spinors and the R-symmetry group is given by \(H_R = \USp(\cN)\), exactly as in five dimensions.
Here every pair of symplectic Majorana spinors carries 16 independent real components, so there is only the half-maximal theory with \(\cN = 2\) and the maximal theory with \(\cN =4\).
The remaining discussion is very similar to the five-dimensional case, so let us only state the essential differences.

The gravitino shift matrix \(A_0\) satisfies
\begin{equation}\label{eq:appD7A0}
(A_0)_{ij} = (A_0)_{[ij]} = (A_0^{ij})^\ast \,.
\end{equation}
Both conditions in combinations imply that \(A^i_{0\,j}\) is hermitian.

The graviphotons \(A^\hal{2}\) as well as the matrices \(B_\hal{2}\) can be obtained from the maximal \(\cN = 4\) theory \cite{Sezgin:1982gi,Pernici:1984xx}.
The graviphotons are given by
\begin{equation}\label{eq:appD7graviphotons}
A^\hal{2} = A^{(ij)}_M \,,
\end{equation}
which is valid for all values of \(\cN\), since with respect to \(\USp(4) \rightarrow \USp(\cN) \times \USp(4-\cN)\) there cannot arise any additional \(\USp(4-\cN)\) singlets from the symmetric representation.
The matrices \(B_\hal{2}\) finally read
\begin{equation}\label{eq:appD7B}
(B_{ij})^k_l = \sqrt{2} \delta^k_{(i} \Omega_{j)l} \,.
\end{equation}
Note that locally
\begin{equation}
\USp(2) = \SU(2) \cong \SO(3) \,,\qquad \USp(4) \cong \SO(5) \,.
\end{equation}
Moreover, the graviphotons transform in the respective adjoint representations, and \eqref{eq:appD7B} is an explicit expression for the generators of \(\USp(\cN)\) in the fundamental representation.


\chapter{Analysis of the Integrability Condition}\label{app:integrability}

In this appendix we outline a computation based on $\Gamma$-matrix manipulations relevant for Chapter~\ref{chap:classification}.%
\footnote{This appendix is based on \cite{Louis:2016tnz}.}
We analyze the integrability condition \eqref{eq:fluxgaugeintegrability} and argue that for all the theories listed in table~\ref{tab:pformfluxes} the term \(\hat\cH_{MN}\) is the only possible term at zeroth order in the \(\Gamma\)-matrices and therefore has to vanish in a maximally supersymmetric background.

Let us first note that all the theories in table~\ref{tab:pformfluxes} only allow for background fluxes \(F^{\hat\alpha_p}\) for one particular value of \(p\), so the expression \eqref{eq:cFM} for \(\cF_M\) simplifies as we do not have to sum over different values for \(p\).
We want to inspect \eqref{eq:fluxgaugeintegrability} term by term.
The Riemann tensor \(R_{MNPQ}\) enters only at the quadratic order in \(\Gamma\), also the third term \(\left(\hat\cD_{[M} \cF_{N]} + \D_{[M} A_0 \Gamma_{N]}\right)\) cannot contain any terms at zeroth order in \(\Gamma\) as can be directly seen from \eqref{eq:cFM} and \eqref{eq:T} with \(p > 1\).
    To analyze the remaining term in \eqref{eq:fluxgaugeintegrability} we notice that this term can only produce something of vanishing order in \(\Gamma\) from the anti-commutator of two equal powers of \(\Gamma\)-matrices, i.e.
\begin{equation}\label{eq:multgammaanticom}
\left\{\Gamma^{M_1\dots M_r},\Gamma_{N_1\dots N_r}\right\} = p!\, \delta^{[M_1}_{N_r} \dots \delta^{M_r]}_{N_1} + \dots \,,
\end{equation}
where the ellipsis denotes terms of higher order in \(\Gamma\).
The corresponding commutator  yields at least a term
quadratic
in \(\Gamma\) and also the (anti-)commutator of two different powers of \(\Gamma\)-matrices cannot give anything at zeroth order.
With this knowledge we can finally compute the last term in \eqref{eq:fluxgaugeintegrability} to find
\begin{equation}\begin{aligned}\label{eq:lastterm}
\Bigl(\left(\cF_M + A_0 \Gamma_M\right)&\left(\cF_N + A_0 \Gamma_N\right) - (M \leftrightarrow N)\Bigr) = \\
&= \bigl[\cF_M,\cF_N\bigr] + \bigl[\cF_M, A_0 \Gamma_N\bigr] - A_0 \bigl[\cF_N, A_0 \Gamma_M\bigr] + 2 A_0 A_0 \Gamma_{MN} \\
&=\frac{p^2 (p-1)!}{8 (p-1)^2} \frac{D-2p}{D-2}\left(\beta_{(p)}^2-p^2\right)\bigl[B_{\hat\alpha_p},B_{\hat\beta_p}\bigr] F^{\hat\alpha_p}_{MP_1\dots P_{p-1}} F^{\hat\beta_p\,P_{p-1}\dots P_1}_N \\
&\qquad+ 2 \, \delta_{p,2} \frac{D-3}{D-2} \bigl[B_{\hat\alpha_2}, A_0\bigr] F^{\hat\alpha_2}_{MN} + \dots \,,
\end{aligned}\end{equation}
where we suppressed the indices \((i,j,\dots)\) and the ellipsis denotes again higher order terms.
For the computation of the commutator \(\bigl[\cF_M,\cF_N\bigr]\) we
used \eqref{eq:cFM}, \eqref{eq:multgammaanticom} and 
\begin{equation}\begin{aligned}
&\bigl[B_{\hat\alpha_p} \Gamma^{M_1\dots M_r}, B_{\hat\beta_p}\Gamma^{N_1\dots N_r} \bigr] \\
&\qquad= \frac{1}{2} \left(\bigl[B_{\hat\alpha_p},B_{\hat\beta_p}\bigr] \left\{\Gamma^{M_1\dots M_r},\Gamma^{N_1\dots N_r}\right\} + \bigl\{B_{\hat\alpha_p},B_{\hat\beta_p}\bigr\} \left[\Gamma^{M_1\dots M_r},\Gamma^{N_1\dots N_r}\right] \right) \,.
\end{aligned}\end{equation}
For all the theories where \(\hat\alpha_p\) can take only one possible value the commutator \(\bigl[B_{\hat\alpha_p},B_{\hat\beta_p}\bigr]\) on the right hand side of \eqref{eq:lastterm} clearly vanishes.
Moreover, in this case \(B_{\hat\alpha_p}\) is proportional to the unit matrix, therefore also the second commutator \(\bigl[B_{\hat\alpha_2}, A_0\bigr]\) vanishes.
The only theory in table~\ref{tab:pformfluxes} for which \(\hat\alpha_p\) can take multiple values is the six-dimensional \(\cN = (2,0)\) theory.
But here \(p = D/2 = 3\) 
so that also in this case the terms at zeroth order in \(\Gamma\) vanish.

It remains to check that in odd dimensions \(D\) there are also no terms of order \(D\) in \(\Gamma\).
These could be dualized into zero order terms using \eqref{eq:gammahodgeodd}.
Since we can restrict the analysis to \(p < \frac{D}{2}\) it is clear
that such terms cannot arise from \(\hat\cD_{[M} \cF_{N]}\) or
\(\bigl[\cF_M A_0, \Gamma_N\bigr]\)
as can be seen from the definition \eqref{eq:cFM}.
The commutator \(\bigl[\cF_M,\cF_N\bigr]\) can, however, produce only terms of even order in \(\Gamma\).

\chapter{Proofs for Chapter~\ref{chap:ads}}\label{app:ads}

In this appendix we give three technical proofs omitted in chapter~\ref{chap:ads}.
In appendix~\ref{app:representationtheory} we describe the representation theoretical constraints on the gauged R-symmetry group \(H^g_R\) following from the formula \eqref{eq:AdSconditionsQP}.
In appendix~\ref{app:symP} we argue that the matrix \({(\cP_{\delta\phi})_\hal{2}}^\tbe{2}\) appearing in the variation \eqref{eq:cVvariation} of the vielbeins \(\cV^\al{2}_I\) is symmetric.
In appendix~\ref{app:Tmoduli} we show that \eqref{eq:Tmoduli} is a sufficient condition for the moduli space \(\cM_{AdS}\) to be a symmetric space of the form \(\cM_{AdS} = G_{AdS}/H_{AdS}\).

\section{Properties of the gauged R-symmetry group \texorpdfstring{$H^g_R$}{Hg\_R}}\label{app:representationtheory}

In this appendix we discuss the implications of the formula \eqref{eq:AdSconditionsQP} on the gauged subalgebra \(\h^g_R\) of the R-symmetry algebra \(\h_R\).
It is self contained and can in principle be read independently from the rest of this thesis.%
\footnote{To keep the notation simple we deviate slightly from the notation used in the main part, e.g. we use \(\alpha\) instead \(\hal{2}\) and \(A\) instead of \(A_0\).}

Let \(\h_R\) a reductive Lie-algebra and let \(\{J_A\}\), \(A= 1,\dots,\dim(\h_R)\) be its generators.
Let \(\mathbf s\) and \(\mathbf v\) be two matrix representations of \(\h_R\), such that the generators in these representations read \((J_A)_i^j\) and \((J_A)_\alpha^\beta\), with \(i,j=1,\dots, \dim({\mathbf s})\) and \(\alpha,\beta = 1, \dots, \dim({\mathbf v})\).
We furthermore demand the existence of \(\dim({\mathbf v})\) linearly independent matrices \((B_\alpha)^j_i\) satisfying
\begin{equation}\label{eq:JB}
(J_A)_\alpha^\beta B_\beta = \bigl[J_A, B_\alpha\bigr] \,,
\end{equation}
where we suppressed the indices \(i\) and \(j\).
This condition implies that \(\mathbf v\) is contained in the tensor product decomposition of \(\mathbf s \otimes \mathbf s^\ast\), where \(\mathbf s^\ast\) denotes the dual representation of \(\mathbf s\).

Let us now assume that there is a matrix \((A)_i^j\) such that \(A^2 = \id\) and such that the matrices \((\cQ_\alpha)_i^j\), defined by
\begin{equation}\label{eq:Qdef}
\cQ_{\alpha} = \bigl\{A, B_\alpha\bigr\} \,,
\end{equation}
are elements of \(\h_R\).
It follows directly from the definition that
\begin{equation}\label{eq:QA}
\bigl[\cQ_\alpha, A\bigr] = 0 \,.
\end{equation}
Moreover, the condition \(\cQ_\alpha \in \h_R\) implies that there is a matrix \(\theta_\alpha^A\) -- usually called the embedding tensor, cf.~chapter~\ref{sec:gauging} -- such that \(\cQ_\alpha = \theta_\alpha^A J_A\).
This yields in combination with \eqref{eq:JB} and \eqref{eq:QA} that
\begin{equation}
\bigl[\cQ_\alpha, \cQ_\beta\bigr] = (\cQ_\alpha)_\beta^\gamma \cQ_\gamma \,,
\end{equation}
and therefore the \(\cQ_\alpha\) span a subalgebra \(\h^g_R \subseteq \h_R\).

Let \(\x\) be the maximal subalgebra of \(\h_R\) such that \([\x,A] = 0\) and let \(\cX_\a\), \(\a = 1, \dots \dim(\x)\), be the generators of \(\x\).
We now decompose the \(\h_R\)-representations \(\mathbf s\) and \(\mathbf v\) into irreducible representations of \(\x\), i.e.
\begin{equation}\label{eq:srdecomp}
{\mathbf s} = \bigoplus_{p = 1}^N {\mathbf s}_p \,,\qquad\text{and}\qquad {\mathbf v} = \bigoplus_{s = 1}^M {\mathbf v}_s \,.
\end{equation}
Analogously, we split the indices \(i\) into \((i_p)\) and \(\alpha\) into \((\alpha_s)\).
In this frame the generators \(\cX_\a\) become block-diagonal and
\begin{equation}
(\cX_\a)_{\alpha_s}^{\beta_s} \cQ_{\beta_s} = \bigl[\cX_\a, \cQ_{\alpha_s}\bigr] \,,
\end{equation}
for every \(s \in {1, \dots, M}\).
This implies that within each irreducible representation \({\mathbf v}_s\) either all the \(\cQ_{\alpha_s}\) vanish or are all non-vanishing and linearly independent.
Therefore \(\h^g_R\) must be a subalgebra of \(\x\) such that its adjoint representation is contained in the decomposition \eqref{eq:srdecomp}.
In other words, if the adjoint representation of the maximal subalgebra \(\mathfrak{z} \subseteq \x\) which satisfies this criterion is given by
\begin{equation}\label{eq:ydecomp}
\mathrm{ad}_\mathfrak{z} = \bigoplus_{s \in Z} {\mathbf v}_s \,,\qquad Z \subseteq \{1, \dots, M\} \,,
\end{equation}
we have
\begin{equation}
\mathrm{ad}_{\h^g_R} = \bigoplus_{s \in H} {\mathbf v}_s \,,\qquad \text{for some}\; H \subseteq Z \,.
\end{equation}

Under certain conditions it is possible to argue that an element \(s \in Z\) is also necessarily in \(H\).
Let \({\mathbf v}_{s}\) be one of the summands in \eqref{eq:ydecomp} (i.e.~\(s \in Z\)) such that 
\begin{equation}\label{eq:maxcriterion}
{\mathbf v}_{s} \notin {\mathbf s}_p \otimes {\mathbf s}^\ast_q \,,\qquad\text{for}\qquad p \neq q
\end{equation}
and therefore
\begin{equation}\label{eq:Bdecomp1}
(B_{\alpha_{s}})_{i_p}^{j_q} = 0 \,,\qquad\text{for}\qquad p \neq q \,.
\end{equation}
On the other hand we must have
\begin{equation}\label{eq:Bdecomp2}
(B_{\alpha_{s}})_{i_{p'}}^{j_{p'}} \neq 0 \,,
\end{equation}
for at least one \(p' \in \{1, \dots N\}\), since we demand all \(B_\alpha\) to be non-vanishing.
Moreover, the condition \([\cX_\a, A] = 0\) enforces (after a possible change of \(i\)-basis)
\begin{equation}\label{eq:Adecomp}
A_{i_p}^{j_q} = \begin{cases}
a_p\, \delta_{i_p}^{i_q} & \text{if}\; p = q \\
0 & \text{if}\; p \neq q
\end{cases} \,,
\end{equation}
where \((a_p)^2 = 1\) for all \(p\).
Inserting \eqref{eq:Bdecomp1}, \eqref{eq:Bdecomp2} and \eqref{eq:Adecomp} into \eqref{eq:Qdef} finally yields
\begin{equation}
\cQ_{\alpha_{s}} \neq 0 \,,
\end{equation}
and therefore \(s \in H\).
Note that \eqref{eq:maxcriterion} is a sufficient criterion for \(s \in H\) but not necessary.

\section{Variation of the vielbeins}\label{app:symP}

In this appendix we show that the variation matrix \({(\cP_{\delta\phi})_\al{2}}^\be{2}\) appearing in the variation \eqref{eq:cVvariation} of the vielbeins \(\cV^\al{2}_I\), i.e.
\begin{equation}\label{eq:appcVvariation}
\cD_{\delta\phi} \cV^\al{2}_I = \cV^\be{2}_I {(\cP_{\delta\phi})_\be{2}}^\al{2} ,
\end{equation}
always satisfies the property
\begin{equation}\label{eq:appsymP}
\bigl(\cP_{\delta\phi}\bigr)_{\hal{2}\tbe{2}} = \bigl(\cP_{\delta\phi}\bigr)_{\tbe{2}\hal{2}} \,,
\end{equation}
where \((\cP_{\delta\phi})_{\hal{2}\tbe{2}} = {(\cP_{\delta\phi})_\hal{2}}^\tga{2} \delta_{\tga{2}\tbe{2}}\) and \((\cP_{\delta\phi})_{\tbe{2}\hal{2}} = {(\cP_{\delta\phi})_\tbe{2}}^\hga{2} \delta_{\hga{2}\tal{2}}\).
To show \eqref{eq:appsymP} we perform a case-by-case analysis and discuss theories with different numbers \(q\) of real supercharges separately.

\subsection*{$q > 16$}

For these theories we do not have any vector multiplets and thus
\begin{equation}
{\bigl(\cP_{\delta\phi}\bigr)_\hal{2}}^\tbe{2} = {\bigl(\cP_{\delta\phi}\bigr)_\tbe{2}}^\hal{2} = 0 \,.
\end{equation}
Therefore \eqref{eq:appsymP} is satisfied trivially.

\subsection*{$q = 16$}

For half-maximal supergravities the duality group \(G\) is of the form
\begin{equation}
G = G^* \times \SO(10-D,n_V) \,,
\end{equation}
where \(n_V\) denotes the number of vector multiplets. In most cases the first factor \(G^\ast\) is given by \(\SO(1,1)\) while in \(D=4\) dimensions it is given by \(\SU(1,1)\), due to electric-magnetic duality.
Moreover, the gauge fields transform in the vector representation of \(\SO(10-D,n_V)\).

As explained in chapter~\ref{sec:adsmoduli} the variation \({(\cP_{\delta\phi})_\al{2}}^\be{2}\) corresponds to a non-compact generator of \(G\).
However, the group \(G^\ast\) does not mix fields from different multiplets, hence it can only give rise to \({(\cP_{\delta\phi})_\hal{2}}^\hbe{2}\) and \({(\cP_{\delta\phi})_\tal{2}}^\tbe{2}\).
This in turns means that the variations \({(\cP_{\delta\phi})_\hal{2}}^\tbe{2}\) and \({(\cP_{\delta\phi})_\tal{2}}^\hbe{2}\), in which we are interested, are elements of \(\so(10-D,n_V)\).
Therefore the split-signature metric
\begin{equation}
\eta_{\al{2}\be{2}} = \begin{pmatrix}
-\delta_{\hal{2}\hbe{2}} & 0 \\
0 & \delta_{\tal{2}\tbe{2}} \\
\end{pmatrix}
\end{equation}
is invariant with respect to \({(\cP_{\delta\phi})_\hal{2}}^\tbe{2}\) and \({(\cP_{\delta\phi})_\tal{2}}^\hbe{2}\), i.e.
\begin{equation}
- {(\cP_{\delta\phi})_\tbe{2}}^\hga{2} \delta_{\hga{2}\hal{2}} + {(\cP_{\delta\phi})_\hal{2}}^\tga{2} \delta_{\tga{2}\tbe{2}} = 0 \,,
\end{equation}
which shows \eqref{eq:appsymP}.

\subsection*{$q = 12$}

Such a theory exists only in $D=4$ dimensions (remember that we restrict our analysis to $D \geq 4$).
The duality group of the four-dimensional \(\cN = 3\) supergravity is given by
\begin{equation}
G = \SU(3,n_V) \,.
\end{equation}
Since \(\SU(3,n_V)\) is a subgroup of \(\SO(6, 2n_V)\) the above arguments also apply here.

\subsection*{$q = 8$}

These theories exist in dimensions \(D = 4,5\) and \(6\).
In six dimensions, however, the vector multiplets do not contain any scalar fields, moreover, the theory does not allow for supersymmetric AdS vacua.
Therefore, it is enough to consider only the cases \(D=4\) and \(D=5\).
We discuss them separately.

In four and five-dimensional \(\cN = 2\) supergravity the scalar field manifold \(\cM\) takes the form of a product
\begin{equation}
\cM = \cM_V \times \cM_H \,,
\end{equation}
where \(\cM_V\) is spanned by the scalar fields in vector multiplets and \(\cM_H\) is spanned by the scalars in hyper multiplets.
The gauge fields \(A^\al{2}_M\) are non-trivial sections only over \(\cM_V\), we can therefore restrict our attention to this space.

In five-dimensions \(\cM_V\) is a \emph{very special real manifold} and can be described as a hypersurface of a \((n_V + 1)\)-dimensional real space with coordinates \(h^{I}\), \(I = 0, \dots, n_V\).%
\footnote{Our presentation follows \cite{Bergshoeff:2004kh}.} 
It is defined as the solution of the cubic polynomial equation
\begin{equation}
C_{IJK} h^{I} h^{J} h^{K} = 1 \,,
\end{equation}
where \(C_{IJK}\) is symmetric and constant.
This construction yields a metric \(M_{IJ}\) on the ambient space,
\begin{equation}
M_{IJ} = -2 C_{IJK} h^{K} + 3 h_{I} h_{J} \,,
\end{equation}
where \(h_{I} = C_{IJK} h^{J} h^{K}\).
This metric appears also as gauge kinetic metric in \eqref{eq:bosonicaction}.
Moreover it induces a metric \(g_{rs}\) on \(\cM_V\) via
\begin{equation}
g_{rs} = h^{I}_r h^{I}_s M_{IJ} \,,
\end{equation}
where \(h^{I}_r\) is defined as the derivatives of \(h^{I}\), i.e.
\begin{equation}\label{eq:hIr}
h^{I}_r = - \sqrt{\tfrac32} \partial_r h^{I} \,.
\end{equation}
The covariant derivatives of \(h^{I}_r\) in turn satisfy
\begin{equation}\label{eq:hIrderiv}
\nabla_r h^{I}_s = - \sqrt{\tfrac32} \left(g_{rs} h^I + T_{rst} h^{I\, t}\right) \,,
\end{equation}
with \(T_{rst} = C_{IJK} h^{I}_r h^{J}_s h^{K}_t\).
We also need the relation
\begin{equation}
M_{I J} = h_{I} h_{J} + g_{rs}  h_{I}^r h_{J}^s \,,
\end{equation}
from which it follows that we can identify the vielbeins \(\cV^\al{2}_I\) introduced in \eqref{eq:kinmatrix} with \(h_{I}\) and \(h^{I}_r\), i.e.
\begin{equation}
\cV^{\hal{2} = 0}_{I} = h_{I} \,,\qquad \cV^{\tal{2} = \al{1}}_{I} = e^{\al{1}}_r h^r_{I} \,,
\end{equation}
where \(e^{\al{1}}_r\) are the vielbeins of the metric \(g_{rs}\) \eqref{eq:scalarvielbeins}.
Notice, that  we can identify the indices \(\tal{2}\) and \(\al{1}\) since there is precisely one scalar field per vector multiplet.
Finally, comparing \eqref{eq:appcVvariation} with \eqref{eq:hIr} and \eqref{eq:hIrderiv} yields
\begin{equation}
{\bigl(\cP_{\delta\phi}\bigr)_{\hal{2} = 0}}^{\tal{2} = \al{1}} = - \sqrt{\tfrac23} \delta\phi^\al{1} \,,
\end{equation}
as well as 
\begin{equation}\label{eq:D5Pdeltaphi}
{\bigl(\cP_{\delta\phi}\bigr)_{\tal{2} = \al{1}}}^{\hal{2} = 0} = - \sqrt{\tfrac23} \delta_{\al{1}\be{1}} \delta\phi^\be{1} \,.
\end{equation}
From this \eqref{eq:appsymP} follows directly.

In four dimensions \(\cM_V\) is a \emph{special K\"ahler manifold} of complex dimension \(n_V\).
It is spanned by the complex scalars \((\phi^r, \bar\phi^{\bar r})\) and we denote its K\"ahler metric by \(g_{r\bar s}\).
A special K\"ahler manifold is characterized by the existence of a symplectic vector bundle over \(\cM_V\) and an holomorphic section \(\Omega\) on this vector bundle,%
\footnote{We follow the presentation and conventions from \cite{Andrianopoli:1996cm}.}
\begin{equation}
\Omega = \begin{pmatrix}
X^{I} \\
F_{I}
\end{pmatrix} \,,
\end{equation}
such that the K\"ahler potential \(\cK\) can be expressed as
\begin{equation}
\cK = - \ln\Bigl[i \left(\bar X^{I} F_{I} - \bar F_{I}  X^{I}\right)\Bigr] \,.
\end{equation}
Moreover one introduces 
\begin{equation}
V = \begin{pmatrix}
L^{I} \\
M_{I}
\end{pmatrix} = e^{\cK/2} \Omega = e^{\cK/2} \begin{pmatrix}
X^{I} \\
F_{I}
\end{pmatrix} \,,
\end{equation}
which satisfies
\begin{equation}\label{eq:appDbarrV}
\cD_{\bar r} V \equiv \Bigl(\partial_{\bar r} - \tfrac12 \partial_{\bar r} \cK \Bigr) V = 0 \,.
\end{equation}
The holomorphic covariant derivatives of \(V\), on the other hand, are not vanishing and one can define
\begin{equation}\label{eq:appUr}
U_r = \cD_r V = \begin{pmatrix}
f^{I}_r \\
h_{I\,r} \,.
\end{pmatrix}
\end{equation}
These objects in turn satisfy
\begin{equation}\label{eq:appDU}
\cD_r U_s = i C_{rst} g^{t \bar u} \bar U_{\bar u} \,,\qquad \cD_{\bar r} U_s = g_{\bar r s} V \,,
\end{equation}
where the precise properties of the completely symmetric tensor \(C_{rst}\) are not relevant for our further discussion.
Moreover, we need to introduce a complex, symmetric matrix \(\cN_{I J}\) which is defined by
\begin{equation}
M_{I} = \cN_{I J} L^{J} \,,\quad h_{I\,r} = \bar \cN_{I J} f^{I}_r \,.
\end{equation}
This matrix is related to the gauge kinetic matrix \(M_{I J}\) \eqref{eq:bosonicaction} via
\begin{equation}
M_{I J} = - \mathrm{Im}\, \cN_{I J} \,.
\end{equation}
The inverse of \(\mathrm{Im}\, \cN_{IJ}\) satisfies
\begin{equation}
- \frac12 \left(\mathrm{Im}\, \cN\right)^{IJ} = \bar L^{I} L^{J} + g^{r\bar s} f^{I}_r \bar f^{J}_{\bar s} \,,
\end{equation}
so we find for the (complex) inverse vielbeins \(\cV^I_\al{2}\),
\begin{equation}
\cV^{I}_{\hal{2} = 0} = \sqrt{2} \bar L^{I} \,,\qquad \cV^{I}_{\hal{2} = \bar 0} = \sqrt{2} L^{I}
\end{equation}
and
\begin{equation}
\cV^{I}_{\tal{2} = \al{1}} = \sqrt{2} e^r_{\al{1}} f^{I}_r \,,\qquad \cV^{I}_{\tal{2} = \bar \alpha_1} = \sqrt{2} \bar e^{\bar r}_{\bar\alpha_1} \bar f^{I}_{\bar r} \,,
\end{equation}
where \(e^r_{\al{1}}\) is a complex vielbein of the inverse metric \(g^{r\bar s}\), i.e.~\(g^{r\bar s} = \delta^{\al{1}\bar\beta_1} e^r_\al{1} \bar e^{\bar s}_{\bar \beta_1}\).
Thus we determine be comparing \eqref{eq:cVvariation} with \eqref{eq:appDbarrV} - \eqref{eq:appDU} that
\begin{equation}
{\bigl(\cP_{\delta\phi}\bigr)_{\hal{2} = \bar 0}}^{\tal{2} = \al{1}} = \delta\phi^\al{1} \,,\qquad {\bigl(\cP_{\delta\phi}\bigr)_{\hal{2} = \bar 0}}^{\tal{2} = \bar\alpha_1} = 0 \,,
\end{equation}
and
\begin{equation}\label{eq:D4Pdeltaphi}
{\bigl(\cP_{\delta\phi}\bigr)_{\tal{2} = \al{1}}}^{\hal{2} = 0} = 0 \,,\qquad {\bigl(\cP_{\delta\phi}\bigr)_{\tal{2} = \bar\alpha_1}}^{\hal{2} = 0} = \delta_{\bar\alpha_1 \be{1}} \delta\phi^{\be{1}} \,,
\end{equation}
as well as the respective relations for the complex conjugates.
This shows \eqref{eq:appsymP}.

\section{Symmetric moduli spaces}\label{app:Tmoduli}

In this appendix we show that the solutions of \eqref{eq:Tmoduli} span a symmetric homogeneous space, even after dividing out possible Goldstone directions.

If the scalar field space is a symmetric space \(\cM = G/H\), the candidates for moduli (denoted by \(\k_{AdS}\) \eqref{eq:fsplit}) of a maximally supersymmetric AdS solution are characterized by the conditions \eqref{eq:cosetmoduli}.
In many examples all elements of \(\k_{AdS}\) satisfy also the stronger condition \eqref{eq:Tmoduli} which in turn guarantees that they are indeed moduli.
However, a priori not every solution of \eqref{eq:cosetmoduli} is necessarily a solution of \eqref{eq:Tmoduli}, in particular the Goldstone bosons \(\k^g\) which all solve \eqref{eq:cosetmoduli} might not all be solutions of \eqref{eq:Tmoduli}.
In the following we show how to divide the space of solutions of \eqref{eq:Tmoduli} by the remaining Goldstone directions and argue that the result corresponds to a symmetric submanifold \(\cM_{AdS} \subseteq \cM\).

Let us denote the set of all solutions of \eqref{eq:Tmoduli} by \(\k^f \subseteq \k\),
\begin{equation}
\k^f = \left\{\cP \in \k : - \cP_\alpha^\beta \cT_\beta + \bigl[\cT_\alpha, \cP\bigr] = 0 \right\} \,,
\end{equation}
where \(\cT_\alpha \in \g^g\) are the generators of the gauge group \(G^g\).
Analogously we define
\begin{equation}
\h^f = \left\{\cQ \in \h : - \cQ_\alpha^\beta \cT_\beta + \bigl[\cT_\alpha, \cQ\bigr] = 0 \right\} \,,
\end{equation}
and
\begin{equation}
\g^f = \h^f \oplus \k^f \,,
\end{equation}
where the direct sum is understood only as a direct sum on the level of vector spaces.
It follows readily from their definitions that \(\h^f\) as well as \(\g^f\) are both closed with respect to the Lie bracket, i.e.~they are subalgebras of \(\h\) and \(\g\) respectively.
(Note that \(\k^f\) itself cannot be a Lie algebra (unless it is abelian) due to \([\k,\k] \subseteq \h\).)

As in \eqref{eq:kg} we define
\begin{equation}
\k^g = \mathrm{span}(\cP_\alpha) \,,\quad \h^g = \mathrm{span}(\cQ_\alpha) \,,
\end{equation}
so \(\k^g\) and \(\h^g\) are the projections of the gauge algebra \(\g^g\) onto \(\k\) and \(\h\).
Note that in general \(\h^g \oplus \k^g\) can be larger than \(\g^g\).
Moreover, as noted in the discussion below \eqref{eq:kg}, \(\k^g\) corresponds to possible Goldstone bosons, so every element in \(\k^g\) which is at the same time also an element of \(\k^f\) must not be counted as a physical modulus and therefore has to be divided out.
Remember that we argued in chapter~\ref{sec:adsmoduli} that every element of \(\k^g\) is a solution of \eqref{eq:cosetmoduli}.
However, the condition \eqref{eq:Tmoduli} is stronger than \eqref{eq:cosetmoduli} and therefore it is possible that not every element of \(\k^g\) is contained in \(\k^f\).
For this reason we furthermore define
\begin{equation}
\k^{fg} = \k^f \cap \k^g \,,\qquad \h^{fg} = \h^f \cap \h^g \,,
\end{equation}
as well as
\begin{equation}
\g^{fg} = \h^{fg} \oplus \k^{fg} \,,
\end{equation}
i.e.~\(\k^{fg}\) corresponds to those Goldstone bosons which are also solutions of \eqref{eq:Tmoduli}.
In the next step we want to show that \(\g^{fg}\) is an ideal of \(\g^f\) and thus can be safely divided out.

Let \(\cP \in \k^f\) and \(\cP' \in \k^{fg}\).
This implies that there is a \(\cQ' \in \h^g\) such that
\begin{equation}
\cT' = \cQ' + \cP' \in \g^g \,.
\end{equation}
It follows from the definition of \(\k^f\) that 
\begin{equation}
\cT'' = \bigl[\cP, \cT'\bigr] \in \g^g \,.
\end{equation}
We split \(\cT''\) according to
\begin{equation}
\cT'' = \cQ'' + \cP'' \,,\quad\mathrm{s.t.}\quad \cQ'' \in \h \,,\; \cP'' \in \k \,.
\end{equation}
Therefore
\begin{equation}
\cQ'' = \bigl[\cP, \cP' \bigr] \in \h^g \,.
\end{equation}
Moreover, \(\cP\) and \(\cP'\) are both elements of \(\g^f\) and thus \(\cQ'' \in \h^{fg}\).
This shows that
\begin{equation}
\bigl[\k^f, \k^{fg}\bigr] \subseteq \h^{fg} \,. 
\end{equation}
Analogously one can show that \(\bigl[\h^f, \k^{fg}\bigr] \subseteq \k^{fg}\), \(\bigl[\k^f, \h^{fg}\bigr] \subseteq \k^{fg}\) and \(\bigl[\h^f, \h^{fg}\bigr] \subseteq \h^{fg}\).
Therefore \(\g^{fg}\) is an ideal of \(\g^f\) and \(\h^{fg}\) is an ideal of \(\h^f\), so we can define
\begin{equation}
\g_{AdS} = \g^f / \g^{fg} \qquad\mathrm{and}\qquad \h_{AdS} = \h^f / \h^{fg} \,.
\end{equation}
If we denote the Lie groups generated by \(\g_{AdS}\) and \(\h_{AdS}\) by \(G_{AdS} \subseteq G\) and \(H_{AdS} \subseteq H\) we find that
\begin{equation}
\cM_{AdS} = \frac{G_{AdS}}{H_{AdS}} 
\end{equation}
is a symmetric space.

\chapter{The \texorpdfstring{$\cN = (1,0)$}{N = (1,0)} Superconformal Algebra and Lorentz-invariant Operators}\label{app:SCFT}

In this appendix we provide supplementary material for chapter~\ref{chap:SCFT}.
In appendix~\ref{app:superconfalgebra} we review the relevant part of the superconformal algebra \(\mathfrak{osp}(6,2|2)\) and in appendix~\ref{app:minlevel} we determine at which levels it is possible to find Lorentz-invariant descendant operators.

\section{The \texorpdfstring{$\cN = (1,0)$}{N = (1,0)} superconformal algebra}
\label{app:superconfalgebra}

In this appendix we review the relevant (anti-)commutator relations
of the six-dimensional \(\cN = (1,0)\) superconformal algebra
\(\mathfrak{osp}(6,2|2)\).
The conformal group \(\SO(6,2)\) is generated by the Lorentz generators \(M_{\mu\nu}\), the momenta \(P_\mu\), the special conformal generators \(K_\mu\) and the dilatation operator \(D\).
The generators of the R-symmetry group \(\SU(2)\) are denoted by \(R_i^j\),
where \(i,j = 1,2\).
In addition there are the supercharges \(Q^i_\alpha\), with \(\alpha = 1,\dots,4\), and the superconformal charges \(S_i^\alpha\), which together span the fermionic part of \(\mathrm{OSp}(6,2|2)\).

It is convenient to use the local isomorphism \(\SO(6) \cong \SU(4)\) to label also the generators of the conformal group in an \(\SU(4)\) covariant way,
i.e. the Lorentz generators become \(M^\alpha_\beta\) (with
$M^\alpha_\alpha=0$) and the momenta and special conformal generators become \(P_{[\alpha\beta]}\) and \(K_{[\alpha\beta]}\), respectively.

Since the commutation relations involving only bosonic operators are not relevant for our analysis and can be found for example in \cite{Minwalla:1997ka},
we only give the fermionic (anti-)commutators.
These are
\begin{equation}\begin{aligned}
\left[D, Q^i_\alpha\right] &= -\tfrac{i}{2}Q^i_\alpha \,, \\
\left[D, S_i^\alpha\right] &= \tfrac{i}{2}S_i^\alpha \,, \\
\left[M^\alpha_\beta, Q^i_\gamma\right] &= -i\left(\delta^\alpha_\gamma Q^i_\beta -\tfrac{1}{4}\delta^\alpha_\beta Q^i_\gamma\right)\,, \\
\left[M^\alpha_\beta, S_i^\gamma\right] &= i\left(\delta^\gamma_\beta S_i^\alpha -\tfrac{1}{4}\delta^\alpha_\beta S_i^\gamma\right) \,, \\
\left[R^i_j, Q^k_\alpha\right] &= -i\left(\delta^k_j Q^i_\alpha - \tfrac{1}{2}\delta^i_j Q^k_\alpha\right) \,, \\
\left[R^i_j, S_k^\alpha\right] &= i\left(\delta^i_k S_j^\alpha - \tfrac{1}{2}\delta^i_j S_k^\alpha\right) \,,
\end{aligned}\end{equation}
and
\begin{subequations}\begin{align}
\left\{Q^i_\alpha, Q^j_\beta\right\} &= \epsilon^{ij} P_{\alpha\beta} \,, \label{eq:qanticom}\\
\left\{S_i^\alpha, S_j^\beta\right\} &= \epsilon_{ij} K^{\alpha\beta} \,, \\
\left\{S_i^\alpha, Q^j_\beta\right\} &= i \left(2\delta_i^j M^\alpha_\beta - 4 \delta^\alpha_\beta R_i^j + \delta^\alpha_\beta \delta_i^j D \right) \,. 
\end{align}\end{subequations}

\section{Level of Lorentz-invariant descendant states}\label{app:minlevel}

In this appendix we discuss at which levels it is possible to find a Lorentz-invariant descendant state, starting from a superconformal primary with given \(\SO(6)\) weights \((h_1, h_2, h_3)\).
Let us denote the minimal level at which this is possible by \(N\) and notice that we will then also find Lorentz invariant states at the levels \(l = N + 4m\), \(m \in \bbN\).

The problem is conveniently analyzed in the language of \(\SU(4)\)
Young tableaux, since here \(N\) corresponds to the number of boxes
that need to be added to the diagram to fill up every of its columns
to the maximal length four. 
More generally, if we switch to an arbitrary \(\SU(n)\) Young tableau
and call the length of its \(i^{\mathrm th}\) row \(r_i\) and the length of its
\(i^{\mathrm th}\) column \(l_i\), \(N\) is given by 
\begin{equation}
N = \sum_{i=1}^{r_1} \left(n - l_i\right) \,,
\end{equation}
where the sum runs over all columns.
If we use the fact that the lengths of the columns and rows are related via
\begin{equation}
l_i = p \qquad \mathrm{for}\quad r_{p+1} < i \le r_p\ , \quad p=
1,\ldots,n-1\ ,
\end{equation}
and that the Dynkin labels \(a_i\) can by read off from the tableau by
\begin{equation}
a_i = r_i - r_{i+1} \,,
\end{equation}
where \(r_n \equiv 0\), we find
\begin{equation}
N = \sum_{i=1}^{n-1}\left(n-i\right)a_i \,.
\end{equation}
Going back to the relevant case \(n=4\) and using that \(a_1 = h_2 - h_3\), \(a_2 = h_1 + h_2\), \(a_3 = h_2 + h_3\), the result reduces to
\begin{equation}
N = 2 \left( h_1 + h_2 + h_3 \right) \,.
\end{equation}

\end{appendix}

\backmatter

\setlength{\parskip}{\oldparskip}

\addcontentsline{toc}{chapter}{\protect Bibliography}
\bibliography{references-arxiv}
\bibliographystyle{utcaps}
\markboth{}{}


\end{document}